\documentclass[a4paper,12pt]{article}
\pdfoutput=1

\usepackage{jheppub}
\usepackage[utf8]{inputenc}
\usepackage[english]{babel}

\usepackage{bm}
\usepackage{amsfonts}
\usepackage{amsthm}
\usepackage{mathrsfs}
\usepackage[mathcal]{euscript}
\usepackage{float}
\usepackage[section]{placeins}
\usepackage{soul}
\usepackage{latexsym}
\usepackage{cancel}

\newcommand{\be}{\begin{equation}}
\newcommand{\ee}{\end{equation}}

\newcommand{\benn}{\begin{equation*}}
\newcommand{\eenn}{\end{equation*}}

\newcommand{\bea}{\begin{equation}\begin{aligned}}
\newcommand{\eea}{\end{aligned}\end{equation}}

\usepackage{mathtools}

\DeclarePairedDelimiterX\braket[2]{\langle}{\rangle}{#1 \delimsize\vert #2}

\newcommand{\eps}{\varepsilon}

\usepackage{comment}
\usepackage{xcolor}

\definecolor{dgreen}{rgb}{0,0.6,0}

\definecolor{brown}{rgb}{0.56,0.22,0.22}

\begin{document}
\begin{flushright}
INR-TH-2025-024
\end{flushright}

\title{One-Loop Renormalization of Anisotropic Two-Scalar Quantum Field Theories}

\author{Dmitry S. Ageev$^{a,b}$ and Yulia A. Ageeva$^{b,c,d}$}

\affiliation{$^{a}$Steklov Mathematical Institute, Russian Academy of Sciences,\\
Gubkin str. 8, 119991 Moscow, Russia\\
  $^{b}$Institute for Theoretical and Mathematical Physics,
  M.V.~Lomonosov Moscow State University, Leninskie Gory 1,
119991 Moscow,
Russia\\
$^{c}$Institute for Nuclear Research of
         the Russian Academy of Sciences,  60th October Anniversary
  Prospect, 7a, 117312 Moscow, Russia\\
 $^{d}$Department of Particle Physics and Cosmology, Physics Faculty, M.V. Lomonosov Moscow State University, Leninskie Gory 1-2, 119991 Moscow, Russia}

\emailAdd{ageev@mi-ras.ru}
\emailAdd{ageeva@inr.ac.ru}

\abstract{
We develop a basis--covariant one--loop renormalization framework for two interacting real scalars in
$D=4-\epsilon$ with the most general two--derivative Lorentz--violating quadratic form, allowing
anisotropic spatial gradients and direction--dependent kinetic mixing, together with general cubic
and quartic interactions forming RG complete set of operators at one-loop.  In dimensional regularization with minimal subtraction we compute the full
set of one--loop UV divergences and obtain closed beta functions for quartic and cubic couplings,
masses.

The pole coefficients admit a universal spectral representation as angular averages over the
direction--dependent eigenvalues and projectors of the UV kinetic matrix; all anisotropy dependence
enters through a single  universal kernel admitting two--particle
phase--space interpretation.   We classify fixed points and fixed
manifolds and show, in particular, that anisotropy restricts the existence of the coupled
Wilson--Fisher--type fixed point. When the cross--gradients are turned on the  coefficients in beta functions are governed by six
phase--space weights  admitting interpretation in terms of encoding ``populations'' and ``coherences'' of the UV normal modes.
}
\maketitle
\newpage

\section{Introduction}

Lorentz invariance is one of the organizing principles of high--energy physics: it underlies the
locality and analyticity properties of relativistic quantum field theory (QFT), it severely constrains the spectrum of
operators, and  provides the kinematic backbone for essentially all perturbative calculations.
At the same time, there are many reasons---both conceptual and practical---to study controlled
departures from exact Lorentz symmetry.  On the one hand, Lorentz invariance may be an
\emph{emergent} symmetry of an interacting many--body system at long distances: in condensed--matter
realizations, different fields typically propagate with different ``speeds'' and the ultraviolet (UV) theory
is anisotropic, while approximate relativistic invariance can appear only in the deep infrared (IR) region
\cite{Chadha:1982ex,Volovik:2003fe,Wen:2004ym}.
On the other hand, in particle physics and quantum gravity it is natural to ask how robust the
Lorentz--invariant fixed points are under Lorentz--violating (LV) perturbations, and whether small
violations can remain technically natural under renormalization
\cite{Myers:2003fd,Collins:2004bp,GrootNibbelink:2004za}
(as explored, for instance, in effective field theory's (EFT) frameworks such as the Standard--Model Extension and its
gravitational generalizations \cite{Kostelecky:1989jp,Colladay:1996iz,Colladay:1998fq,Kostelecky:2003fs,Kostelecky:2008ts}
and in the broader phenomenological literature \cite{Mattingly:2005re,Jacobson:2005bg,Liberati:2013xla}).

There are (at least) two complementary ways in which Lorentz invariance can be violated in a
continuum field theory.  The first is to change the \emph{scaling} structure by introducing
higher spatial derivatives (Lifshitz--type theories with dynamical exponent $z\neq 1$\footnote{See for example \cite{Kachru:2008yh} and references therein.},
including the Ho\v{r}ava--Lifshitz construction in gravity)
\cite{Lifshitz:1941,Horava:2009uw,Sotiriou:2009gy,Blas:2009qj,Anselmi:2008fk,Barvinsky:2015kil,Barvinsky:2021ubv,Barvinsky:2023uir}.
The second---and in some sense the
minimal one---is to keep the standard two--derivative structure (so that the canonical scaling
is still relativistic, $z=1$) but to allow different fields, or different linear combinations of
fields, to propagate on different effective ``light cones'' (or, in Euclidean language, with
different positive--definite quadratic forms in spatial gradients and/or time derivatives).
This second class of models is ubiquitous: it appears as the low--energy theory of coupled order
parameters with different stiffnesses, as multi--field EFTs with distinct sound speeds
\cite{Garriga:1999vw,Cheung:2007st,Weinberg:2008hq},
and as toy models for ``multi--metric'' matter sectors in Lorentz--violating extensions of relativistic QFT
\cite{Coleman:1998ti,Jacobson:2000xp}.

Lorentz violation has been systematized from several complementary angles. 
In the EFT approach, the Standard--Model Extension provides a general local operator basis
\cite{Kostelecky:1989jp,Colladay:1996iz,Colladay:1998fq,Kostelecky:2003fs},
while simplified ``different limiting velocities'' parametrizations were emphasized in
\cite{Coleman:1998ti,Iengo:2009ix,Anber:2011xf}. See also classic LV electrodynamics examples such as \cite{Carroll:1989vb}; more about LV in such setups maybe also found in \cite{Rubtsov:2012kb,Satunin:2017wmk}.
Renormalization and operator mixing in LV gauge theories have been investigated in a number of sectors,
including one--loop analysis in Lorentz--violating scalar-spinor interactions \cite{Altschul:2006jj,Ferrero:2011yu,Altschul:2012xu,Altschul:2014gqa},  quantum electrodynamic (QED) -- type settings \cite{Kostelecky:2001jc, Altschul:2022isc} and quantum chronodynamics (QCD) -- in \cite{Altschul:2023vnf}.
Reviews and continually updated experimental constraints can be found in
\cite{Mattingly:2005re,Jacobson:2005bg,Liberati:2013xla,Kostelecky:2008ts}.
Naturalness issues for small LV under radiative corrections have been highlighted in
\cite{Myers:2003fd,Collins:2004bp}, with possible protection mechanisms in supersymmetric settings
\cite{GrootNibbelink:2004za}.
On the many--body side, early renormalization group (RG) arguments for the emergence of a common effective light cone go back to
\cite{Chadha:1982ex} (see also \cite{Volovik:2003fe,Wen:2004ym}),
and in Lorentz--violating gravity, controlled preferred--frame frameworks include Einstein--aether and
Ho\v{r}ava--Lifshitz models \cite{Jacobson:2000xp,Jacobson:2007ve,Horava:2009uw,Blas:2009qj,Sotiriou:2009gy}. In \cite{Affleck:2021vzo} authors consider the renormalization of velocities in non-Lorentz invariant sigma model, showing that the velocity differences flow to zero at low enough energies and it is possible to restore the Lorentz invariance.

Related setups -- where scalar perturbations propagate with reduced or field‑\\dependent sound speeds -- arise in single‑ and multi‑field inflation \cite{Cheung:2007st,Armendariz-Picon:1999hyi,Kobayashi:2010cm} and in other contexts \cite{Mironov:2018oec,Ageeva:2022asq}. The study of multi‑scalar theories with different sound speeds was proposed in Refs.~\cite{Ageeva:2022byg,Ageev:2025fqv}, where the authors analyze partial‑wave unitarity.

The present work complements these studies by providing a compact, basis--covariant one--loop renormalization
machinery tailored to multi--metric two--derivative scalar theories with direction--dependent mixing. Related renormalization analyses for spatially anisotropic QFTs include Refs.~\cite{Arefeva:1994qr,Periwal:1995wm}. Another similar setup is $O(N)$ self-interacting scalar field theory with Lorentz violation from \cite{Carvalho:2013wsa}, where a mass renormalization up to two-loop level has been proceeded.

A basic structural question in this setting is: \emph{how do interactions renormalize when the
quadratic kinetic operator is anisotropic and mixes fields in a direction--dependent way?}
Even for scalar theories, a fully systematic answer requires some care.
First, the propagator is matrix--valued in field space, and kinetic mixing terms imply that the
normal modes are not the original fields but momentum--dependent linear combinations. For example, in \cite{Bijnens:2018rqw} there is a similar to ours setup with both kinetic and mass mixing: it was investigated whether the kinetic term can be kept diagonal under renormalization group evolution, even though the fields mix.
Second, in dimensional regularization and minimal subtraction
\cite{Bollini:1972ui,Hooft:1972fi,Hooft:1973mf}
one must separate UV poles from
possible IR subtleties of massless vacuum diagrams.
Third, if one wants results that are robust under field redefinitions, it is desirable to express
UV divergences not in a particular diagonal basis but directly in terms of invariant spectral data
of the quadratic operator \cite{Kamefuchi:1961sb}.

One might naively think just to introduce anisotropy  in  spatial derivatives and study such theory with kinetic term as follows
$$
\mathcal L_{kin}^{0}
=\frac12\,(\partial_0\bm\varphi)^{\top}(\partial_0\bm\varphi)
+\frac12\,(\nabla\phi)^{\top}\mathbf C_{\phi}(\nabla\phi)
+\frac12\,(\nabla\chi)^{\top}\mathbf C_{\chi}(\nabla\chi)\nonumber.
$$
However to keep the  theory RG complete (in the presence of parity-violating couplings)  one must   consider more complicated version of such theory with addition of cross derivative term and keeping arbitrary normalization, i.e. add to the theory terms of the form
\begin{align}
\frac{1}{2}\left(\partial_0 \boldsymbol{\varphi}\right)^{\top} \mathbf{Z}_t\left(\partial_0 \boldsymbol{\varphi}\right)+(\nabla \phi)^{\top} \mathbf{Y}(\nabla \chi).\label{new terms added}
\end{align}
The latter term with gradient mixing may be not optional and arises in different physical systems, for example in superfluids solutions \cite{Andreev}.
In this work we develop a  one--loop renormalization of the theory for the simplest interacting
model that  contains all of these ingredients: two real scalar fields $\phi$ and $\chi$
in $D=4-\epsilon$ Euclidean dimensions with the most general quadratic form built from two derivatives,
including time--derivative mixing and cross--gradient mixing, together with the most general cubic
and quartic interactions.

Concretely, the kinetic terms are governed by
(i) a positive--definite $2\times 2$ matrix $\mathbf Z_t$ in field space multiplying
$\partial_0\varphi_i\,\partial_0\varphi_j$, and
(ii) three real symmetric $d\times d$ matrices $\mathbf C_\phi$, $\mathbf C_\chi$, $\mathbf Y$
multiplying spatial gradients, including the cross term
$Y^{ab}\,\partial_a\phi\,\partial_b\chi$.
We also allow a general symmetric mass matrix $\mathbf M^2$, and we keep linear couplings $j_i$
because they are generated whenever cubic couplings and masses are present.
The resulting quadratic operator defines, for every direction $\hat{\bm n}\in S^{d-1}$,
a positive--definite $2\times 2$ matrix in field space controlling the UV dispersion 
with positive eigenvalues $s_a(\hat{\bm n})$ and orthogonal projectors $\Pi_a(\hat{\bm n})$.
Physically, $v_a(\hat{\bm n})\equiv \sqrt{s_a(\hat{\bm n})}$ are the \emph{directional sound speeds}
(or, in Lorentzian signature, limiting velocities) of the UV normal modes.

We show that the one--loop UV poles of the theory can be
written in a remarkably universal form.  At $D=4-\epsilon$ all one--loop divergences are controlled by
two ``master integrals'': the tadpole (which carries one insertion of the mass operator in minimal subtraction (MS) renormalization scheme) and the
bubble at zero external momentum.  After performing the frequency integral, both reduce to angular
averages over $\hat{\bm n}$ dressed by a single positive kernel
\begin{equation}
\mathcal K(s_a,s_b)
\;\equiv\;
\frac{1}{\sqrt{s_a}\sqrt{s_b}\big(\sqrt{s_a}+\sqrt{s_b}\big)}
=
\frac{1}{v_a\,v_b\,(v_a+v_b)}.
\label{eq:intro_kernel}
\end{equation}
This kernel admits a direct physical interpretation: it is the two--particle energy denominator (and,
equivalently, the two--particle phase--space density per logarithmic interval in the RG scale) for intermediate UV modes with
linear dispersion $E_a(\mathbf p)=|\mathbf p|\,v_a(\hat{\bm n})$.
Anisotropy and mixing enter only through the directional velocities $v_a(\hat{\bm n})$ and through the
``coherence factors'' encoded in $\Pi_a(\hat{\bm n})$, which determine how strongly the UV normal modes
overlap with the original fields.

\subsection*{Summary of results}

The paper contains both general structural results and explicit fixed--point analyses in consistent
one--loop truncations.  The main results are:

\begin{enumerate}
\item \textbf{General one--loop MS renormalization for a two--scalar anisotropic QFT.}
We compute the complete set of one--loop UV divergences in dimensional regularization with minimal
subtraction near $D=4-\epsilon$ for the theory \eqref{eq:renorm L}--\eqref{eq:L3L4}.
All counterterms and beta functions are expressed in compact tensor form in terms of two finite objects:
the tadpole residue $\mathcal T_{ij}$ and the bubble residue $\mathcal B_{ij;kl}$.

\item \textbf{Spectral representation of UV pole coefficients.}
We show that both $\mathcal T$ and $\mathcal B$ can be written as angular averages over the UV spectral
data of some operator $\mathbf S(\hat{\bm n})$:
\begin{equation}
\mathcal B_{ij;kl}
=
\frac{1}{4\pi^2}\Big\langle
\sum_{a,b}\mathcal K\!\big(s_a(\hat{\bm n}),s_b(\hat{\bm n})\big)\;
\mathbf P_a(\hat{\bm n})_{ik}\,\mathbf P_b(\hat{\bm n})_{jl}
\Big\rangle_{S^2},
\qquad
\mathbf P_a\equiv \mathbf Z_t^{-1/2}\Pi_a\,\mathbf Z_t^{-1/2},
\end{equation}
and similarly for $\mathcal T$ with a single insertion of the mass operator (see
eqs.~\eqref{T answer} and \eqref{bubble answer} in the main text).
This makes the dependence on the anisotropy data $(\mathbf Z_t,\mathbf C_{\phi,\chi},\mathbf Y)$ explicit
and basis--covariant.

\item\textbf{Locality of one--loop poles and non--running of anisotropy matrices in MS.}
A key structural simplification is that in $D=4-\epsilon$ the one--loop UV poles in two--point functions
are momentum--independent (compare the standard Lorentz--invariant $\phi^4$ case, where wavefunction
renormalization starts at two loops in MS; see e.g.~\cite{ZinnJustin:2002ru,Amit:2005cd}).
As a consequence, in dimensional regularization (DR)+MS at one loop
\begin{equation}
\beta_{\mathbf Z_t}=\beta_{\mathbf C_\phi}=\beta_{\mathbf C_\chi}=\beta_{\mathbf Y}=0,
\qquad
\text{(one loop, MS)}
\end{equation}
while the couplings $(\lambda_{ijkl},h_{ijk},M^2_{ij},j_i)$ do run. Notice, that while there are no divergent parts to get non-vanishing beta functions  finite pieces may be generated at one-loop level and this is the reason to include them obtaining RG closed set of operators.
In other words, at this order the anisotropy data act as \emph{external parameters} that weight the RG
flow of interactions but do not themselves evolve in MS.  Finite parts, other schemes, or higher loops
can induce running of these matrices; our formalism is designed to extend to those settings.

\item \textbf{Closed form beta functions in terms of $\mathcal B$ and $\mathcal T$.}
We obtain universal tensor beta functions,
\begin{align}
\beta_{\lambda_{ijkl}}
&=
-\epsilon\,\lambda_{ijkl}
+\nonumber\\
&+\frac12\Big(
\lambda_{ijmn}\,\mathcal B_{mn;pq}\,\lambda_{klpq}
+\lambda_{ikmn}\,\mathcal B_{mn;pq}\,\lambda_{jlpq}
+\lambda_{ilmn}\,\mathcal B_{mn;pq}\,\lambda_{jkpq}
\Big),\\
\beta_{h_{ijk}}
&=
-\Bigl(1+\frac{\epsilon}{2}\Bigr)h_{ijk}\nonumber\\
&+\frac12\Big(
\lambda_{ijmn}\,\mathcal B_{mn;pq}\,h_{kpq}
+\lambda_{ikmn}\,\mathcal B_{mn;pq}\,h_{jpq}
+\lambda_{jkmn}\,\mathcal B_{mn;pq}\,h_{ipq}
\Big),
\end{align}
together with the mass and linear beta functions \eqref{eq:beta_m2} and \eqref{eq:beta_j}.
These expressions make it transparent how anisotropy enters: \emph{only} through the residues
$\mathcal B$ and $\mathcal T$.

\item \textbf{Three ``weights'' for $\mathbf Z_t=\mathbf 1$ and $\mathbf Y=0$ and their meaning.}
In the truncation with no kinetic mixing in field space,
\begin{equation}
\mathbf Z_t=\mathbf 1,\qquad \mathbf Y=0,
\end{equation}
the field--space projectors are constant and all anisotropy dependence collapses into three scalar
weights
\begin{align}
&J_{11}=\Big\langle(\hat{\bm n}^{\top}\mathbf C_1\hat{\bm n})^{-3/2}\Big\rangle_{S^2},\qquad
J_{22}=\Big\langle(\hat{\bm n}^{\top}\mathbf C_2\hat{\bm n})^{-3/2}\Big\rangle_{S^2},\\
&J_{12}=2\Big\langle \mathcal K(s_1(\hat{\bm n}),s_2(\hat{\bm n}))\Big\rangle_{S^2},
\end{align}
with $\mathbf C_1\equiv \mathbf C_\phi$ and $\mathbf C_2\equiv \mathbf C_\chi$.
They have a clean interpretation: they are the \emph{UV two--particle phase--space weights} of the
normal modes, averaged over directions, and they reduce to the familiar isotropic factors when
$\mathbf C_a=c_a\mathbf 1$.
We show that in $d=3$ the diagonal weights are purely geometric,
$J_{aa}=(\det\mathbf C_a)^{-1/2}$, while the mixed weight admits an interpolating determinant
representation
\begin{equation}
J_{12}=\int_0^1 d\tau\,\Big(\det\big[(1-\tau)\mathbf C_1+\tau\mathbf C_2\big]\Big)^{-1/2},
\end{equation}
and obeys the universal inequality $J_{12}^2\le J_{11}J_{22}$.
Introducing the mismatch ratio
$\rho\equiv J_{12}/\sqrt{J_{11}J_{22}}\in(0,1]$ makes many RG statements manifest. In the RG closed sector corresponding to parity-symmetric quartic couplings case  beta functions have simple form 
$$
\begin{aligned}
\beta_{\lambda_1} & =-\epsilon \lambda_1+\frac{3}{16 \pi^2}\left[J_{11} \lambda_1^2+J_{22} \lambda_3^2\right], \\
\beta_{\lambda_2} & =-\epsilon \lambda_2+\frac{3}{16 \pi^2}\left[J_{22} \lambda_2^2+J_{11} \lambda_3^2\right], \\
\beta_{\lambda_3} & =-\epsilon \lambda_3+\frac{1}{16 \pi^2}\left[J_{11} \lambda_1 \lambda_3+4 J_{12} \lambda_3^2+J_{22} \lambda_2 \lambda_3\right],
\end{aligned}
$$
structurally similar to the isotropic case.
\item \textbf{Fixed points, existence bounds, and  cubic couplings.}
In the parity--symmetric quartic sector ($\lambda_{4,5}=0$) we find coupled fixed points generalizing the
standard two--scalar Wilson--Fisher point \cite{Wilson:1971dc,Wilson:1971vs,WilsonFisher:1971dc}
and the coupled--order--parameter multicritical analyses in $O(n_1)\oplus O(n_2)$--type theories
\cite{Aharony:1973zz,Pelissetto:2000ek,Calabrese:2002bm}.
Their very existence depends on anisotropy through $\rho$:
the IR--stable coupled point exists only for sufficiently small mismatch,
$\rho\ge \sqrt{3}/2$ (equivalently $4J_{12}^2\ge 3J_{11}J_{22}$), and disappears when the mismatch is
too large.
In the fully coupled quartic sector ($\lambda_{4,5}\neq 0$) the one--loop equations admit a fixed line
(a marginal direction at this order), which we analyze and show to be of saddle type.
Cubic couplings are relevant already at tree level in $4-\epsilon$ dimensions, and at one loop the cubic
flow remains linear in $h$ (see e.g.~\cite{Amit:2005cd,ZinnJustin:2002ru}).
In particular, within the perturbative $4-\epsilon$ expansion there is no
interacting fixed point with $h^*\neq 0$; cubic perturbations destabilize the quartic critical loci
unless they are tuned to zero (or forbidden by a symmetry).

\item \textbf{Six ``weights'' when $\mathbf Y\neq 0$ and the interpretation of mixing.}
When cross--gradients are turned on ($\mathbf Z_t=\mathbf 1$, $\mathbf Y\neq 0$), the UV eigenmodes are
direction--dependent mixtures of $\phi$ and $\chi$ and the bubble residue no longer reduces to three
numbers.  Instead, six independent angular weights (eq.~\eqref{eq:Jsix_def}) control the quartic flow.
These weights admit a natural interpretation as phase--space--weighted moments of \emph{populations} and
\emph{coherences} of the UV normal modes (the projector entries), and they quantify precisely how kinetic
mixing opens additional virtual channels in one--loop scattering.  In the isotropic homogeneous limit
$\mathbf C_{\phi,\chi}\propto \mathbf 1$, $\mathbf Y\propto \mathbf 1$ they reduce to elementary functions
of the two eigen--speeds and the global mixing angle.
\end{enumerate}

$\,$

Beyond providing a compact set of beta functions for a particular two--scalar model, we view the main
utility of the present analysis as structural.

The spectral--decomposition method packages anisotropy into invariant data
$\{s_a(\hat{\bm n}),\Pi_a(\hat{\bm n})\}$ and isolates a universal kernel \eqref{eq:intro_kernel}.
This is well suited for systematic generalizations: more fields, other interaction tensors, and
ultimately gauge or fermionic sectors where species propagate on different effective quadratic forms.

Also many continuum descriptions of phase transitions involve multiple scalar order parameters with distinct
stiffness matrices and gradient mixing.  Our weights provide a sharp measure of ``phase--space overlap''
between such sectors and determine when coupled Wilson--Fisher--type fixed points can exist.

$\,$

An important thing to notice is that even though in MS at one loop the anisotropy matrices do not run, the interaction flow is already
nontrivially reshaped by anisotropy through the weights.
This clarifies what part of ``emergent Lorentz'' questions is scheme-- or loop--order--sensitive: the
\emph{distribution of UV phase space} among channels is already visible at one loop, while the actual
running of velocities is deferred to higher loops and/or non--minimal schemes.

The fact that the same kernel $\mathcal K$ controls both tadpoles and bubbles makes it easy to build
intuition: $J$--weights are not arbitrary integrals but encode how many UV intermediate states are
available in each channel, dressed by mixing coherence factors.  This viewpoint is especially helpful
when $\mathbf Y\neq 0$, where direction--dependent mixing makes a naive ``two speeds'' picture inadequate.

$\,$

This paper is organized as follows. In Sec.~\ref{sec:master} we evaluate the one--loop master integrals (tadpole and bubble) in a form adapted
to a matrix--valued anisotropic propagator and extract their UV pole parts in DR+MS
\cite{Bollini:1972ui,Hooft:1972fi,Hooft:1973mf}.
In Sec.~\ref{sec:beta_MS} we assemble these results into compact tensor beta functions for the general
theory.  In Sec.~\ref{sec:partic} we analyze several consistent truncations, including the decoupled
gradient limit $\mathbf Y=0$ (where anisotropy enters through three weights) and the mixed gradient case
$\mathbf Y\neq 0$ (where six weights appear), and we discuss the associated fixed points and their
stability.  Technical derivations and useful identities for the weights are collected in the appendices.

\section{Setup: theory and its decomposition}
\subsubsection*{The Lagrangian}
In this paper we consider the theory of two real interacting scalar fields $\phi$ and $\chi$ in $D \equiv d+1$ Euclidean spacetime dimensions. The Lagrangian under consideration is given by
\begin{align}
\label{eq:renorm L}
\mathcal L
&=\frac12\,(\partial_0\bm\varphi)^{\top}\,\mathbf Z_t\,(\partial_0\bm\varphi)
+\frac12\,(\nabla\phi)^{\top}\mathbf C_{\phi}(\nabla\phi)
+\frac12\,(\nabla\chi)^{\top}\mathbf C_{\chi}(\nabla\chi)\nonumber\\
&\quad+(\nabla\phi)^{\top}\mathbf Y(\nabla\chi)
+\frac12\,\bm\varphi^{\top}\mathbf M^2\,\bm\varphi
+\mathcal L_1+\mathcal L_3+\mathcal L_4\nonumber\\
&\equiv \mathcal L_{\rm free}+\mathcal L_{\rm int},
\end{align}
where\footnote{Equivalently, the quadratic derivative terms could be written in index notation as
$
\mathcal L_{kin}
=\frac12\,Z_t^{ij}\,\partial_0\varphi_i\,\partial_0\varphi_j
+\frac12\,C_{\phi}^{ab}\,\partial_a\phi\,\partial_b\phi
+\frac12\,C_{\chi}^{ab}\,\partial_a\chi\,\partial_b\chi
+Y^{ab}\,\partial_a\phi\,\partial_b\chi$.
} $\bm\varphi\equiv(\phi,\chi)^{\top}$ and the interactions have the form
\begin{subequations}
\label{eq:L3L4}
\begin{align}
\mathcal L_1 &= \mu^{\frac{D+2}{2}}\,(j_\phi\,\phi + j_\chi\,\chi),\\
\mathcal L_3 &=\mu^{\frac{6-D}{2}}\Big(\frac{h_1}{3!}\,\phi^3 + \frac{h_2}{3!}\,\chi^3
+\frac{h_3}{2}\,\phi^2\chi + \frac{h_4}{2}\,\phi\chi^2\Big),\\
\mathcal L_4 &=\mu^{4-D}\Big(\frac{\lambda_1}{4!}\,\phi^4 + \frac{\lambda_2}{4!}\,\chi^4
+\frac{\lambda_3}{4}\,\phi^2\chi^2 + \frac{\lambda_4}{6}\,\phi^3\chi + \frac{\lambda_5}{6}\,\phi\chi^3\Big).
\end{align}
\end{subequations}
Interaction part could be also rewritten using compact notation
\begin{equation}
\label{renorm int}
\mathcal{L}_1+\mathcal{L}_3+\mathcal{L}_4
=\mu^{\frac{D+2}{2}}\, j_i\,\varphi_i
+\frac{\mu^{\frac{6-D}{2}}}{3!} \,h_{i j k}\,\varphi_i \varphi_j \varphi_k
+\frac{\mu^{4-D}}{4!}\,\lambda_{i j k l}\,\varphi_i \varphi_j \varphi_k \varphi_l .
\end{equation}
The counterterms for renormalization in this theory have the form
\begin{gather}\label{counter_terms}
\mathcal{L}_{\mathrm{ct}}=
\frac{1}{2}(\partial_0 \bm\varphi)^{\top} \delta \mathbf{Z}_t(\partial_0 \bm\varphi)
+\frac{1}{2}(\nabla \phi)^{\top} \delta \mathbf{C}_{\phi}(\nabla \phi)
+\nonumber\\+\frac{1}{2}(\nabla \chi)^{\top} \delta \mathbf{C}_{\chi}(\nabla \chi)
+(\nabla \phi)^{\top} \delta \mathbf{Y}(\nabla \chi)
+\frac12\,\bm\varphi^{\top}\delta \mathbf{M}^2\,\bm\varphi
+\nonumber\\+\mu^{\frac{D+2}{2}}\,\delta j_i\,\varphi_i
+\frac{\mu^{\frac{6-D}{2}}}{3!}\,\delta h_{ijk}\,\varphi_i\varphi_j\varphi_k
+\frac{\mu^{4-D}}{4!}\,\delta\lambda_{ijkl}\,\varphi_i\varphi_j\varphi_k\varphi_l .
\end{gather}

We use $i,j\in\{1,2\}$ for field-space indices and $a,b\in\{1,\dots,d\}$ for spatial indices.  The matrices in \eqref{eq:renorm L} characterize the anisotropy and read
\begin{itemize}
\item \(\mathbf{Z}_t\) is a \textbf{\(2\times 2\)} positive definite matrix in field space. Positivity is required for a well‑defined Euclidean path integral and for stability. This matrix is defined as
\begin{align}
   \mathbf{Z}_t \equiv \begin{pmatrix}
        z_{\phi\phi}\;\;\; z_{\phi \chi}\\
        z_{\phi \chi}\;\;\; z_{\chi \chi}
    \end{pmatrix},
    \label{Z matrix def}
\end{align}
in a field space mixing the time derivatives  and  allowing cross time‑derivative mixing \(\partial_0\phi\,\partial_0\chi\). We denote the eigenvalues of this matrix as $z_1$ and $z_2$ in what follows. 
\item \(\mathbf{C}_{\phi},\mathbf{C}_{\chi}\) are  ``sound‑speed''\footnote{We tacitly  assume that they are sound-speed matrices in Lorentzian signature; in the Euclidean signature they could be called stiffness matrices.} real symmetric positive‑definite \textbf{\(d\times d\)} matrices acting on spatial indices. For example, if one considers $d=2$ case, then $\mathbf{C}_{\phi}$ is
\begin{align}
\label{C matrix def}
    \mathbf{C}_{\phi} = \begin{pmatrix}
        C^{(\phi)}_{xx} \;\;\; C^{(\phi)}_{xy}\\
        C^{(\phi)}_{xy}\;\;\; C^{(\phi)}_{yy}
    \end{pmatrix}.
\end{align}
\item $\mathbf Y$ is a real symmetric $d\times d$ matrix that mixes spatial gradients, i.e.
$Y^{ab}\,\partial_a\phi\,\partial_b\chi$.

\item \(M_{ij}^2\) is a real symmetric \textbf{\(2\times 2\)} mass matrix and it has similar to \eqref{Z matrix def} form.
\item $j_i$ are linear couplings. They are generated by renormalization whenever cubic couplings
and masses are present (one-point 1PI divergences), so they must be included formally for closure of RG flow.

\end{itemize}
 We keep the most general quadratic form, including unusual terms like $\mathbf{Z}_t$ and $\mathbf{Y}$, because it provides RG closed basis at least at one loop level. 

$\,$

Even if the spatial-gradient mixing matrix $\mathbf{Y}$ is set to zero at tree level, it is generically generated at one loop as a finite contribution to the off-diagonal two-point function. In particular, mixed cubic interactions $\phi^2 \chi$ and $\phi \chi^2$ induce a momentum-dependent selfenergy $\Sigma_{\phi \chi}(p)$ whose low-momentum expansion contains a term proportional to $p_a p_b$, corresponding to the operator $(\nabla \phi)^{\top} \mathbf{Y}_{\text {eff }}(\nabla \chi)$. Near $D=4-\varepsilon$, the UV divergent part of $\Sigma_{\phi \chi}$ is momentum independent and renormalizes only the mass matrix, while the gradient mixing arises solely from finite terms. In the special case of equal velocities and isotropic kinetic terms, the generated contribution is proportional to $p^2=p_0^2+\mathbf{p}^2$ and can be removed by a linear field redefinition; however, in the general anisotropic theory it represents an independent operator. Consequently, inclusion of $\mathbf{Y}$ is required for closure of the effective action under renormalization.

$\,$

 In the present one-loop DR+MS analysis near $D=4-\eps$, the UV $1/\eps$ poles in two-point functions are
momentum independent, hence they renormalize only $\mathbf M^2$ and not
$\mathbf Z_t$, $\mathbf C_{\phi,\chi}$, or $\mathbf Y$.
 The form of counterterms matrices $\delta \mathbf{Z}_t$, $\delta \mathbf{C}_{\phi,\chi}$, $\delta \mathbf{Y}$, and $\delta M_{ij}^2$ is defined by the term which this counterterm renormalize.

Next we turn to the interaction part of the renormalized Lagrangian \eqref{renorm int}. Here we have the totally symmetric tensors \(h_{ijk}=h_{(ijk)}\) and \(\lambda_{ijkl}=\lambda_{(ijkl)}\), where \(i,j,k,l\in\{1,2\}\). If we write \eqref{renorm int} explicitly, we arrive at  expression \eqref{eq:L3L4}
where tensor notations of \eqref{renorm int} is related to \eqref{eq:L3L4} as
\begin{subequations}
\begin{align}
h_{111}\equiv h_1,\;\; h_{222}\equiv h_2,\;\; h_{112}=h_{121}=h_{211}\equiv h_3,
\;\; h_{122}=h_{212}=h_{221}\equiv h_4,
\end{align}
and
\begin{equation}
\lambda_{1111}=\lambda_1,\quad \lambda_{2222}=\lambda_2,\quad
\lambda_{1122}=\lambda_3,\quad
\lambda_{1112}=\lambda_4,\quad
\lambda_{1222}=\lambda_5,
\end{equation}
\end{subequations}
with all components related by permutations of $(i,j,k,l)$ are equal by total symmetry of $\lambda_{ijkl}$.

\subsubsection*{The propagator}
 For Fourier transform of each field component given by
\begin{equation}
\varphi_i(t,\mathbf x)=\int\!\frac{d\omega}{2\pi}\,\frac{d^d\mathbf p}{(2\pi)^d}\,e^{i(\omega t+\mathbf p\cdot\mathbf x)}\,\varphi_i(\omega,\mathbf p)\equiv\int_p e^{ip\cdot x}\varphi_i(p),
\end{equation}
where $p=(\omega,\mathbf p)$ and $\int_p\equiv\int\frac{d^Dp}{(2\pi)^D}$   the action with the free Lagrangian from \eqref{eq:renorm L} reads as
\begin{align}
\mathcal{S}_{\text{free}}
=\int d^Dx\,\mathcal{L}_{\text{free}}
=\frac12\int_p \bm\varphi(-p)^{\top}\,\mathbf D(p)\,\bm\varphi(p).
\end{align}
Here the \textbf{\(2\times2\)} field space kernel is
\begin{equation}
\label{eq:kernelD}
\mathbf D(p)\equiv\omega^2\,\mathbf Z_t\;+\;\mathbf Q(\mathbf p)\;+\;\mathbf M^2,
\end{equation}
where \(\mathbf Q(\mathbf p)\) contains all spatial-gradient contributions in field space
\begin{equation}
\mathbf Q(\mathbf p)\equiv\sum_{a,b=1}^d p_a p_b
\begin{pmatrix}
C^{(\phi)}_{ab} & Y_{ab} \\
Y_{ab} & C^{(\chi)}_{ab}
\end{pmatrix}.
\end{equation}
For later use it is convenient to define \(\hat{\mathbf n}\equiv \mathbf p/|\mathbf p|\), $\hat{\mathbf{n}} \in S^{d-1}$ for each direction, so that
\begin{equation}
\mathbf C(\hat{\mathbf n})\;\equiv\;\hat n_a
\begin{pmatrix}
C^{(\phi)}_{ab} & Y_{ab} \\
Y_{ab} & C^{(\chi)}_{ab}
\end{pmatrix}\hat n_b,\quad \mathbf Q(\mathbf p)=|\mathbf p|^2\,\mathbf C(\hat{\mathbf n}),
\end{equation}
where for every direction $\hat{\mathbf{n}}$ the matrix $\mathbf C(\hat{\mathbf n})$ is also symmetric positive-definite to avoid gradient instabilities. For each direction $\hat{\bm n}\in S^{d-1}$ we assume the $2\times2$ matrix
$\mathbf C(\hat{\bm n})$ is positive-definite, i.e.
\be
\hat{\bm n}^{\top}\mathbf C_\phi\,\hat{\bm n}>0,\qquad \hat{\bm n}^{\top}\mathbf C_\chi\,\hat{\bm n}>0 
\qquad
\det \mathbf C(\hat{\bm n})>0.
\ee
Thus, within these notations the propagator has the form
\be
\label{eq:Gmatr}
\mathbf{G}(p)=\mathbf{Z}_t^{-1 / 2}\left(\omega^2 \mathbf{1}+\mathbf{K}\right)^{-1} \mathbf{Z}_t^{-1 / 2},\;\;\; \text{with}\;\;\;  \mathbf{K}(\mathbf p)\equiv\mathbf Z_t^{-1/2}[\mathbf Q(\mathbf p)+\mathbf M^2]\mathbf Z_t^{-1/2},
\ee 
with $\mathbf{Z}_t^{1 / 2}$ being a square root of the matrix $\mathbf{Z}_t$, which is well-defined since $\mathbf{Z}_t$ is symmetric and positive-definite, as it was mentioned above.  Here $\mathbf 1$ denotes the $2\times2$ identity in field space. We take $\mathbf Z_t^{1/2}$ to be the principal symmetric square root, so that
$\mathbf Z_t^{1/2}=\mathbf Z_t^{1/2\,\top}$.


\subsubsection*{Eigenvalue decomposition of propagator}
The general form of propagator \eqref{eq:Gmatr} is cumbersome for loop computations and
it is therefore useful to diagonalize the field-space kernel and
it is convenient to do this via the spectral decomposition of the $2\times2$ field-space operator. To this end, we first introduce the following matrices
\be
\mathbf{S}(\hat{\mathbf{n}}) \equiv \mathbf{Z}_t^{-1 / 2} \mathbf{C}(\hat{\mathbf{n}}) \mathbf{Z}_t^{-1 / 2},
\ee
as well as
\begin{align}
    \mathbf{L} \equiv \mathbf{Z}_t^{-1 / 2} \mathbf{M}^2 \mathbf{Z}_t^{-1 / 2},
\end{align}
so that $\mathbf{K}$ from \eqref{eq:Gmatr} reads
\begin{equation}
\label{K in S and L}
\mathbf{K}(\mathbf{p})=|\mathbf{p}|^2 \mathbf{S}(\hat{\mathbf{n}})+\mathbf{L}.
\end{equation}
The matrix $\mathbf K(\mathbf p)$ is real and symmetric for every $\mathbf p$.
Under the stability assumptions $\mathbf Z_t=\mathbf Z_t^{\top}>0$ and
$\mathbf C(\hat{\mathbf n})=\mathbf C(\hat{\mathbf n})^{\top}>0$ for all directions
$\hat{\mathbf n}\in S^{d-1}$, the matrix
$\mathbf S(\hat{\mathbf n})=\mathbf Z_t^{-1/2}\mathbf C(\hat{\mathbf n})\mathbf Z_t^{-1/2}$
is positive definite, and therefore
\begin{equation}
\mathbf K(\mathbf p)=|\mathbf p|^2\,\mathbf S(\hat{\mathbf n})+\mathbf L
\qquad(\hat{\mathbf n}=\mathbf p/|\mathbf p|)
\end{equation}
This UV regime is the only one relevant for the extraction of one--loop MS poles\footnote{If the IR spectrum of $\mathbf K(\mathbf p)$ is not strictly positive for some parameters
(e.g.\ due to $\mathbf M^2$), one may temporarily work in the symmetric phase where $\mathbf M^2>0$,
or introduce an auxiliary IR regulator $\mathbf L\to \mathbf L+\delta^2\mathbf 1$ with $\delta>0$.
The MS pole parts are local and UV--controlled, hence independent of the IR regulator and can be
analytically continued back.}.
Accordingly, we may use the spectral decomposition of $\mathbf K(\mathbf p)$
in the UV to define matrix functions such as $\mathbf K^{-1/2}$.
There exist orthonormal eigenvectors $\mathbf{u}_a(\mathbf{p})$ and positive eigenvalues $\kappa_a(\mathbf{p})>0$ (for $a=1,2$, since $\mathbf{K}(\mathbf{p})$ is a $2\times 2$ matrix in a field space) such that

\begin{equation}
\mathbf{K}(\mathbf{p}) \mathbf{u}_a(\mathbf{p})=\kappa_a(\mathbf{p}) \mathbf{u}_a(\mathbf{p}), \quad \mathbf{u}_a(\mathbf{p})^{\top} \mathbf{u}_b(\mathbf{p})=\delta_{a b}, 
\end{equation}
and one may form the Euclidean spectral projectors using  eigenvectors $\mathbf{u}_a$ as
\be
\tilde{\boldsymbol{\Pi}}_a(\mathbf{p}) \equiv \mathbf{u}_a(\mathbf{p}) \mathbf{u}_a(\mathbf{p})^T, \quad \sum_{a=1}^2 \tilde{\boldsymbol{\Pi}}_a=\mathbf{1}, \quad \tilde{\boldsymbol{\Pi}}_a \tilde{\boldsymbol{\Pi}}_b=\delta_{a b} \tilde{\boldsymbol{\Pi}}_a .
\ee
Then
\be
\label{K}
\mathbf{K}(\mathbf{p})=\sum_{a=1}^2 \kappa_a(\mathbf{p}) \tilde{\boldsymbol{\Pi}}_a(\mathbf{p}),
\ee
and so for the propagator one has\footnote{The matrix function $f(\mathbf K)$  (e.g.\ $\mathbf K^{-1/2}$ or $e^{-t\mathbf K}$),
 is defined by 
if $\mathbf K(\mathbf p)=\sum_{a=1}^2 \kappa_a(\mathbf p)\,\tilde{\boldsymbol{\Pi}}_a(\mathbf p)$,
then
$
f(\mathbf K(\mathbf p))=\sum_{a=1}^2 f\!\big(\kappa_a(\mathbf p)\big)\,\tilde{\boldsymbol{\Pi}}_a(\mathbf p).$
In particular, in the UV region (large $|\mathbf p|$) the eigenvalues $\kappa_a(\mathbf p)$ are positive
under the assumptions $\mathbf Z_t>0$ and $\mathbf C(\hat{\mathbf n})>0$.
}
\begin{align}
\label{resolut G}
\mathbf{G}(p)=\left[\sum_{a=1}^2 \frac{\mathbf{Z}_t^{-1 / 2}\tilde{\mathbf{\Pi}}_a(\mathbf{p})\mathbf{Z}_t^{-1 / 2}}{\omega^2+\kappa_a(\mathbf{p})}\right] \equiv \sum_{a=1}^2 \frac{\tilde{\mathbf{P}}_a(\mathbf{p})}{\omega^2+\kappa_a(\mathbf{p})},
\end{align}
where we introduce for the simplicity
\begin{align}
\label{eq:Ptilde}
\tilde{\mathbf{P}}_a(\mathbf{p}) \equiv \mathbf{Z}_t^{-1 / 2}\tilde{\mathbf{\Pi}}_a(\mathbf{p})\mathbf{Z}_t^{-1 / 2}.
\end{align}
Note that $\tilde{\mathbf P}_a(\mathbf p)=\mathbf Z_t^{-1/2}\tilde{\bm\Pi}_a(\mathbf p)\mathbf Z_t^{-1/2}$
is rank~1 but is not a projector in general:
$\tilde{\mathbf P}_a^2\neq \tilde{\mathbf P}_a$.
However it satisfies
\be
\sum_{a=1}^2 \tilde{\mathbf P}_a(\mathbf p)=\mathbf Z_t^{-1}.
\ee
In the massless UV limit $\mathbf L=0$ one has
$\kappa_a(\mathbf p)=|\mathbf p|^2 s_a(\hat{\bm n})$ and
$\tilde{\bm\Pi}_a(\mathbf p)=\bm\Pi_a(\hat{\bm n})$,
so it is convenient to define the UV projectors
\be
\mathbf P_a(\hat{\bm n})\equiv \mathbf Z_t^{-1/2}\,\bm\Pi_a(\hat{\bm n})\,\mathbf Z_t^{-1/2}.
\ee

For the further purposes we also consider the massless case, i.e. we put $\mathbf{L} = 0$ in \eqref{K}. For the latter in the right-hand side we have overall factor $|\mathbf{p}|^2$ multiplying $\mathbf{S}(\hat{\mathbf{n}})$, hence we find eigenvalues and vectors for $\mathbf{S}(\hat{\mathbf{n}})$ only, 
\be
\mathbf{S}(\hat{\mathbf{n}}) \mathbf{v}_a(\hat{\mathbf{n}})=s_a(\hat{\mathbf{n}}) \mathbf{v}_a(\hat{\mathbf{n}}), \quad \mathbf{v}_a^T(\hat{\mathbf{n}}) \mathbf{v}_b(\hat{\mathbf{n}})=\delta_{a b},
\ee
and  projectors are
\be
\boldsymbol{\Pi}_a(\hat{\mathbf{n}}) \equiv \mathbf{v}_a(\hat{\mathbf{n}}) \mathbf{v}_a(\hat{\mathbf{n}})^T, \quad \sum_{a=1}^2 \boldsymbol{\Pi}_a=\mathbf{1}, \quad \boldsymbol{\Pi}_a \boldsymbol{\Pi}_b=\delta_{a b} \boldsymbol{\Pi}_a ,
\ee
so
\be
\label{S}
\mathbf{S}(\hat{\mathbf{n}})=\sum_{a=1}^2 s_a(\hat{\mathbf{n}}) \boldsymbol{\Pi}_a(\hat{\mathbf{n}}).
\ee
Also notice that in the UV ($|\mathbf p|\to\infty$) one has
$\kappa_a(\mathbf p)=|\mathbf p|^2\,s_a(\hat{\mathbf n})+\mathcal O(1)$, and the corresponding
spectral projectors approach those of $\mathbf S(\hat{\mathbf n})$ (the $\mathbf L$-dependence
affects only IR/finite parts and does not modify MS poles).

The resolution for  $(\omega^2 \mathbf{1}+|\mathbf{p}|^2 \mathbf{S}(\hat{\mathbf{n}}))^{-1}$ then reads 
\be
(\omega^2 \mathbf{1}+|\mathbf{p}|^2 \mathbf{S}(\hat{\mathbf{n}}))^{-1}=\sum_{a=1}^2 \frac{\boldsymbol{\Pi}_a(\hat{\mathbf{n}})}{\omega^2+|\mathbf{p}|^2 s_a(\hat{\mathbf{n}})},
\ee
and finally conjugating with $\mathbf{Z}_t^{-1 / 2}$ we get
\be
\mathbf{G}(p)=\sum_{a=1}^2 \frac{\mathbf{Z}_t^{-1 / 2} \mathbf{\Pi}_a(\hat{\mathbf{n}}) \mathbf{Z}_t^{-1 / 2}}{\omega^2+|\mathbf{p}|^2 s_a(\hat{\mathbf{n}})}.
\ee
In such form  $\mathbf{G}(p)$ it is clear that $s_a(\hat{\mathbf{n}})$ are ``directional sound speeds'' of normal modes of such a system.

\section{One-loop diagrams calculation}
\label{sec:master}

In this section we calculate all one-loop diagrams necessary for the extraction of beta functions.  First let us introduce two ``master integrals'' which are needed for mentioned purposes. So far having the propagator in the form \eqref{resolut G} and considering the interactions \eqref{eq:L3L4}, we need integrals for tadpole and bubble diagrams involving different field species. First let us consider tadpole master integral and then the bubble  integral and work out the final answers for pole necessary to calculate beta function expressed via data like $\mathbf{Z}_t,\mathbf{Y}$, $\mathbf{C_{\phi,\chi}}$.\footnote{
In pure dimensional regularization, massless vacuum integrals at zero external momentum are scaleless
and vanish, mixing UV and IR. To extract the UV MS pole parts unambiguously we introduce an auxiliary
IR regulator $\delta_{\rm IR}>0$; the $1/\eps$ poles are UV-local and independent of $\delta_{\rm IR}$.
}
\subsection*{Tadpole integral}
The first master integral reads
\begin{align}
\label{eq:T begin}
\mathbf{T}\equiv \int \frac{d \omega}{2 \pi} \frac{d^d \mathbf{p}}{(2 \pi)^d} \mathbf{G}(p)=\int \frac{d \omega}{2 \pi} \frac{d^d \mathbf{p}}{(2 \pi)^d} \mathbf{Z}_t^{-1 / 2}\left(\omega^2 \mathbf{1}+\mathbf{K}(\mathbf{p})\right)^{-1} \mathbf{Z}_t^{-1 / 2},
\end{align}
where we substitute \eqref{eq:Gmatr}. Let us first work out the integration over energy $\omega$ in \eqref{eq:T begin} as follows:
\begin{align}
\label{eigen for tad}
\int_{-\infty}^{\infty} \frac{d \omega}{2 \pi}\left(\omega^2 \mathbf{1}+\mathbf{K}\right)^{-1}=\frac{1}{2} \mathbf{K}^{-1 / 2} ,
\end{align}
with the use of \eqref{K}, so \eqref{eq:T begin}  reduces to
$$
\mathbf{T}=\frac{1}{2} \int \frac{d^d \mathbf{p}}{(2 \pi)^d} \mathbf{Z}_t^{-1 / 2} \mathbf{K}(\mathbf{p})^{-1 / 2} \mathbf{Z}_t^{-1 / 2}.
$$
The leading term of $\mathbf T$ in the UV (obtained by setting $\mathbf L=0$) is a
scaleless power-divergent integral. In dimensional regularization such scaleless integrals vanish,
so it produces no $1/\epsilon$ pole in the DR+MS scheme. Therefore the MS pole starts at
$\mathcal O(\mathbf L)$ (a mass insertion).

It is more convenient to rewrite the integral over momentum in spherical coordinates as
\begin{align}
\label{ddp}
\int \frac{d^d \mathbf{p}}{(2 \pi)^d}=\frac{1}{(2 \pi)^d} \int_0^{\infty} d \rho \rho^{d-1} \int_{S^{d-1}} d \Omega_{d-1}=\frac{\Omega_{d-1}}{2(2 \pi)^d} \int_0^{\infty} d x x^{\frac{d}{2}-1}\langle\cdots\rangle_{\hat{\mathbf{n}}},
\end{align}
where $\mathbf{p}\equiv \rho \hat{\mathbf{n}}$ with $\rho \equiv |\mathbf{p}| \geq 0$, and $x \equiv \rho^2; \Omega_{d-1}$ is the solid angle on the unit sphere and $\Omega_{d-1}=2 \pi^{d / 2} / \Gamma\left(\frac{d}{2}\right)$. The notation $\langle \ldots\rangle_{\hat{\mathbf{n}}}$ is given by 
\begin{align}
\label{eq:n average}
\langle f(\hat{\mathbf{n}})\rangle_{\hat{\mathbf{n}}} \equiv \frac{1}{\Omega_{d-1}} \int_{S^{d-1}} d \Omega_{d-1} f(\hat{\mathbf{n}}).
\end{align}
In dimensional regularization with $D=4-\eps$ one has $d=3-\eps$.
The $1/\eps$ poles in the master integrals come from the radial $|\mathbf p|$ integration
(e.g.\ $\int_0^\infty dx\,x^{-1-\eps/2}\sim 1/\eps$), while the angular averages over $S^{d-1}$
remain finite as $\eps\to0$ under our positivity assumptions.
Therefore, for MS pole parts we may evaluate all angular averages at $d=3$ (i.e.\ over $S^2$);
keeping the exact $S^{d-1}$ measure affects only finite terms.
Hence, we arrive at 
\begin{align}
\label{eq:T begin 2}
    \mathbf{T}=\frac{1}{4} \frac{\Omega_{d-1}}{(2 \pi)^d} \int_0^{\infty} d x \;x^{\frac{d}{2}-1}\left\langle\mathbf{Z}_t^{-1 / 2} \mathbf{K}(x, \hat{\mathbf{n}})^{-1 / 2} \mathbf{Z}_t^{-1 / 2}\right\rangle_{\hat{\mathbf{n}}}.
\end{align}
To work out this integral and extract divergence, we use the form of $\mathbf{K}$ matrix as \eqref{K in S and L}, as well as eigenvalue decomposition \eqref{K}. Moreover, to regularize the UV divergence in \eqref{eq:T begin 2}, we use the expansion at large momentum, i.e. at large $x$.
We put all explicit evaluations related to the integral \eqref{eq:T begin 2} in App.~\ref{app:tad} and here we write the final answer for the part of \eqref{eq:T begin}, which includes the pole only
\begin{align}
\label{T answer}
&\mathbb{T}\equiv [\mathbf{T}]_{\text{pole}}=-\frac{1}{4\pi^2 \epsilon}\left[ \left\langle \sum_{a, b} \mathcal{K}(s_a,s_b)\mathbf{Z}_t^{-1 / 2}  \mathbf{\Pi}_a \mathbf{L} \mathbf{\Pi}_b \mathbf{Z}_t^{-1 / 2} \right\rangle_{\hat{\mathbf{n}}}\right], 
\end{align}
where we also introduce 
\begin{align}
\label{K_sa_sb}
    \mathcal{K}\left(s_a, s_b\right)\equiv \frac{1}{\sqrt{s_a} \sqrt{s_b}\left(\sqrt{s_a}+\sqrt{s_b}\right)} .
\end{align}

\subsection*{Bubble integral}
\label{sec:bubble}
For the diagrams with two propagators we need another master integral, which reads
\begin{align}
\label{bubble main}
    &\mathbf{B}_{i j ; k l}(k) \equiv \int\frac{d\omega}{2\pi}\frac{d^d \mathbf{p}}{(2 \pi)^d} \mathbf{G}_{i k}(p) \mathbf{G}_{j l}(p+k) \nonumber\\
    &=\int\frac{d\omega}{2\pi}\frac{d^d \mathbf{p}}{(2 \pi)^d}  \Big[\mathbf{Z}_t^{-1 / 2}\left(\omega^2 \mathbf{1}+\mathbf{K}(\mathbf{p})\right)^{-1} \mathbf{Z}_t^{-1 / 2}\Big]_{ik}\nonumber\\
    &\times\Big[\mathbf{Z}_t^{-1 / 2}\left((\omega+\Omega)^2 \mathbf{1}+\mathbf{K}(\mathbf{p+k})\right)^{-1} \mathbf{Z}_t^{-1 / 2}\Big]_{jl},
\end{align}
where $k \equiv (\Omega, \mathbf{k})$ is an external energy-momentum.  
At large loop momentum $p=(\omega,\mathbf p)$ one has $\mathbf G(p)=\mathcal O(p^{-2})$
because $\mathbf K(\mathbf p)\sim |\mathbf p|^2\,\mathbf S(\hat{\mathbf n})$.
Hence $\mathbf B(k)$ is superficially logarithmically divergent in $D=4-\eps$,
$\mathbb B(k)\sim \int d^Dp\,p^{-4}$, so any UV divergence must be local.
Expanding the integrand in the external momentum $k$ produces additional inverse powers of $p$
and yields UV-finite contributions. Therefore the $1/\eps$ pole is momentum-independent and is
fully captured by $\mathbf B_{ij;kl}(0)$: 
\begin{align}
\label{eq:B0_def}
&\mathbf{B}_{i j ; k l}(0)
\equiv \int\frac{d\omega}{2\pi}\frac{d^d \mathbf{p}}{(2 \pi)^d}\,
\mathbf{G}_{i k}(p)\,\mathbf{G}_{j l}(p)\nonumber\\
&=\int\frac{d\omega}{2\pi}\frac{d^d \mathbf{p}}{(2 \pi)^d}
\Big[\mathbf{Z}_t^{-1 / 2}\left(\omega^2 \mathbf{1}+\mathbf{K}(\mathbf{p})\right)^{-1} \mathbf{Z}_t^{-1 / 2}\Big]_{ik}
\Big[\mathbf{Z}_t^{-1 / 2}\left(\omega^2 \mathbf{1}+\mathbf{K}(\mathbf{p})\right)^{-1} \mathbf{Z}_t^{-1 / 2}\Big]_{jl}\nonumber\\
&=\int\frac{d^d \mathbf{p}}{(2 \pi)^d}\sum_{a,b}
\tilde{\mathbf{P}}_a(\mathbf{p})_{i k}\tilde{\mathbf{P}}_b(\mathbf{p})_{j l}
\int\frac{d\omega}{2\pi}\frac{1}{(\omega^2+\kappa_a(\mathbf p))(\omega^2+\kappa_b(\mathbf p))}\nonumber\\
&=\int\frac{d^d \mathbf{p}}{(2 \pi)^d}\sum_{a,b}
\frac{\tilde{\mathbf{P}}_a(\mathbf{p})_{i k}\tilde{\mathbf{P}}_b(\mathbf{p})_{j l}}
{2\sqrt{\kappa_a(\mathbf p)\kappa_b(\mathbf p)}\big(\sqrt{\kappa_a(\mathbf p)}+\sqrt{\kappa_b(\mathbf p)}\big)}.
\end{align}

where in the last equality we take $\omega$ integral with the use of decomposition \eqref{K} as well as \eqref{eq:Ptilde}. Next, using \eqref{eq:n average}, we rewrite integration over spatial $\mathbf{p}$ as
\begin{align}
    \mathbf{B}_{i j ; k l}(0) = \frac{\Omega_{d-1}}{2(2 \pi)^d}  \int_0^{\infty} d x \;x^{\frac{d}{2}-1}\Big\langle\sum_{a,b}\frac{\tilde{\mathbf{P}}_a(\hat{\mathbf{n}},x)_{i k}\tilde{\mathbf{P}}_b(\hat{\mathbf{n}},x)_{j l}}{2\sqrt{\kappa_a(\hat{\mathbf{n}},x)\kappa_b(\hat{\mathbf{n}},x)}\Big(\sqrt{\kappa_a(\hat{\mathbf{n}},x)}+\sqrt{\kappa_b(\hat{\mathbf{n}},x)}\Big)}\Big\rangle_{\hat{\mathbf{n}}}.\nonumber
\end{align}
Performing the expansion at large momentum $|\mathbf{p}|$ after some algebra one arrives to the following answer for the divergent part of bubble
\begin{align}
\label{bubble answer}
    &\mathbb{B}_{i j ; k l}\equiv[\mathbf{B}_{i j ; k l}]_{\text{pole}}=\frac{1}{4\pi^2\epsilon}  \Big\langle\sum_{a,b}\mathcal{K}(s_a,s_b)\mathbf{P}_a(\hat{\mathbf{n}})_{i k}\mathbf{P}_b(\hat{\mathbf{n}})_{j l}\Big\rangle_{\hat{\mathbf{n}}},
\end{align}
where we also have used \eqref{K_sa_sb}. We put all detailed and step-by-step calculations of \eqref{bubble answer} into the App.~\ref{app:tad}. 

\subsection*{Diagrams calculation}
\begin{figure}[t!]
\centering
\includegraphics[width=0.49\textwidth]{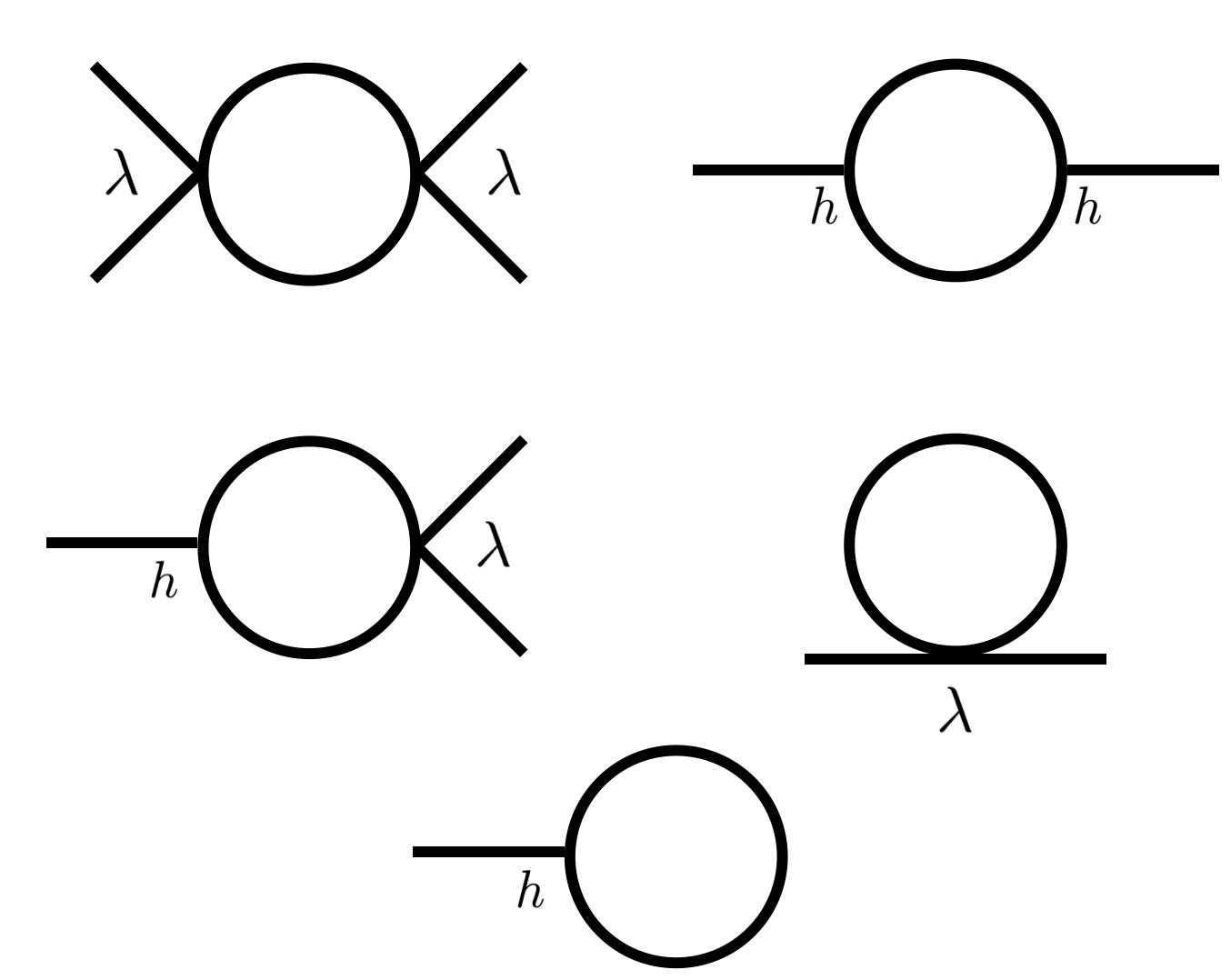}
\caption{All possible one loop divergent diagrams for the model with the interaction Lagrangian \eqref{eq:L3L4}. The notations $\lambda$ and $h$ are just schematic and we specify couplings for each concrete diagram in the text. Since there are $\delta j_i \phi_i$ terms in the Lagrangian \eqref{eq:L3L4}, we add the bottom diagram with one external leg, one propagator and $h_{ijk}$-type vertex as well.}
\label{fig:all 1 loop}
\end{figure}

Finally let us express all divergent parts of diagrams in compact notation via tadpole and bubble master integral contributions. Diagrams relevant for our calculation  are well known and depicted in Fig.\ref{fig:all 1 loop}. 

In compact notation contributions from fish diagrams in different channels are given by 
\begin{align}
&\Delta \Gamma_{i j ; k l}^{(4)}= \\
&= \frac12\sum_{m,n,p,q}(\lambda_{i j m n} \mathbb{B}_{m n ; p q} \lambda_{k l p q} +\lambda_{i k m n} \mathbb{B}_{m n ; p q} \lambda_{j l p q} +\lambda_{i l m n} \mathbb{B}_{m n ; p q} \lambda_{j k p q}) ,\nonumber
\end{align}
where each term corresponds to $s, t, u$ channels, respectively.\footnote{Let us briefly clarify our notations: for example, in the case of incoming and outgoing $\phi \phi$-pair we write for s-channel
\begin{align}
    \Delta \Gamma_{1 1; 11}^{(4)} =\frac12\Big( \lambda_1^2 \mathbb{B}_{11 ; 11}+\lambda_4^2 \mathbb{B}_{12 ; 12}+\lambda_3^2 \mathbb{B}_{22 ; 22}+2 \lambda_1 \lambda_4 \mathbb{B}_{11 ; 12}+2 \lambda_1 \lambda_3 \mathbb{B}_{11 ; 22}+2 \lambda_4 \lambda_3 \mathbb{B}_{12 ; 22} \Big),\nonumber
\end{align}
where $i=j=k=l=1$ 
and coefficients $2$ comes from two possibilities such as $\mathbb{B}_{11 ; 12}$ and $\mathbb{B}_{12 ; 11}$.}
All indices we have: external ones are $i, j, k, l \in\{1,2\}$, while internal are $m, n, p, q \in\{1,2\}$ and the factor $1/2$ comes from the overall 
$\frac{1}{2!}$ in the second‑order expansion.  
Next, decay diagrams contribution is given by 
\begin{align}
\label{h_l}
\Delta \Gamma_{i j k}^{(3)}=\frac12 \sum_{m,n,p,q}\Big(\lambda_{i j m n} \mathbb{B}_{m n ; p q} h_{k p q}+\lambda_{i k m n} \mathbb{B}_{m n ; p q} h_{j p q}+\lambda_{j k m n} \mathbb{B}_{m n ; p q} h_{i p q}\Big),
\end{align}
and for all indices we have: external ones are $i, j, k \in\{1,2\}$, while internal are $m, n, p, q \in\{1,2\}$.

The bubble diagram reads
\begin{align}
\label{h2}
    \Sigma_{i j}^{\left(h^2\right)}(\Omega, \mathbf{k})= \frac12 \sum_{m,n,p,q} h_{i m n} \mathbb{B}_{m n ; p q}(\Omega, \mathbf{k}) h_{j p q} ,
\end{align}
where external indices are $i, j \in\{1,2\}$, while internal ones are $m, n, p, q \in\{1,2\}$.

Finally, tadpole diagram contributes as the sum
\begin{align}
\label{sigma_l}
    \Sigma_{i j}^{(\lambda)}= \frac12\sum_{m,n}\lambda_{i j m n} \mathbb{T}_{m n} ,
\end{align}
with external indices are $i, j \in\{1,2\}$, while internal ones are $m, n \in\{1,2\}$.
If $h_{ijk}\neq 0$ and $M^2_{ij}\neq 0$, there is an additional one-loop divergent 1PI diagram with one
external leg (a tadpole). Its contribution is just
\begin{align}
\label{Gamma1}
\Delta \Gamma^{(1)}_{i} = \frac12\, \sum_{m,n}h_{i m n}\,\mathbb{T}_{m n},
\end{align}
where $\mathbb{T}_{mn}$ is the tadpole master integral (and the factor $1/2$
comes from the combinatorics $3/3!$ at the cubic vertex). This divergence requires the linear counterterm
$\mu^{\frac{D+2}{2}}\,\delta j_i\,\varphi_i$ in \eqref{counter_terms}.

The interaction Lagrangian \eqref{eq:L3L4} contains explicit factors of $\mu$ which make
$h_{ijk}$ and $\lambda_{ijkl}$ dimensionless in $D=4-\epsilon$.
In the diagrammatic expressions we write only the
\emph{reduced} tensor contractions built from $(h,\lambda)$ and the master integrals
$(\mathbb T,\mathbb B)$. The corresponding full one--loop 1PI contributions that enter
the renormalization of couplings are obtained by restoring the overall $\mu$--factors
from the vertices:
\begin{align}
\Gamma^{(4)}_{\text{1-loop}} &\;=\;\mu^{2(4-D)}\,\Delta\Gamma^{(4)},\\
\Gamma^{(3)}_{\text{1-loop}} &\;=\;\mu^{(4-D)+\frac{6-D}{2}}\,\Delta\Gamma^{(3)},\\
\Sigma^{(\lambda)}_{\text{1-loop}} &\;=\;\mu^{4-D}\,\Sigma^{(\lambda)},\\
\Sigma^{(h^2)}_{\text{1-loop}} &\;=\;\mu^{6-D}\,\Sigma^{(h^2)},\\
\Gamma^{(1)}_{\text{1-loop}} &\;=\;\mu^{\frac{6-D}{2}}\,\Delta\Gamma^{(1)}.
\end{align}
For quartic and cubic counterterms the extra factor $\mu^{4-D}=\mu^{\epsilon}=1+\mathcal O(\epsilon)$
does not affect the MS pole part. In contrast, the self--energy generated by two cubic vertices
carries $\mu^{6-D}=\mu^{2+\epsilon}$, and the overall $\mu^2$ is essential for dimensional consistency
of the mass counterterm.

\section{General expressions for beta functions}
\label{sec:beta_MS}

Now summarize what we obtained to get  the renormalization--group flow in our compact notation.  
All one--loop UV divergences of the theory have been analyzed in detail in
Sec.~\ref{sec:master}, where the relevant master integrals were evaluated, and the tensor
structure of the one--particle--irreducible (1PI) vertices was classified.
Here we only assemble those results into a compact and systematic set of beta functions.

At fixed bare couplings beta functions are defined
\be
\beta_g \equiv \mu \frac{d g}{d\mu}\Big|_{g_0\ \mathrm{fixed}} .
\ee
All one--loop UV poles are controlled by two
objects:
\begin{itemize}
\item the tadpole matrix $\mathbb T_{mn}$,
\item the zero--momentum bubble tensor $\mathbb B_{ij;kl}(0)$.
\end{itemize}
Their pole parts define finite tensors
\be
\mathcal T_{mn} \equiv \eps\, \mathbb T_{mn}, \qquad
\mathcal B_{ij;kl} \equiv \eps\, \mathbb B_{ij;kl}(0),
\ee
whose explicit expressions, in terms of the UV spectral data of the kinetic operator and
angular averages over $\hat{\bm n}$, are given in
eqs.~\eqref{T answer} and \eqref{bubble answer}.

Importantly, the UV poles are local and momentum--independent.
As a consequence:
\begin{itemize}
\item only the couplings $(\lambda,h,M^2,j)$ are renormalized at one loop,
\item the kinetic and anisotropy matrices do \emph{not} run in MS at this order.
\end{itemize}

\subsection*{Minimal subtraction and counterterms}

In the MS scheme all counterterms are chosen to cancel only the UV pole parts in
$\eps=4-D$ and to contain no finite contributions.  Concretely, for every renormalized
parameter $g$ we write the corresponding bare one as
\be
g_0=\mu^{d_g}\,\big(g+\delta g\big),
\ee
where $d_g$ is the canonical mass dimension (in $D=4-\eps$) of the bare
coupling $g_0$ in our normalization, fixed by the explicit powers of $\mu$ in the interaction
Lagrangian \eqref{renorm int}.  For the couplings in \eqref{eq:L3L4} this means
\be
\lambda^{(0)}_{ijkl}=\mu^{4-D}\big(\lambda_{ijkl}+\delta\lambda_{ijkl}\big)
=\mu^{\eps}\big(\lambda_{ijkl}+\delta\lambda_{ijkl}\big),
\qquad
h^{(0)}_{ijk}=\mu^{\frac{6-D}{2}}\big(h_{ijk}+\delta h_{ijk}\big)
,
\ee
\be
j^{(0)}_{i}=\mu^{\frac{D+2}{2}}\big(j_{i}+\delta j_{i}\big)
=\mu^{3-\eps/2}\big(j_{i}+\delta j_{i}\big),
\qquad
(M^2_{ij})^{(0)}=M^2_{ij}+\delta M^2_{ij},
\ee
in agreement with the counterterm Lagrangian \eqref{counter_terms}.

Denoting  the simple-pole coefficients as
\be
\delta\lambda_{ijkl}=\frac{A_{ijkl}}{\eps},\qquad
\delta h_{ijk}=\frac{B_{ijk}}{\eps},\qquad
\delta j_i=\frac{C_i}{\eps},\qquad
\delta M^2_{ij}=\frac{D_{ij}}{\eps},
\ee
where 
\bea
A_{ijkl} &= \Big[\Delta\Gamma^{(4)}_{ij;kl}\Big]_{\rm pole},\\
B_{ijk}  &= \Big[\Delta\Gamma^{(3)}_{ijk}\Big]_{\rm pole},\\
D_{ij}   &= \Big[\mu^{4-D}\Sigma^{(\lambda)}_{ij}+\mu^{6-D}\Sigma^{(h^2)}_{ij}\Big]_{\rm pole},\\
C_i      &= \Big[\mu^{-\frac{D+2}{2}}\,\Gamma^{(1)}_{i,\;{\rm 1\mbox{-}loop}}\Big]_{\rm pole}
= -\,\Big[\mu^{2-D}\Delta\Gamma^{(1)}_i\Big]_{\rm pole},
\eea
then the expressions for beta functions have the form 
\begin{gather}
\beta_{\lambda_{ijkl}}=-\eps\,\lambda_{ijkl}+A_{ijkl},\qquad
\beta_{h_{ijk}}=-(1+\eps/2)\,h_{ijk}+B_{ijk},\\
\beta_{j_i}=-(3-\eps/2)\,j_i+C_i,
\qquad
\beta_{M^2_{ij}}=D_{ij},
\end{gather}
and, for the dimensionless masses $m^2_{ij}\equiv M^2_{ij}/\mu^2$,
\be
\beta_{m^2_{ij}}=-2\,m^2_{ij}+\frac{1}{\mu^2}\,\beta_{M^2_{ij}}.
\ee
Thus, once the pole residues $(A,B,C,D)$ are known from the 1PI diagrams, the beta functions are
fixed immediately without keeping any finite parts. Now let us express beta functions in terms of tadpoles and bubbles.

\subsection*{Quartic couplings}

The quartic vertex receives contributions from the three one--loop fish diagrams.
Using the pole structure of the four--point function derived in
Sec.~\ref{sec:master}, the beta function takes the universal tensor form
\bea 
&\beta_{\lambda_{ijkl}}
=
-\eps\,\lambda_{ijkl}\\
&+\frac{1}{2}\Big(
\lambda_{ijmn}\,\mathcal B_{mn;pq}\,\lambda_{klpq}
+\lambda_{ikmn}\,\mathcal B_{mn;pq}\,\lambda_{jlpq}
+\lambda_{ilmn}\,\mathcal B_{mn;pq}\,\lambda_{jkpq}
\Big).
\label{eq:beta-lambda}
\eea
All anisotropy dependence enters exclusively through the bubble residue
$\mathcal B$.

\subsection*{Cubic couplings}

The cubic couplings renormalize only in the presence of quartic interactions.
At one loop the flow is linear in $h$, reflecting the absence of UV-divergent purely cubic
one--loop diagrams near $D=4$.
From the three--point function pole computed in Sec.~\ref{sec:master} one finds
\bea
&\beta_{h_{ijk}}
=
-\Bigl(1+\frac{\eps}{2}\Bigr) h_{ijk}\\
&+\frac{1}{2}\Big(
\lambda_{ijmn}\,\mathcal B_{mn;pq}\,h_{kpq}
+\lambda_{ikmn}\,\mathcal B_{mn;pq}\,h_{jpq}
+\lambda_{jkmn}\,\mathcal B_{mn;pq}\,h_{ipq}
\Big).
\label{eq:beta-h}
\eea

\subsection*{Mass matrix}

The quadratic couplings are conveniently expressed in terms of the dimensionless
mass matrix
\be
m^2_{ij} \equiv \frac{M^2_{ij}}{\mu^2}.
\ee
Its beta function receives a homogeneous contribution proportional to $m^2$, as well as
inhomogeneous terms induced by quartic and cubic interactions.
Combining the tadpole and bubble contributions obtained in Sec.~\ref{sec:master}, one finds
\be 
\beta_{m^2_{ij}}
=
-2\,m^2_{ij}
-\frac{1}{2}\,\lambda_{ijmn}\,\widetilde{\mathcal T}_{mn}
+\frac{1}{2}\,h_{imn}\,\mathcal B_{mn;pq}\,h_{jpq},
\label{eq:beta_m2}
\ee
where $\widetilde{\mathcal{T}}_{i j} \equiv \frac{\mathcal{T}_{i j}}{\mu^2} $ denotes the dimensionless tadpole residue, while $B_{mn;pq}$ is already dimensionless.
In particular, for $h\neq0$ the surface $m^2=0$ is not preserved by the RG flow.

\subsection*{Linear couplings}

When both cubic couplings and masses are present, one--loop tadpole diagrams generate
a divergence linear in the fields.
As a result, the beta function for the linear couplings reads
\be
\beta_{j_i}
=
-\Bigl(3-\frac{\eps}{2}\Bigr) j_i
-\frac{1}{2}\,h_{imn}\,\widetilde{\mathcal T}_{mn}.
\label{eq:beta_j}
\ee
Thus the hyperplane $j_i=0$ is RG invariant only if either $h=0$ or $m^2=0$.

\subsection*{Absence of one--loop running of anisotropy matrices}

Finally, since all one--loop UV divergences in the two--point function are local and
momentum--independent, they renormalize only the mass operator.
Therefore, in DR+MS,
\be
\beta_{\mathbf Z_t}
=
\beta_{\mathbf C_\phi}
=
\beta_{\mathbf C_\chi}
=
\beta_{\mathbf Y}
=0
\qquad
\text{(one loop)} .
\ee
Possible running of these matrices arises only beyond one loop or in non--minimal schemes.

\vspace{2mm}
The remainder of the paper is devoted to evaluating the tensors
$\mathcal B$ and $\mathcal T$ in specific truncations of the anisotropy data and to
analyzing the resulting RG flows and fixed points.

\section{Explicit form of beta functions, fixed points and physical interpretation}\label{sec:partic}

So far we obtained the general expression for beta functions when non-trivial velocity tensors $\mathbf{C_{1,2}}$ as well as $\mathbf{Z}_t$ and $\mathbf{Y}$  are non-trivial. Now let us explore several consistent truncations of the general one-loop beta functions. Some truncations are closed at one loop in MS (because no momentum-dependent poles are generated),
but they are not guaranteed to remain closed beyond one loop or in other schemes.
 In addition, if cubic couplings $h_{ijk}$ and masses $M^2_{ij}$ are non-zero, the RG flow generates
linear (tadpole) couplings $j_i$, so in this section we include their beta functions as well.

\subsection{Cross-gradients turned off: $\mathbf{Y}=0$}
In this subsection we set
\begin{equation}
\mathbf Z_t=\mathbf 1,\qquad \mathbf Y=0,
\end{equation}
so that the UV kinetic operator is diagonal in the field space and all anisotropy
is encoded solely in the two matrices $\mathbf C_1\equiv \mathbf C_\phi$
and $\mathbf C_2\equiv \mathbf C_\chi$. The details of calculation for formulae from this subsection can be found in \ref{sec:threeweights}.
In this limit the direction-dependent matrix $\mathbf C(\hat{\bm n})$ is block-diagonal in the field space,
so its spectral projectors reduce to the constant matrices
$\Pi_1=\mathrm{diag}(1,0)$ and $\Pi_2=\mathrm{diag}(0,1)$.
As a result, all one-loop pole coefficients depend on anisotropy only through three scalar
angular weights $J_{11},J_{22},J_{12}$:
\be
\begin{aligned}
& J_{11}=\left\langle\left[\hat{n}^{\top} \mathbf{C}_1 \hat{n}\right]^{-3 / 2}\right\rangle_{S^2}, \quad J_{22}=\left\langle\left[\hat{n}^{\top} \mathbf{C}_2 \hat{n}\right]^{-3 / 2}\right\rangle_{S^2}, \\
& J_{12}= \int_0^1\left(\operatorname{det}\left[(1-\tau) \mathbf{C}_1+\tau \mathbf{C}_2\right]\right)^{-1 / 2} d \tau,
\end{aligned}
\ee
with the kernels 
$\mathcal{K}_{m n ; p q}$  given by the expression
\begin{gather}
\mathcal B_{mn;pq}\equiv \eps\,\mathbb B_{mn;pq}
\nonumber\\=\frac{1}{8\pi^2}\Big(
J_{11}\delta_{m1}\delta_{n1}\delta_{p1}\delta_{q1}
+J_{22}\delta_{m2}\delta_{n2}\delta_{p2}\delta_{q2}
+J_{12}(\delta_{m1}\delta_{n2}\delta_{p1}\delta_{q2}+\delta_{m2}\delta_{n1}\delta_{p2}\delta_{q1})
\Big),
\end{gather}
and 
\be
\mathcal T \equiv \eps\,\mathbf T
=-\frac{1}{8\pi^2}
\begin{pmatrix}
J_{11} M_{11}^2 & J_{12} M_{12}^2\\
J_{12} M_{12}^2 & J_{22} M_{22}^2
\end{pmatrix}.
\ee
Throughout this subsection we set $\eps\to0$ in angular averages and use
$\langle\cdot\rangle_{S^2}$ for the normalized average over $S^2$.

After some algebra one can obtain the explicit form of $J_{ij}$ in terms of eigenvalues data of $\mathbf{C_a}$. 
In dimensional regularization with $d=3-\epsilon$,
the MS pole coefficients require the angular averages only at $\epsilon=0$
(equivalently, one may keep $S^{d-1}$ and set $\epsilon\to 0$ at the end; the difference contributes only to finite parts).

One can show that one can simplify the integrals in the form  $\left\langle\left(\hat{n}^{\top} \mathbf{C} \hat{n}\right)^{-3 / 2}\right\rangle_{S^2}=(\operatorname{det} \mathbf{C})^{-1 / 2}$ in our case.
Let $\mathbf C_a$ have (positive) eigenvalues $\{c_{a,1},c_{a,2},c_{a,3}\}$.
Then
\[
J_{aa}=(\det\mathbf C_a)^{-1/2}=(c_{a,1}c_{a,2}c_{a,3})^{-1/2}.
\]
while the cross-term $J_{12}$ is defined by the generalized eigenvalues of the pair $\left(\mathbf{C}_2, \mathbf{C}_1\right)$ denoted by  $\gamma_1, \gamma_2, \gamma_3>0$, i.e.

$$
\mathbf{C}_2 v=\gamma \mathbf{C}_1 v \quad \Longleftrightarrow \quad \mathbf{A} \equiv \mathbf{C}_1^{-1 / 2} \mathbf{C}_2 \mathbf{C}_1^{-1 / 2}=U \operatorname{diag}\left(\gamma_i\right) U^{\top} .
$$
Then
$$
J_{12}=\left(\operatorname{det} \mathbf{C}_1\right)^{-1 / 2} \int_0^1 \prod_{i=1}^3\left((1-\tau)+\tau \gamma_i\right)^{-1 / 2} d \tau,
$$
and this integral can be expressed as Lauricella $F_D$ hypergeometric function
$$
J_{12}=\left(\operatorname{det} \mathbf{C}_1\right)^{-1 / 2} F_D^{(3)}\left(1 ; \frac{1}{2}, \frac{1}{2}, \frac{1}{2} ; 2 ; 1-\gamma_1, 1-\gamma_2, 1-\gamma_3\right) .
$$
These weights arise from the universal kernel
$\mathcal K(s_a,s_b)=1/(\sqrt{s_a}\sqrt{s_b}(\sqrt{s_a}+\sqrt{s_b}))$ via
$J_{12}=2\langle \mathcal K(s_1(\hat{\bm n}),s_2(\hat{\bm n}))\rangle_{S^2}$,
and how these formulae follows from simple $\tau$-representation \eqref{eq:J12_tau_app} is derived in App.~\ref{app:Jweights}.

Also, let us introduce 
\be 
\rho^2=\frac{J_{12}^2}{J_{11} J_{22}}, \,\,\,\, \rho \in (0,1].
\ee 
Another important thing which holds for general positive-definite matrices $C_{ij}$ inequality taking place for $J_{ij}$ in this case, namely 
\be\label{eq:Jrest}
J_{12}^2 \leq J_{11} J_{22},
\ee 
and that is why $\rho \in (0,1]$; the derivation of this \eqref{eq:Jrest} inequality is presented in App.~\ref{app:J_ineq}.
In these notations we obtain that beta functions can be written down explicitly in the form resembling the well-known answer for two-fields beta function, but weighted by $J_{ij}$ in appropriate places. Namely, for $\lambda_i$ set of constants we obtain 
\begin{subequations}
\label{betah_lambda_Y}
\begin{align}
& \beta_{\lambda_1}=-\epsilon \lambda_1+\frac{3}{16 \pi^2}\left[J_{11} \lambda_1^2+2 J_{12} \lambda_4^2+J_{22} \lambda_3^2\right], \\
& \beta_{\lambda_2}=-\epsilon \lambda_2+\frac{3}{16 \pi^2}\left[J_{22} \lambda_2^2+2 J_{12} \lambda_5^2+J_{11} \lambda_3^2\right], \\
& \beta_{\lambda_3}=-\epsilon \lambda_3\nonumber\\
&+\frac{1}{16 \pi^2}\left[J_{11}\left(\lambda_1 \lambda_3+2 \lambda_4^2\right)+4 J_{12} \lambda_3^2+J_{22}\left(\lambda_2 \lambda_3+2 \lambda_5^2\right)+2 J_{12} \lambda_4 \lambda_5\right], \\
& \beta_{\lambda_4}=-\epsilon \lambda_4+\frac{3}{16 \pi^2}\left[J_{11} \lambda_1 \lambda_4+2 J_{12} \lambda_3 \lambda_4+J_{22} \lambda_3 \lambda_5\right], \\
& \beta_{\lambda_5}=-\epsilon \lambda_5+\frac{3}{16 \pi^2}\left[J_{22} \lambda_2 \lambda_5+2 J_{12} \lambda_3 \lambda_5+J_{11} \lambda_3 \lambda_4\right] ,
\end{align}
\end{subequations}
with $h_i$ running described by
\begin{subequations}
\label{beta h y0 z1}
\begin{align}
& \beta_{h_1}=-\left(1+\frac{\epsilon}{2}\right) h_1+\frac{3}{16 \pi^2}\left[J_{11} \lambda_1 h_1+2 J_{12} \lambda_4 h_3+J_{22} \lambda_3 h_4\right], \\
& \beta_{h_2}=-\left(1+\frac{\epsilon}{2}\right) h_2+\frac{3}{16 \pi^2}\left[J_{22} \lambda_2 h_2+2 J_{12} \lambda_5 h_4+J_{11} \lambda_3 h_3\right], \\
& \beta_{h_3}=-\left(1+\frac{\epsilon}{2}\right) h_3\nonumber\\
&+\frac{1}{16 \pi^2}\left[J_{11}\left(\lambda_1 h_3+2 \lambda_4 h_1\right)+4 J_{12} \lambda_3 h_3+2 J_{12} \lambda_4 h_4+J_{22}\left(\lambda_3 h_2+2 \lambda_5 h_4\right)\right], \\
& \beta_{h_4}=-\left(1+\frac{\epsilon}{2}\right) h_4\nonumber\\
&+\frac{1}{16 \pi^2}\left[J_{22}\left(\lambda_2 h_4+2 \lambda_5 h_2\right)+4 J_{12} \lambda_3 h_4+2 J_{12} \lambda_5 h_3+J_{11}\left(\lambda_3 h_1+2 \lambda_4 h_3\right)\right] .
\end{align}
\end{subequations}
Beta functions for masses also obtain a similar form
\begin{subequations}
\begin{align}
 \beta_{m_{11}^2}&=-2\,m^2_{11}
+\frac{1}{16 \pi^2}\left(\lambda_1 J_{11} m_{11}^2+2 \lambda_4 J_{12} m_{12}^2+\lambda_3 J_{22} m_{22}^2\right)\nonumber\\
&+\frac{1}{16 \pi^2}\left(J_{11} h_1^2+2 J_{12} h_3^2+J_{22} h_4^2\right), \\
 \beta_{m_{22}^2}&=-2\,m^2_{22}
+\frac{1}{16 \pi^2}\left(\lambda_2 J_{22} m_{22}^2+2 \lambda_5 J_{12} m_{12}^2+\lambda_3 J_{11} m_{11}^2\right)\nonumber\\
&+\frac{1}{16 \pi^2}\left(J_{22} h_2^2+2 J_{12} h_4^2+J_{11} h_3^2\right), \\
 \beta_{m_{12}^2}&=-2\,m^2_{12}
+\frac{1}{16 \pi^2}\left(\lambda_4 J_{11} m_{11}^2+\lambda_5 J_{22} m_{22}^2+2 \lambda_3 J_{12} m_{12}^2\right)\nonumber\\
&+\frac{1}{16 \pi^2}\left(J_{11} h_1 h_3+J_{22} h_2 h_4+2 J_{12} h_3 h_4\right),
\end{align}
\end{subequations}
and, finally, for 
$j_\phi\equiv j_1$ and $j_\chi\equiv j_2$:
\begin{subequations}
\begin{align}
\beta_{j_\phi} &=
-\left(3-\frac{\epsilon}{2}\right) j_\phi
+\frac{1}{16\pi^2}\Big(J_{11}\,h_1\,m_{11}^2 + 2J_{12}\,h_3\,m_{12}^2 + J_{22}\,h_4\,m_{22}^2\Big),\\
\beta_{j_\chi} &=
-\left(3-\frac{\epsilon}{2}\right) j_\chi
+\frac{1}{16\pi^2}\Big(J_{11}\,h_3\,m_{11}^2 + 2J_{12}\,h_4\,m_{12}^2 + J_{22}\,h_2\,m_{22}^2\Big),
\end{align}
\end{subequations}
where we have introduced 
$m^2_{ij}\equiv M^2_{ij}/\mu^2$.
In particular, for $h\neq 0$ and $M^2\neq 0$, the plane $j_\phi=j_\chi=0$ is not RG-invariant.

\subsubsection*{Parity-symmetric sector $\lambda_{4,5}=0$ and $h_i=0$}
Now consider renormalization group in such a model. Let us specialize first to the simplest case keeping only $\lambda_{1,2,3}$ and setting $\lambda_{4,5}=0$. In this case beta functions have the form
$$
\begin{aligned}
\beta_{\lambda_1} & =-\epsilon \lambda_1+\frac{3}{16 \pi^2}\left[J_{11} \lambda_1^2+J_{22} \lambda_3^2\right], \\
\beta_{\lambda_2} & =-\epsilon \lambda_2+\frac{3}{16 \pi^2}\left[J_{22} \lambda_2^2+J_{11} \lambda_3^2\right], \\
\beta_{\lambda_3} & =-\epsilon \lambda_3+\frac{1}{16 \pi^2}\left[J_{11} \lambda_1 \lambda_3+4 J_{12} \lambda_3^2+J_{22} \lambda_2 \lambda_3\right],
\end{aligned}
$$
and formal solutions to  set equations $\beta_{\lambda_i}=0$ corresponds to different fixed points of our model. Namely, we have two solutions we call  $S_\pm$ (corresponding to $\pm$ sign choices):
\begin{subequations}
\label{eq:fixed-WF}
\begin{align}
&\lambda_1^*=\frac{\frac{16 \pi^2}{3} \epsilon \cdot 3-4 J_{12} \lambda_3^*}{2 J_{11}}, \\ &\lambda_2^*=\frac{\frac{16 \pi^2}{3} \epsilon \cdot 3-4 J_{12} \lambda_3^*}{2 J_{22}},\\
&\lambda_3^*=\frac{\frac{16 \pi^2}{3} \epsilon\left(4 J_{12} \pm \sqrt{4 J_{12}^2-3 J_{11} J_{22}}\right)}{2\left(4 J_{12}^2+J_{11} J_{22}\right)},
\end{align}
\end{subequations}
which are real-valued only when  $4 J_{12}^2 \geq3J_{11}J_{22}$. 
Besides the coupled solutions $S_\pm$ with $\lambda_3^*\neq 0$, the equations
$\beta_{\lambda_1}=\beta_{\lambda_2}=\beta_{\lambda_3}=0$ also admit the standard fixed points:
\begin{itemize}
\item Gaussian: $(\lambda_1^*,\lambda_2^*,\lambda_3^*)=(0,0,0)$.
\item Single-field Wilson--Fisher points (with $\lambda_3^*=0$):
\[
(\lambda_1^*,\lambda_2^*,\lambda_3^*)=\Big(\frac{16\pi^2}{3}\frac{\eps}{J_{11}},\,0,\,0\Big),
\qquad
(\lambda_1^*,\lambda_2^*,\lambda_3^*)=\Big(0,\,\frac{16\pi^2}{3}\frac{\eps}{J_{22}},\,0\Big).
\]
\item Decoupled two-field Wilson--Fisher point (with $\lambda_3^*=0$):
\[
(\lambda_1^*,\lambda_2^*,\lambda_3^*)=
\Big(\frac{16\pi^2}{3}\frac{\eps}{J_{11}},\,\frac{16\pi^2}{3}\frac{\eps}{J_{22}},\,0\Big).
\]
\end{itemize}
We focus below on the coupled fixed points with $\lambda_3^*\neq 0$.

Combining with \eqref{eq:Jrest} we obtain that this fixed point exists when mismatch ratio $\rho$ is restricted to 
\be 
\rho^2=\frac{J_{12}^2}{J_{11} J_{22}} \in[3/4,1].
\ee
while in general it belongs to $\rho\in (0,1]$.
Solution corresponding to $S_-$ is the IR-stable attractor generalizing ordinary Wilson-Fisher fixed point and reducing to it when all velocities are set to unity and for appropriate choice of constants.   The coupled fixed point exists only when the mismatch is sufficiently small, $\rho \geq \sqrt{3} / 2$. When $\rho<\sqrt{3} / 2$ the coupled solution becomes complex and disappears from the real coupling space. The solution $S_+$ is  the  IR fixed point of saddle type containing one repulsive direction.

The last formal solution of $\beta_{\lambda_i}=0$  is given by
\begin{subequations}
    \begin{align}
     &\lambda_1^*=\frac{8 \pi^2}{3} \frac{\epsilon}{J_{11}}\left(1+\sqrt{1-\frac{J_{11} J_{22}}{J_{12}^2}}\right), \\ &\lambda_2^*=\frac{8 \pi^2}{3} \frac{\epsilon}{J_{22}}\left(1-\sqrt{1-\frac{J_{11} J_{22}}{J_{12}^2}}\right),\\
     &\lambda_3^*=\frac{8 \pi^2}{3} \frac{\epsilon}{J_{12}},
    \end{align}
\end{subequations}
which is complex-valued due to restriction \eqref{eq:Jrest}.
A necessary condition for a stable Euclidean path integral is that the quartic potential is bounded
from below. In the $\lambda_{4,5}=0$ sector this requires $\lambda_1>0$, $\lambda_2>0$, and
$\lambda_3>-\frac13\sqrt{\lambda_1\lambda_2}$ (in our normalization).
We will implicitly restrict to this domain when interpreting fixed points.

\subsubsection*{Fully coupled quartic sector $\lambda_{4,5}\neq0$ and $h_i=0$ }
Turning on $\lambda_{4,5}$ always admits the trivial solution
$\lambda_4^*=\lambda_5^*=0$, which reduces to the parity-symmetric sector.
To obtain a genuinely parity-breaking fixed point with $\lambda_4^*\neq 0$ and
$\lambda_5^*\neq 0$, one must impose \emph{both} conditions
$\beta_{\lambda_4}=0$ and $\beta_{\lambda_5}=0$. Dividing these equations by
$\lambda_4$ and $\lambda_5$ yields a homogeneous linear system for $(\lambda_4,\lambda_5)$,
which has a nontrivial solution iff its determinant vanishes, i.e. $\Delta=0$ below.

 At one loop this produces a one-parameter family (a fixed line), reflecting a marginal direction in the
quartic sector. Beyond one loop this degeneracy is expected to be lifted.
 In other words, for  $\lambda_3$ satisfying $\Delta_3 \geq 0$ (determinant $\Delta_3$ will also be defined below) with $\lambda_4^2, \lambda_5^2 \geq 0$ one can find $\lambda_{1,2,4,5}$ such that $\beta_{\lambda_i}=0$. Introducing the notation
\be 
\label{XandY}
\mathcal{X} \equiv J_{11}\lambda_1+2J_{12}\lambda_3-c,\qquad
\mathcal{Y} \equiv J_{22}\lambda_2+2J_{12}\lambda_3-c,
\ee 
we rewrite equations   $\beta_{\lambda_{4,5}}=0$ as
\be 
\mathcal{X}\,\lambda_4+J_{22}\lambda_3\lambda_5=0,\qquad
\mathcal{Y}\,\lambda_5+J_{11}\lambda_3\lambda_4=0,
\ee 
and condition $\beta_{\lambda_4}\cdot\beta_{\lambda_5}=0$  can be identically rewritten as   $\Delta=0$, where
\be \label{eq:det-1}
\Delta \equiv \mathcal{X} \mathcal{Y}-J_{11}J_{22} \lambda_3^2 ,\,\,\,\,c \equiv \frac{16 \pi^2}{3} \epsilon.
\ee
Thus we see that a fully parity-breaking fixed point (with $\lambda_4, \lambda_5 \neq 0$ ) must lie on the codimension-1 surface $\mathcal{X} \mathcal{Y}= J_{11} J_{22} \lambda_3^2$ in $\left(\lambda_1, \lambda_2, \lambda_3\right)$-space.
From $\beta_{\lambda_1}=\beta_{\lambda_2}=0$  we get the relations between different couplings at fixed point
\be\label{eq:addw}
\lambda_4^2=\frac{c \lambda_1-J_{11} \lambda_1^2-J_{22} \lambda_3^2}{2 J_{12}}, \quad \lambda_5^2=\frac{c \lambda_2-J_{22} \lambda_2^2-J_{11} \lambda_3^2}{2 J_{12}}
\ee
which also immediately imply constraints 
$$
c \lambda_1 \geq J_{11} \lambda_1^2+J_{22} \lambda_3^2, \quad c \lambda_2 \geq J_{22} \lambda_2^2+J_{11} \lambda_3^2 .
$$
Finally, using $\beta_{\lambda_3}=0$ and $\Delta=0$ one can obtain different branches of solution for $\lambda_{1,2}$ as functions of $\lambda_3$. 

\textit{(i)} For the first branch, the equation $\beta_{\lambda_3}=0$ together with $\Delta=0$ gives:
\begin{equation}
\mathcal{X}+\mathcal{Y}=2 J_{12} \lambda_3-c ,
\end{equation}
and
\begin{align}
    \mathcal{X} \mathcal{Y}= J_{11} J_{22} \lambda_3^2.
\end{align}
Then picking a real parameter $\lambda_3 \neq 0$ for  $\sigma= \pm 1$ we solve these two equations together and also define
$$
\begin{gathered}
\Delta_3\left(\lambda_3\right) \equiv\left(c-2 J_{12} \lambda_3\right)^2-4 J_{11} J_{22} \lambda_3^2 \geq 0, \\
\mathcal{X}_\sigma=\frac{-\left(c-2 J_{12} \lambda_3\right)+\sigma \sqrt{\Delta_3\left(\lambda_3\right)}}{2}, \quad \mathcal{Y}_\sigma=\frac{-\left(c-2 J_{12} \lambda_3\right)-\sigma \sqrt{\Delta_3\left(\lambda_3\right)}}{2},
\end{gathered}
$$
then $\mathcal{X}_\sigma \mathcal{Y}_\sigma=J_{11} J_{22} \lambda_3^2$ and
$$
\lambda_1=-\frac{\mathcal{Y}_\sigma}{J_{11}}, \quad \lambda_2=-\frac{\mathcal{X}_\sigma}{J_{22}}.
$$

\textit{(ii)} The second branch of formal solution has the following  form (with $\lambda_3 \neq 0$ and $\sigma= \pm 1$):
$$
\mathcal{X}_\sigma=\lambda_3\left(J_{12}+\sigma \sqrt{J_{12}^2-J_{11} J_{22}}\right), \quad \mathcal{Y}_\sigma=\lambda_3\left(J_{12}-\sigma \sqrt{J_{12}^2-J_{11} J_{22}}\right),
$$
so $\mathcal{X}_\sigma \mathcal{Y}_\sigma=J_{11} J_{22} \lambda_3^2$ and $\mathcal{X}_\sigma+\mathcal{Y}_\sigma=2 J_{12} \lambda_3$. Then, using \eqref{XandY}
$$
\lambda_1=\frac{\mathcal{X}_\sigma+c-2 J_{12} \lambda_3}{J_{11}}, \quad \lambda_2=\frac{\mathcal{Y}_\sigma+c-2 J_{12} \lambda_3}{J_{22}} ,
$$
and this branch is also complex-valued due to \eqref{eq:Jrest}.

The fully coupled fixed points for $\lambda_{4,5}>0$ do not come as isolated points but as a family, i.e. there is a marginal eigen-direction at one loop.

Also inequalities \eqref{eq:addw} imposes  for $\lambda_1, \lambda_2$ to lie in finite intervals for any given $\lambda_3$. So if such fixed points exist, they live in a bounded region of coupling space (at order $\epsilon$). The analysis of stability for these branches repeat the same line of reasoning as for isotropic case and is not affected significantly by anisotropy. In other words, the line of fixed points corresponding to the fully interacting $\lambda_i$ set of couplings in general has the same stability properties as the isotropic one, changing only location of lines. Anisotropy only restricts the existence of fixed points and 
shifts their location in coupling space. In particular, for generic initial conditions the RG flow runs away from the fully coupled fixed line.

\subsubsection*{Effect of cubic couplings}
\label{sec:effect_cubic}

The cubic interactions in \eqref{eq:L3L4} are parametrized by the dimensionless couplings $h_i$
(or, equivalently, $h_{ijk}$). In $D=4-\eps$ they are already \emph{relevant} at tree level:
their canonical RG contribution is
\be
\beta_{h_{ijk}}\supset -\frac{6-D}{2}\,h_{ijk}=-(1+\eps/2)\,h_{ijk}.
\ee
At one loop in MS the beta functions remain \emph{linear} in $h$ (
in the $\mathbf Z_t=\mathbf 1$, $\mathbf Y=0$ limit, the explicit formulas for $\beta_{h_1},\dots,\beta_{h_4}$
are given by \eqref{beta h y0 z1}).  Therefore the hyperplane $h_{ijk}=0$ is RG-invariant, but it is not IR-attractive:
generic infinitesimal cubic perturbations grow toward the IR and destabilize the quartic critical
manifolds unless $h$ is tuned to zero (or forbidden by a $\mathbb Z_2$ symmetry).


In the decoupled-gradient limit $\mathbf Z_t=\mathbf 1$, $\mathbf Y=0$, the one-loop pole coefficients
depend on anisotropy only through the three angular weights $J_{11},J_{22},J_{12}$ (or equivalently
through $\rho=J_{12}/\sqrt{J_{11}J_{22}}$).  In particular, after the constant (i.e.\ $\mu$-independent)
rescalings used in App.~\ref{app:cubic-sector}, the cubic RG flow can be written as a linear system
whose coefficients depend on the quartic couplings only via the rescaled variables $u_i$ on the
quartic fixed line (App.~\ref{app:stability_full}).

Concretely, introduce the rescaled quartics 
\begin{gather}
g_1\equiv J_{11}\lambda_1,\quad 
g_2\equiv J_{22}\lambda_2,\quad 
g_3\equiv \sqrt{J_{11}J_{22}}\,\lambda_3,\\
g_4\equiv \sqrt{J_{11}J_{12}}\,\lambda_4,\quad
g_5\equiv \sqrt{J_{22}J_{12}}\,\lambda_5, \quad
\rho\equiv \frac{J_{12}}{\sqrt{J_{11}J_{22}}}\in(0,1],
\end{gather}
and $u_i\equiv g_i/c$ with $c\equiv \frac{16\pi^2}{3}\eps$.  Next, rescale the cubics by the $\mu$-independent
transformation (App.~\ref{app:H-rescaling})
\begin{gather}
H_1 \equiv  h_1,\qquad
H_2 \equiv  \sqrt{\frac{J_{12}J_{22}}{J_{11}^2}}\,h_2,\qquad
H_3 \equiv  \sqrt{\frac{J_{12}}{J_{11}}}\,h_3,\\
H_4 \equiv  \sqrt{\frac{J_{22}}{J_{11}}}\,h_4,
\qquad
\bm H\equiv (H_1,H_2,H_3,H_4)^{\sf T}.
\end{gather}
Then the one-loop cubic flow takes the compact matrix form (App.~\ref{app:cubic-sector})
\be
\beta_{\bm H}
=
-\Bigl(1+\frac{\eps}{2}\Bigr)\bm H
+\frac{\eps}{3}\,\mathbb M_h(u,\rho)\,\bm H,
\label{eq:betaH_matrix_maintext}
\ee
with the explicit $4\times4$ matrix $\mathbb M_h$ given in \eqref{app:Mh-matrix}.


Evaluating $\mathbb M_h$ on the fully coupled quartic fixed line $u=u^\ast(r)$ (App.~\ref{app:cubic-mass-fixedline})
one finds a strong universality statement:
\be
\mathrm{spec}\bigl(\mathbb M_h(u^\ast,\rho)\bigr)=\{0,0,1,3\},
\ee
where ``spec'' means spectrum and it is
independent of the fixed-line parameter $r$ and also independent of the anisotropy ratio $\rho$
(App.~\ref{app:cubic-sector}).  The corresponding RG exponents are therefore
\be
\omega_h(x)= -\Bigl(1+\frac{\eps}{2}\Bigr)+\frac{\eps}{3}\,x,
\qquad x\in\{0,0,1,3\},
\ee
i.e.
\be
\omega_h=\Bigl\{-1-\frac{\eps}{2},\,-1-\frac{\eps}{2},\,-1-\frac{\eps}{6},\,-1+\frac{\eps}{2}\Bigr\}.
\label{eq:omega_h_list_maintext}
\ee
With the convention $\beta_g=\mu\frac{dg}{d\mu}$ (with fixed bare coupling), the IR corresponds to $\mu\to0$, so a perturbation scales
as $\delta h\sim \mu^{\omega_h}$ and thus grows in the IR when $\omega_h<0$.
Hence all cubic couplings are \emph{relevant} perturbations of the quartic fixed line for $\eps>0$.
In particular, within the perturbative $4-\eps$ expansion there is no interacting fixed point with
$h_i^*\neq0$: at one loop the only fixed locus in the cubic sector is $h^*=0$.

Two technical remarks sharpen the interpretation:
(i) the two zero eigenvalues of $\mathbb M_h$ do \emph{not} signal marginal cubic couplings; they merely
mean that two cubic combinations receive no quartic correction at one loop, yet they remain relevant due
to the canonical term $-(1+\eps/2)\bm H$;
(ii) anisotropy affects the \emph{eigenvectors} (which linear combinations of the original $h_i$ form the
scaling operators) through the $J$-dependent rescaling above, but it does not change the one-loop
\emph{relevance} pattern or the universal set of exponents.


In the isotropic limit $\mathbf C_1=\mathbf C_2=\mathbf 1$ one has $J_{11}=J_{22}=J_{12}=1$ and $\rho=1$,
so all rescalings become trivial.  The cubic beta functions then reduce to the familiar isotropic
two-scalar results, and the linearized cubic exponents around the fully coupled quartic fixed line are
exactly the same numbers \eqref{eq:omega_h_list_maintext}.  Thus, at one loop in $4-\eps$,
\emph{anisotropy does not generate new cubic fixed points and does not change the qualitative fate of
nonzero cubic couplings}: it only changes the map between the microscopic couplings $h_i$ and the
scaling eigenoperators.

The same conclusion is robust around the generalized Wilson--Fisher fixed point in the parity-symmetric
quartic sector: since $\lambda^*=\mathcal O(\eps)$, the loop mixing in $\beta_h$ is $\mathcal O(\eps)\,h$
and cannot overcome the order-one canonical term $-(1+\eps/2)h$ for small $\eps$, so cubic perturbations
remain relevant near $D=4$.


Allowing $h\neq 0$ also activates the quadratic/linear sectors.  First, the mass beta functions contain
an \emph{inhomogeneous} contribution $\beta_{m^2}\supset +h^2$ already at one loop, so the critical surface
$m^2=0$ is not RG-invariant unless $h=0$.  Second, when both $h\neq 0$ and $m^2\neq 0$ are present, the RG
flow generates linear (tadpole) couplings $j_i$ through the one-point divergence discussed around
\eqref{Gamma1}.  Therefore a quartic critical fixed point/fixed line describes the critical manifold
only after tuning to the symmetry-preserving surface
\be
h_i=0,\qquad m^2_{ij}=0,\qquad j_i=0,
\ee
while generic flows with $h\neq 0$ depart from criticality by generating $m^2$ (and then $j$).
For completeness, the linearized mass spectrum around the fixed line at $h=0$ is
$\omega_m=\{-2,-2,-2-\eps/3\}$ (App.~\ref{app:mass-sector}), so all quadratic deformations are relevant,
as expected for a critical theory in $4-\eps$ dimensions.
\subsection{Turning on cross-gradients: $\mathbf Z_t=\mathbf 1$ and $\mathbf Y\neq 0$}
\label{sec:partic_Yneq0}

In the previous subsection we imposed $\mathbf Y=0$ and obtained a drastic simplification:
the UV kinetic operator was diagonal in field space and all direction dependence collapsed into
three scalar weights $J_{11},J_{22},J_{12}$.
Here we relax this truncation and keep
\be
\mathbf Z_t=\mathbf 1,\qquad \mathbf Y\neq 0,
\ee
while still working in the one--loop DR+MS scheme near $D=4-\eps$.
We also set $h_i=0$ throughout this subsection (the cubic sector can be reinstated later by
substituting the same bubble/tadpole residues into the general formulae of the previous section).


With $\mathbf Z_t=\mathbf 1$ the UV spatial kinetic matrix equals $\mathbf S(\hat{\bm n})=\mathbf C(\hat{\bm n})$.
For every direction $\hat{\bm n}\in S^2$ we define the three directional scalars
\be
c_\phi(\hat{\bm n})\equiv \hat{\bm n}^{\top}\mathbf C_\phi\,\hat{\bm n},\qquad
c_\chi(\hat{\bm n})\equiv \hat{\bm n}^{\top}\mathbf C_\chi\,\hat{\bm n},\qquad
y(\hat{\bm n})\equiv \hat{\bm n}^{\top}\mathbf Y\,\hat{\bm n},
\ee
so that
\be
\mathbf C(\hat{\bm n})=
\begin{pmatrix}
c_\phi(\hat{\bm n}) & y(\hat{\bm n})\\
y(\hat{\bm n}) & c_\chi(\hat{\bm n})
\end{pmatrix}.
\ee
The stability assumptions (no gradient instabilities) require $\mathbf C(\hat{\bm n})>0$ for all $\hat{\bm n}$,
i.e.
\be
c_\phi(\hat{\bm n})>0,\qquad c_\chi(\hat{\bm n})>0,\qquad
c_\phi(\hat{\bm n})\,c_\chi(\hat{\bm n})-y(\hat{\bm n})^2>0
\qquad(\forall\,\hat{\bm n}\in S^2).
\ee


Since $\mathbf C(\hat{\bm n})$ is a real symmetric $2\times2$ matrix, it can be diagonalized explicitly.
Introduce
\be
\Delta(\hat{\bm n})\equiv c_\phi(\hat{\bm n})-c_\chi(\hat{\bm n}),\qquad
R(\hat{\bm n})\equiv \sqrt{\Delta(\hat{\bm n})^2+4\,y(\hat{\bm n})^2}\;>\;0,
\ee
then the two (positive) directional eigenvalues are
\be
s_\pm(\hat{\bm n})
=\frac12\Big(c_\phi(\hat{\bm n})+c_\chi(\hat{\bm n})\pm R(\hat{\bm n})\Big),
\qquad s_\pm(\hat{\bm n})>0.
\label{eq:splusminus_Y}
\ee
Equivalently, one may parameterize the diagonalization by a mixing angle $\theta(\hat{\bm n})$ defined by
\be
\tan\big(2\theta(\hat{\bm n})\big)=\frac{2y(\hat{\bm n})}{c_\phi(\hat{\bm n})-c_\chi(\hat{\bm n})},
\qquad
\cos\big(2\theta\big)=\frac{\Delta}{R},\qquad
\sin\big(2\theta\big)=\frac{2y}{R}.
\label{eq:theta_Y}
\ee
Writing $c(\hat{\bm n})\equiv \cos\theta(\hat{\bm n})$ and $s(\hat{\bm n})\equiv \sin\theta(\hat{\bm n})$,
the (direction-dependent) orthogonal projectors onto the eigenmodes are
\begin{gather}
\Pi_+(\hat{\bm n})=
\begin{pmatrix}
c^2 & cs\\
cs & s^2
\end{pmatrix},
\qquad
\Pi_-(\hat{\bm n})=
\begin{pmatrix}
s^2 & -cs\\
-cs & c^2
\end{pmatrix},
\\
\Pi_\pm^2=\Pi_\pm,\quad \Pi_+\Pi_-=0,\quad \Pi_++\Pi_-=\mathbf 1.
\label{eq:Pi_pm_Y}
\end{gather}
In contrast to the $\mathbf Y=0$ truncation, the projectors now \emph{vary} with $\hat{\bm n}$, and this is
precisely what complicates the RG coefficients.


For the MS pole parts, it is sufficient to evaluate all angular averages at $d=3$ (i.e.\ over $S^2$),
as explained in Sec.~\ref{sec:master}. With $\mathbf Z_t=\mathbf 1$ the general one-loop bubble residue
\eqref{bubble answer} becomes
\begin{gather}
\mathcal B_{ij;kl}\;\equiv\;\eps\,\mathbb B_{ij;kl}
=\frac{1}{4\pi^2}\Big\langle
\sum_{a,b\in\{+,-\}}
\mathcal K\!\big(s_a(\hat{\bm n}),s_b(\hat{\bm n})\big)\;
\Pi_a(\hat{\bm n})_{ik}\,\Pi_b(\hat{\bm n})_{jl}
\Big\rangle_{S^2}.
\label{eq:Bpole_Y}
\end{gather}
Because $\Pi_\pm(\hat{\bm n})$ are no longer constant, the tensor $\mathcal B_{ij;kl}$ does not reduce to
three weights. Instead, in the original $(\phi,\chi)$ basis one finds that six components are generally
independent, and all others follow from the index symmetries
$\mathcal B_{ij;kl}=\mathcal B_{ji;lk}=\mathcal B_{kl;ij}=\mathcal B_{kj;il}=\mathcal B_{il;kj}$.

To keep the notation parallel to the $\mathbf Y=0$ discussion, we package them into six angular weights
(with the conventional $8\pi^2$ normalization):
\be
\begin{gathered}
J_{11}^{(Y)}\equiv 8\pi^2\,\mathcal B_{11;11},\qquad
J_{22}^{(Y)}\equiv 8\pi^2\,\mathcal B_{22;22},\qquad
J_{12}^{(Y)}\equiv 8\pi^2\,\mathcal B_{12;12},\\[2pt]
J_{1122}^{(Y)}\equiv 8\pi^2\,\mathcal B_{11;22},\qquad
J_{1112}^{(Y)}\equiv 8\pi^2\,\mathcal B_{11;12},\qquad
J_{2212}^{(Y)}\equiv 8\pi^2\,\mathcal B_{22;12}.
\end{gathered}
\label{eq:Jsix_def}
\ee
They can be written explicitly as $S^2$ averages of elementary functions of $s_\pm(\hat{\bm n})$ and
$\theta(\hat{\bm n})$. Introducing the shorthand
\be
\mathcal K_{++}\equiv \mathcal K(s_+,s_+)=\frac{1}{2\,s_+^{3/2}},\qquad
\mathcal K_{--}\equiv \frac{1}{2\,s_-^{3/2}},\qquad
\mathcal K_{+-}\equiv \mathcal K(s_+,s_-),
\ee
one finds \emph{pointwise} in $\hat{\bm n}$ the integrands
\be
\begin{aligned}
J_{11}^{(Y)}&=2\Big\langle \mathcal K_{++}\,c^4+\mathcal K_{--}\,s^4+2\mathcal K_{+-}\,c^2s^2\Big\rangle_{S^2},\\
J_{22}^{(Y)}&=2\Big\langle \mathcal K_{++}\,s^4+\mathcal K_{--}\,c^4+2\mathcal K_{+-}\,c^2s^2\Big\rangle_{S^2},\\
J_{1122}^{(Y)}&=2\Big\langle \big(\mathcal K_{++}+\mathcal K_{--}-2\mathcal K_{+-}\big)\,c^2s^2\Big\rangle_{S^2},\\
J_{12}^{(Y)}&=2\Big\langle \mathcal K_{+-}+\big(\mathcal K_{++}+\mathcal K_{--}-2\mathcal K_{+-}\big)\,c^2s^2\Big\rangle_{S^2},\\
J_{1112}^{(Y)}&=2\Big\langle cs\Big(\mathcal K_{++}\,c^2-\mathcal K_{--}\,s^2+\mathcal K_{+-}(s^2-c^2)\Big)\Big\rangle_{S^2},\\
J_{2212}^{(Y)}&=2\Big\langle cs\Big(\mathcal K_{++}\,s^2-\mathcal K_{--}\,c^2+\mathcal K_{+-}(c^2-s^2)\Big)\Big\rangle_{S^2}.
\end{aligned}
\label{eq:Jsix_explicit}
\ee
When $\mathbf Y\to 0$ the mixing angle becomes $\theta\to 0$ (or $\pi/2$) and therefore
$cs\to 0$ and $c^2s^2\to 0$; in this limit
\be
J_{1112}^{(Y)}\to 0,\qquad J_{2212}^{(Y)}\to 0,\qquad J_{1122}^{(Y)}\to 0,
\ee
while $J_{11}^{(Y)},J_{22}^{(Y)},J_{12}^{(Y)}$ reduce precisely to the three weights $J_{11},J_{22},J_{12}$
of the $\mathbf Y=0$ truncation.


With $h_i=0$ the one-loop flow of quartic couplings is closed.
For practical analysis it is convenient to absorb the overall ``anisotropy weights''
by a constant rescaling of couplings.
Define the rescaled quartics
\be
\begin{aligned}
g_1&\equiv J_{11}^{(Y)}\,\lambda_1,\qquad
g_2\equiv J_{22}^{(Y)}\,\lambda_2,\qquad
g_3\equiv \sqrt{J_{11}^{(Y)}J_{22}^{(Y)}}\,\lambda_3,\\
g_4&\equiv \sqrt{J_{11}^{(Y)}\big(J_{12}^{(Y)}+J_{1122}^{(Y)}\big)}\,\lambda_4,\qquad
g_5\equiv \sqrt{J_{22}^{(Y)}\big(J_{12}^{(Y)}+J_{1122}^{(Y)}\big)}\,\lambda_5,
\end{aligned}
\label{eq:rescaled_gY}
\ee
and the dimensionless ratios
\be
\rho\equiv \frac{J_{12}^{(Y)}}{\sqrt{J_{11}^{(Y)}J_{22}^{(Y)}}},\qquad
\sigma\equiv \frac{J_{1122}^{(Y)}}{\sqrt{J_{11}^{(Y)}J_{22}^{(Y)}}},\qquad
\widetilde\rho\equiv\rho+\sigma=\frac{J_{12}^{(Y)}+J_{1122}^{(Y)}}{\sqrt{J_{11}^{(Y)}J_{22}^{(Y)}}},
\label{eq:ratios_rhosigma}
\ee
together with
\be
\alpha\equiv \frac{J_{1112}^{(Y)}}{\sqrt{J_{11}^{(Y)}\big(J_{12}^{(Y)}+J_{1122}^{(Y)}\big)}},
\qquad
\beta\equiv \frac{J_{2212}^{(Y)}}{\sqrt{J_{22}^{(Y)}\big(J_{12}^{(Y)}+J_{1122}^{(Y)}\big)}},
\qquad
\kappa\equiv \frac{\rho+3\sigma}{\rho+\sigma}.
\label{eq:ratios_alphabeta}
\ee
In terms of $(g_1,\dots,g_5)$ the one-loop beta functions take the compact form
\begin{subequations}
\label{eq:beta_g_Y}
\begin{align}
\beta_{g_1}=&-\eps\,g_1+\frac{3}{16\pi^2}\Big[
g_1^2+g_3^2+2g_4^2+2\sigma\,g_1g_3+4\alpha\,g_1g_4+4\beta\,g_3g_4
\Big],\\
\beta_{g_2}=&-\eps\,g_2+\frac{3}{16\pi^2}\Big[
g_2^2+g_3^2+2g_5^2+2\sigma\,g_2g_3+4\beta\,g_2g_5+4\alpha\,g_3g_5
\Big],\\
\beta_{g_3}=&-\eps\,g_3+\frac{1}{16\pi^2}\Big[
(g_1+g_2)\,g_3+\sigma\,g_1g_2+(4\rho+5\sigma)\,g_3^2
+\frac{2}{\widetilde\rho}(g_4^2+g_5^2)\nonumber\\
&+2\kappa\,g_4g_5+2\alpha\,(g_1g_5+5g_3g_4)+2\beta\,(g_2g_4+5g_3g_5)
\Big],\\
\beta_{g_4}=&-\eps\,g_4+\frac{3}{16\pi^2}\Big[
g_1g_4+\sigma\,g_1g_5+(2\rho+3\sigma)\,g_3g_4+g_3g_5\nonumber\\
&+2\widetilde\rho\,\alpha\,g_1g_3+2\alpha\,g_4^2
+2\widetilde\rho\,\beta\,g_3^2+2\beta\,g_4g_5
\Big],\\
\beta_{g_5}=&-\eps\,g_5+\frac{3}{16\pi^2}\Big[
g_2g_5+\sigma\,g_2g_4+g_3g_4+(2\rho+3\sigma)\,g_3g_5\nonumber\\
&+2\widetilde\rho\,\beta\,g_2g_3+2\beta\,g_5^2
+2\widetilde\rho\,\alpha\,g_3^2+2\alpha\,g_4g_5
\Big].
\end{align}
\end{subequations}

Undoing the rescaling \eqref{eq:rescaled_gY} reproduces the explicit $\beta_{\lambda_i}$
expressions \eqref{betah_lambda_Y}.
The $\mathbf Y=0$ limit is obtained by setting
$J_{1112}^{(Y)}=J_{2212}^{(Y)}=J_{1122}^{(Y)}=0$, i.e.\ $\sigma=\alpha=\beta=0$
and $\widetilde\rho=\rho$, which reduces \eqref{eq:beta_g_Y} to the earlier
$\mathbf Y=0$ system.

In this limit the rescaled couplings reduce smoothly to the variables used in the
$\mathbf Y=0$ analysis,
\begin{align*}
g_1=&J_{11}\lambda_1,\qquad
g_2=J_{22}\lambda_2,\qquad
g_3=\sqrt{J_{11}J_{22}}\lambda_3,\\
&g_4=\sqrt{J_{11}J_{12}}\lambda_4,\qquad
g_5=\sqrt{J_{22}J_{12}}\lambda_5,
\end{align*}
and the beta functions written in terms of $g_i$ reduce exactly to the
$\mathbf Y=0$ flow depending only on $J_{ij}$ weights.

\subsection*{Interpretation of weights and effect of anisotropy. Physical interpretation of the $Y\neq 0$ weights.}
In the $Z_t=\mathbf 1$ truncation the cross--gradient matrix $\mathbf Y$ does not run at one loop in
DR+MS (no momentum--dependent poles), but it \emph{does} enter the RG coefficients through the
bubble residue
$\mathcal B_{ij;kl}\equiv \epsilon\,\mathbb B_{ij;kl}(0)$.
When $\mathbf Y\neq 0$ the matrix $\mathbf C(\hat{\bm n})$ is non--diagonal in the $(\phi,\chi)$ basis for
generic directions $\hat{\bm n}\in S^2$, so the one--loop coefficients are governed not by three but by
six independent angular weights, defined in \eqref{eq:Jsix_def}.
Below we give a Wilsonian/phase--space interpretation of these weights.

\medskip
\noindent
\textit{Bubble pole as a sum over intermediate two--particle states.}
At zero external momentum the bubble master integral is
\be
\label{bubble (i)}
\mathbb B_{ij;kl}(0)=\int\!\frac{d\omega}{2\pi}\frac{d^3\mathbf p}{(2\pi)^3}\,
G_{ik}(\omega,\mathbf p)\,G_{jl}(\omega,\mathbf p),
\ee
and in the UV (massless) regime we may use the spectral representation of the
directional field--space matrix $\mathbf C(\hat{\bm n})$:
\be
\mathbf C(\hat{\bm n})=\sum_{a=\pm}s_a(\hat{\bm n})\,\Pi_a(\hat{\bm n}),
\qquad s_a(\hat{\bm n})>0,\qquad \Pi_a\Pi_b=\delta_{ab}\Pi_a,\qquad \Pi_++\Pi_-=\mathbf 1.
\ee
Defining the directional velocities $v_a(\hat{\bm n})\equiv \sqrt{s_a(\hat{\bm n})}$, the UV energies of
the normal modes are
\be
E_a(\mathbf p)=|\mathbf p|\,v_a(\hat{\bm n}),\qquad \hat{\bm n}\equiv \mathbf p/|\mathbf p|,
\ee
and the propagator decomposes as
\be
G(\omega,\mathbf p)=\sum_{a=\pm}\frac{\Pi_a(\hat{\bm n})}{\omega^2+E_a(\mathbf p)^2}.
\ee
Substituting this into the bubble \eqref{bubble (i)} and integrating over $\omega$ gives the standard two--particle energy
denominator:
\be
\int_{-\infty}^{\infty}\frac{d\omega}{2\pi}\,
\frac{1}{(\omega^2+E_a^2)(\omega^2+E_b^2)}
=\frac{1}{2E_aE_b(E_a+E_b)}.
\ee
Therefore the bubble is a sum over intermediate two--particle normal--mode states $(a,b)$,
weighted by the overlaps (``coherence factors'') encoded in the projectors:
\be
\label{eq:bubble_Edenom}
\mathbb B_{ij;kl}(0)=
\sum_{a,b=\pm}\int\!\frac{d^3\mathbf p}{(2\pi)^3}\;
\frac{\Pi_a(\hat{\bm n})_{ik}\,\Pi_b(\hat{\bm n})_{jl}}
{2\,E_a(\mathbf p)\,E_b(\mathbf p)\,\big(E_a(\mathbf p)+E_b(\mathbf p)\big)}.
\ee
This form makes the physical content transparent: the one--loop correction to a four--point amplitude is
a second--order quantum process in which two external fields create an intermediate fast pair with total
energy $E_a+E_b$, which is then annihilated back into the external fields; the factor
$1/(E_a+E_b)$ is precisely the energy denominator of that virtual two--particle state.

\medskip
\noindent
\textit{Spectral (K{\"a}ll{\'e}n--Lehmann--type) viewpoint.}
It is useful to view the bubble \eqref{bubble main} as a Euclidean two--point function of a bilinear field \textit{density}.
To make the spectral structure explicit, consider the bubble \eqref{bubble main} with a purely Euclidean external energy
$k_0$ and vanishing external spatial momentum:
\be
\label{b (ii)}
\mathbf B_{ij;kl}(k_0)\;\equiv\;
\int\!\frac{d\omega}{2\pi}\frac{d^3\mathbf p}{(2\pi)^3}\,
G_{ik}(\omega,\mathbf p)\,G_{jl}(\omega+k_0,\mathbf p).
\ee
In the massless UV regime we may use the normal--mode decomposition
$$G(\omega,\mathbf p)=\sum_a \frac{\Pi_a(\hat{\bm n})}{(\omega^2+E_a(\mathbf p)^2)},$$ where
$E_a(\mathbf p)=|\mathbf p|\,v_a(\hat{\bm n})$ and $\hat{\bm n}=\mathbf p/|\mathbf p|$.
Performing the $\omega$ integral in \eqref{b (ii)} gives 
\be
\int_{-\infty}^{\infty}\frac{d\omega}{2\pi}\,
\frac{1}{(\omega^2+E_a^2)\big((\omega+k_0)^2+E_b^2\big)}
=
\frac{E_a+E_b}{2E_aE_b\big(k_0^2+(E_a+E_b)^2\big)},
\ee
thus we arrive at
\be
\mathbf{B}_{i j ; k l}\left(k_0\right)=\sum_{a, b} \int \frac{d^3 \mathbf{p}}{(2 \pi)^3} \Pi_a(\hat{\mathbf{n}})_{i k} \Pi_b(\hat{\mathbf{n}})_{j l} \frac{E_a+E_b}{2 E_a E_b\left(k_0^2+\left(E_a+E_b\right)^2\right)} .
\ee
Next, introduce the positive two--particle spectral density 
\bea
&\rho^{(2)}_{ij;kl}(\Omega)
\;\equiv\;
\sum_{a,b}\int\!\frac{d^3\mathbf p}{(2\pi)^3}\,
\frac{\Pi_a(\hat{\bm n})_{ik}\,\Pi_b(\hat{\bm n})_{jl}}{4E_a(\mathbf p)E_b(\mathbf p)}\,
\delta\!\big(\Omega-(E_a(\mathbf p)+E_b(\mathbf p))\big),
\nonumber\\ &\rho_{ij;kl}^{(2)}(\Omega)\succeq 0,
\eea
and the positivity is immediate, since $\Pi_a$ are Hermitian projectors and $E_a>0$.
Then the bubble \eqref{b (ii)} admits the following K{\"a}ll{\'e}n--Lehmann--type representation
\be
\mathbf B_{ij;kl}(k_0)
=
\int_0^\infty d\Omega^2\,
\frac{\rho^{(2)}_{ij;kl}(\Omega)}{k_0^2+\Omega^2}\,.
\label{eq:bubble_spectral_Stieltjes}
\ee
At $k_0=0$ this reduces to
$\mathbf B_{ij;kl}(0)=2\int_0^\infty \frac{d\Omega}{\Omega}\,\rho^{(2)}_{ij;kl}(\Omega)$,
which makes explicit the logarithmic UV sensitivity ($\int d\Omega/\Omega$) that becomes a $1/\epsilon$
pole in DR+MS.

Evaluating $\rho_{ij;kl}^{(2)}$ in spherical variables
$\mathbf p=\rho\,\hat{\bm n}$ gives an $\Omega$--independent integrand and reproduces the universal
kernel:
\bea
&\rho^{(2)}_{ij;kl}
=
\frac{1}{8\pi^2}\Big\langle
\sum_{a,b}\Pi_a(\hat{\bm n})_{ik}\,\Pi_b(\hat{\bm n})_{jl}\,
\mathcal K\!\big(s_a(\hat{\bm n}),s_b(\hat{\bm n})\big)
\Big\rangle_{S^2},
\\
&\mathcal K(s_a,s_b)=\frac{1}{v_av_b(v_a+v_b)}.
\eea
Thus $\mathcal K$ controls the spectral weight of intermediate two--particle states, while the projector
entries provide the coherence factors (overlaps with the microscopic fields). Moreover, using \eqref{eq:bubble_spectral_Stieltjes} one may straightforwardly show, that 
 in a Lorentz's signature continuation
$\rho_{ij;kl}^{(2)}$ becomes the discontinuity across the corresponding two--particle cut.

\medskip
\noindent
\textit{Direction--dependent mixing and coherence factors.}
Turning on $\mathbf Y$ makes $\mathbf C(\hat{\bm n})$ non--diagonal in $(\phi,\chi)$, so the fast normal
modes are direction--dependent linear combinations of $\phi$ and $\chi$.
It is convenient to parametrize the projectors by a mixing angle $\theta(\hat{\bm n})$:
\be
\tan 2\theta(\hat{\bm n})=
\frac{2y(\hat{\bm n})}{c_\phi(\hat{\bm n})-c_\chi(\hat{\bm n})},
\qquad
c_{\phi,\chi}(\hat{\bm n})=\hat{\bm n}^{\top}\mathbf C_{\phi,\chi}\hat{\bm n},
\quad
y(\hat{\bm n})=\hat{\bm n}^{\top}\mathbf Y\,\hat{\bm n},
\ee
so that
\be
\Pi_+(\hat{\bm n})=
\begin{pmatrix}
\cos^2\theta & \sin\theta\cos\theta\\
\sin\theta\cos\theta & \sin^2\theta
\end{pmatrix},
\quad
\Pi_-(\hat{\bm n})=
\begin{pmatrix}
\sin^2\theta & -\sin\theta\cos\theta\\
-\sin\theta\cos\theta & \cos^2\theta
\end{pmatrix}.
\ee
Define the ``populations'' $p_a(\hat{\bm n}), \; q_a(\hat{\bm n})$ and ``coherence'' $r_a(\hat{\bm n})$ of mode $a$ by
\be
p_a(\hat{\bm n})\equiv (\Pi_a)_{11},\qquad
q_a(\hat{\bm n})\equiv (\Pi_a)_{22},\qquad
r_a(\hat{\bm n})\equiv (\Pi_a)_{12}.
\ee
Then $p_a,q_a\in[0,1]$ measure how much of $\phi$ or $\chi$ participates in the fast mode $a$,
while $r_a=\pm\frac12\sin 2\theta$ measures the off--diagonal overlap, i.e. coherence.
Under $\mathbf Y\to-\mathbf Y$ one has $\theta\to-\theta$, hence $p_a,q_a$ are even while $r_a$ is odd.
This parity property will immediately explain which weights are even/odd in $\mathbf Y$.

\medskip
\noindent
\textit{The six weights as phase--space--weighted moments of populations and coherence.}
In the $Z_t=\mathbf 1$ truncation, the residue $\mathcal B$ can be viewed as a symmetric $3\times3$
matrix acting on the space of symmetric bilinears
$(\phi^2,\chi^2,\sqrt2\,\phi\chi)$.
Equivalently, the six independent entries of $\mathcal B$ may be packaged into the six scalar weights
defined in \eqref{eq:Jsix_def}. A convenient explicit form is
\bea
J_{11}^{(Y)}&=2\Big\langle\sum_{a,b} \mathcal K(s_a,s_b)\,p_a\,p_b\Big\rangle_{S^2},\qquad
J_{22}^{(Y)}=2\Big\langle\sum_{a,b} \mathcal K(s_a,s_b)\,q_a\,q_b\Big\rangle_{S^2},\nonumber\\
J_{12}^{(Y)}&=2\Big\langle\sum_{a,b} \mathcal K(s_a,s_b)\,p_a\,q_b\Big\rangle_{S^2},\qquad
J_{1122}^{(Y)}=2\Big\langle\sum_{a,b} \mathcal K(s_a,s_b)\,r_a\,r_b\Big\rangle_{S^2},\nonumber\\
J_{1112}^{(Y)}&=2\Big\langle\sum_{a,b} \mathcal K(s_a,s_b)\,p_a\,r_b\Big\rangle_{S^2},\qquad
J_{2212}^{(Y)}=2\Big\langle\sum_{a,b} \mathcal K(s_a,s_b)\,q_a\,r_b\Big\rangle_{S^2}.
\label{eq:J_moments_phase_space}
\eea
These formulae make the interpretation immediate:
\begin{itemize}
\item $J^{(Y)}_{11},J^{(Y)}_{22}$ are \emph{population--population} moments: they count how much
two--particle UV phase space is ``seen'' as $(\phi,\phi)$ or $(\chi,\chi)$ on the internal lines.
They control the strength of purely $\phi$ and purely $\chi$ loop renormalizations in
$\beta_{\lambda_1}$ and $\beta_{\lambda_2}$.

\item $J^{(Y)}_{12}$ is the \emph{cross} population moment: it counts how often an intermediate fast
pair looks $\phi$--like on one internal line and $\chi$--like on the other. It generalizes the familiar
$Y=0$ weight $J_{12}$.

\item $J^{(Y)}_{1122}$ is the basic \emph{coherence--coherence} moment: it measures the amount of
two--particle phase space carried by genuinely mixed fast modes (neither purely $\phi$ nor purely
$\chi$). It is even in $\mathbf Y$.

\item $J^{(Y)}_{1112},J^{(Y)}_{2212}$ are \emph{population--coherence interference} moments: they
measure a directed bias correlating the sign of mixing (the sign of $r_a\propto \sin2\theta$) with the
$\phi$-- or $\chi$--content of the fast modes. They are odd in $\mathbf Y$ and vanish whenever the
phase--space average has no net mixing sign bias.
\end{itemize}

\medskip
\noindent
\textit{Positivity and useful inequalities.}
Since $\mathcal K(s_a,s_b)>0$ and the projectors are real, $\mathcal B$ defines a positive
bilinear form on symmetric bilinears, hence its $3\times3$ matrix is positive semidefinite.
As a result, the $J$--weights satisfy Cauchy--Bunyakovsky--Schwarz--type inequalities (useful as consistency checks),
for example
\be
\big(J^{(Y)}_{12}\big)^2\le J^{(Y)}_{11}\,J^{(Y)}_{22},\qquad
\big(J^{(Y)}_{1112}\big)^2\le J^{(Y)}_{11}\,J^{(Y)}_{1122},\qquad
\big(J^{(Y)}_{2212}\big)^2\le J^{(Y)}_{22}\,J^{(Y)}_{1122}.
\ee
Moreover, $J_{1122}^{(Y)}\ge 0$ is manifest from \eqref{eq:J_moments_phase_space} together with the
pointwise identity (with $w_{ab}\equiv \mathcal K(s_a,s_b)$ and $v_a=\sqrt{s_a}$)
\be
\sum_{a,b} w_{ab}\,r_a r_b
=\frac{\sin^2 2\theta}{4}\,\Big(w_{++}+w_{--}-2w_{+-}\Big),
\ee
where
\be
w_{++}+w_{--}-2w_{+-}
=\frac{(v_+-v_-)^2\,(v_+^2+3v_+v_-+v_-^2)}{2v_+^3v_-^3(v_++v_-)}\ge 0.
\ee
Thus $J^{(Y)}_{1122}$ is enhanced both when mixing is strong ($|\sin2\theta|\sim 1$) and when the two
mode velocities are split ($|v_+-v_-|\neq 0$), while it vanishes if either mixing is absent
($\sin2\theta=0$) \emph{or} if the dispersions are degenerate ($v_+=v_-$), reflecting the fact that
``coherence'' becomes unobservable when the kinetic operator is proportional to the identity in field
space.

\medskip
\noindent
\textit{How the weights enter $\beta$--functions (diagrammatic meaning).}
At one loop the quartic beta functions arise from fish diagrams and are quadratic in $\lambda$.
At the tensor level
\be
\beta_{\lambda_{ijkl}}=-\epsilon\,\lambda_{ijkl}
+\frac12\Big(\lambda_{ijmn}\mathcal B_{mn;pq}\lambda_{klpq}
+\lambda_{ikmn}\mathcal B_{mn;pq}\lambda_{jlpq}
+\lambda_{ilmn}\mathcal B_{mn;pq}\lambda_{jkpq}\Big),
\ee
so every monomial in $\beta$ can be read as a specific \emph{two--step scattering
process} weighted by a corresponding entry of $\mathcal B$.
In particular:
\begin{itemize}
\item Terms weighted by $J^{(Y)}_{11},J^{(Y)}_{22},J^{(Y)}_{12}$ are the familiar contributions from
intermediate pairs that are predominantly $\phi\phi$, $\chi\chi$, or mixed $\phi\chi$ in the UV.

\item Terms weighted by $J^{(Y)}_{1122}$ originate from loops in which at least one internal contraction
uses the off--diagonal propagator component $G_{\phi\chi}$, i.e.\ from intermediate fast pairs that
propagate as direction--dependent linear combinations of $\phi$ and $\chi$.
These are new quantum channels that are absent when $\mathbf Y=0$.

\item Terms weighted by $J^{(Y)}_{1112}$ and $J^{(Y)}_{2212}$ are interference contributions between
``population'' and ``coherence'' channels. Being odd in $\mathbf Y$, they quantify the extent to which
the UV mixing texture has a directed bias and therefore can feed parity--breaking couplings in the
beta functions.
\end{itemize}
In this sense the six weights provide a compact bookkeeping of how the kinetic mixing $\mathbf Y$
opens additional virtual intermediate states in one--loop quantum processes and redistributes the
phase--space weight among them.

\medskip
\noindent
\textit{Isotropic simplification and the role of field redefinitions.}
If $\mathbf C_\phi=c_\phi\,\mathbf 1$, $\mathbf C_\chi=c_\chi\,\mathbf 1$, $\mathbf Y=y\,\mathbf 1$, then
$\mathbf C(\hat{\bm n})$ is independent of $\hat{\bm n}$, so $v_\pm$ and $\theta$ are constants and all
angular averages become trivial. The weights \eqref{eq:J_moments_phase_space} reduce to explicit algebraic
combinations of the three numbers $w_{++},w_{--},w_{+-}$ and the constant angle $\theta$.
Moreover, in this isotropic case there exists a \emph{global} $SO(2)$ field rotation that diagonalizes
$\mathbf C$ once and for all, removing $y$ from the kinetic term; in that rotated basis the propagator
is diagonal and the one--loop coefficients reduce to the $\mathbf Y=0$ structure (three weights), while
the effect of mixing is transferred entirely to the interaction tensors $(h,\lambda)$ by the same
rotation.
For generic anisotropic matrices $\mathbf C_{\phi,\chi}$ and $\mathbf Y$, however,
$\theta(\hat{\bm n})$ depends on direction and no single global rotation diagonalizes
$\mathbf C(\hat{\bm n})$ for all $\hat{\bm n}$; the six weights in \eqref{eq:Jsix_def} are then genuine
invariants of the UV kinetic data and encode the physically relevant directional mixing of fast modes.

$\,$
\subsection*{Isotropic limit for $\mathbf{Y} \neq 0$}
A useful consistency check is the fully homogeneous (isotropic) case
\be
\mathbf C_\phi=c_\phi\,\mathbf 1,\qquad
\mathbf C_\chi=c_\chi\,\mathbf 1,\qquad
\mathbf Y=y\,\mathbf 1,
\qquad c_\phi>0,\ c_\chi>0,\ c_\phi c_\chi-y^2>0.
\ee
Then for any $\hat{\bm n}\in S^2$ one has
$c_\phi(\hat{\bm n})=c_\phi$, $c_\chi(\hat{\bm n})=c_\chi$, $y(\hat{\bm n})=y$, and therefore
the field--space matrix is direction--independent:
\be
\mathbf C(\hat{\bm n})\equiv \mathbf C=
\begin{pmatrix}
c_\phi & y\\
y & c_\chi
\end{pmatrix}.
\ee
Its eigenvalues are obtained from $\det(\mathbf C-s\mathbf 1)=0$:
\be
s_\pm=\frac{c_\phi+c_\chi}{2}\pm \frac12\Delta,
\qquad
\Delta\equiv\sqrt{(c_\phi-c_\chi)^2+4y^2},
\qquad v_\pm\equiv \sqrt{s_\pm}>0.
\ee
A single global $SO(2)$ rotation diagonalizes $\mathbf C$,
$R(\theta)^{\top}\mathbf C R(\theta)=\mathrm{diag}(s_+,s_-)$, with
\be
\tan2\theta=\frac{2y}{c_\phi-c_\chi},
\qquad
\sin2\theta=\frac{2y}{\Delta},
\qquad
\cos2\theta=\frac{c_\phi-c_\chi}{\Delta}.
\ee
In terms of $c\equiv\cos\theta$, $s\equiv\sin\theta$ the constant projectors read
\be
\Pi_+=
\begin{pmatrix}
c^2 & sc\\
sc & s^2
\end{pmatrix},
\qquad
\Pi_-=
\begin{pmatrix}
s^2 & -sc\\
-sc & c^2
\end{pmatrix}.
\ee
Since $s_\pm$ and $\Pi_\pm$ are constant, all angular averages trivialize and the six weights
\eqref{eq:Jsix_def} reduce to elementary combinations of the three kernel values
\begin{align}
&w_{++}=\mathcal K(s_+,s_+)=\frac{1}{2s_+^{3/2}},\\
&w_{--}=\mathcal K(s_-,s_-)=\frac{1}{2s_-^{3/2}}, \\ &w_{+-}=K(s_+,s_-) = \frac{1}{\sqrt{s_{+}} \sqrt{s_{-}}\left(\sqrt{s_{+}}+\sqrt{s_{-}}\right)}.
\end{align}
Explicitly one finds
\bea
J_{11}^{(Y)}&=2\Big(w_{++}c^4+w_{--}s^4+2w_{+-}s^2c^2\Big),\qquad
J_{22}^{(Y)}=2\Big(w_{++}s^4+w_{--}c^4+2w_{+-}s^2c^2\Big),\nonumber\\
J_{1122}^{(Y)}&=2s^2c^2\Big(w_{++}+w_{--}-2w_{+-}\Big)
=\frac12\sin^2(2\theta)\Big(w_{++}+w_{--}-2w_{+-}\Big),\nonumber\\
J_{12}^{(Y)}&=2\Big(w_{+-}+s^2c^2(w_{++}+w_{--}-2w_{+-})\Big)
=2w_{+-}+J_{1122}^{(Y)},\nonumber\\
J_{1112}^{(Y)}&=2sc\Big(c^2(w_{++}-w_{+-})-s^2(w_{--}-w_{+-})\Big),\\
J_{2212}^{(Y)}&=2sc\Big(s^2(w_{++}-w_{+-})-c^2(w_{--}-w_{+-})\Big).
\label{eq:Jhomogeneous}
\eea
These identities immediately imply nontrivial relations among the six weights:
\be
J_{12}^{(Y)}-J_{1122}^{(Y)}=2w_{+-},\qquad
J_{11}^{(Y)}+J_{22}^{(Y)}+2J_{1122}^{(Y)}=2(w_{++}+w_{--})=s_+^{-3/2}+s_-^{-3/2},
\ee
showing that in the homogeneous case the apparent six parameters are in fact rigidly determined
by the three kinetic constants $(c_\phi,c_\chi,y)$ (equivalently $(s_+,s_-,\theta)$).

For linear dispersions $E_\pm(\mathbf p)=|\mathbf p|\,v_\pm$ the constant--energy surfaces
$\Sigma_\Omega^{ab}=\{\mathbf p:\ E_a(\mathbf p)+E_b(\mathbf p)=\Omega\}$ are spheres of radius
$|\mathbf p|=\Omega/(v_a+v_b)$. A direct evaluation gives the two--particle phase--space identity
\be
\int\frac{d^3\mathbf p}{(2\pi)^3}\,\frac{1}{4E_aE_b}\,
\delta\!\big(\Omega-(E_a+E_b)\big)
=\frac{1}{8\pi^2}\,\frac{1}{v_av_b(v_a+v_b)}
=\frac{1}{8\pi^2}\,\mathcal K(s_a,s_b),
\ee
so the kernel $\mathcal K$ is literally the (direction--independent) two--particle density of states per logarithmic interval in the RG scale,
while the weights \eqref{eq:Jhomogeneous} are obtained by dressing this phase space by the constant
coherence factors in the projectors $\Pi_\pm$.

In this isotropic case one may diagonalize the kinetic term by the global field rotation
$(\varphi_+,\varphi_-)^{\top}=R(\theta)^{\top}(\phi,\chi)^{\top}$, which removes $y$ from the quadratic
action:
\(
\frac12\,c_\phi(\nabla\phi)^2+\frac12\,c_\chi(\nabla\chi)^2+y\,\nabla\phi\!\cdot\!\nabla\chi
=\frac12\,s_+(\nabla\varphi_+)^2+\frac12\,s_-(\nabla\varphi_-)^2.
\)
In the eigenbasis the bubble residue is of the three--weight $Y=0$ form with constants
$J_+=s_+^{-3/2}$, $J_-=s_-^{-3/2}$ and
$J_{+-}=2/(\sqrt{s_+}\sqrt{s_-}(\sqrt{s_+}+\sqrt{s_-}))$,
while the interaction couplings are simply rotated by the same $R(\theta)$.
In particular, the homogeneous limit $y\to0$ gives $\theta\to0$ and
$J_{1122}^{(Y)}=J_{1112}^{(Y)}=J_{2212}^{(Y)}\to0$, and one recovers the isotropic $Y=0$ weights.

\section{Conclusion}

In this work we have presented a systematic one--loop renormalization analysis of a two--scalar
quantum field theory with the most general two--derivative but Lorentz--violating quadratic
structure.  The theory allows for anisotropic spatial gradients, field--space mixing in time
derivatives, and cross--gradient terms, while retaining standard ($z=1$) power counting.
Even in this minimal setting, the renormalization structure turns out to be considerably richer
than in Lorentz--invariant scalar theories.

  Although Lorentz violation is introduced exclusively through
the kinetic sector, interactions inevitably generate an enlarged but closed set of couplings under
renormalization, including masses and linear (tadpole) terms.  As a consequence, a consistent
renormalization--group treatment cannot be formulated solely in terms of modified kinetic terms or
``running velocities''.  Instead, an RG--complete operator basis compatible with the reduced symmetry
must be retained from the outset.

Technically, we have shown that in dimensional regularization with minimal subtraction near
$D=4-\epsilon$ all one--loop ultraviolet divergences are governed by two universal master objects:
the tadpole with a single mass insertion and the zero--momentum bubble.  Using a spectral
decomposition of the anisotropic quadratic operator, we derived compact, basis--covariant expressions
for their pole parts in terms of the direction--dependent eigenvalues and projectors of the kinetic
matrix.  Both master residues are dressed by a single positive kernel, which admits a transparent
interpretation as a two--particle energy denominator, or equivalently as a two--particle phase--space
weight per logarithmic interval in the RG scale.  This representation makes the physical origin of all anisotropy--dependent
coefficients in the beta functions explicit.

A key structural consequence of locality at one loop is that, in the MS scheme, all two--point UV
poles are momentum independent.  As a result, the anisotropy matrices in the quadratic action do not
run at one loop, while masses, linear couplings, and interaction tensors do.  The full set of one--loop
beta functions for cubic and quartic couplings was obtained in closed tensor form in terms of the
master residues.

We then explored several consistent truncations of the general flow.  In the absence of kinetic
mixing in field space, all anisotropy dependence collapses into three geometric/phase--space weights,
which control the existence and stability of interacting fixed points.  In particular, we found that
the coupled Wilson--Fisher--type fixed point survives only when the mismatch between the two kinetic
sectors is sufficiently small, and disappears beyond a universal threshold.  In the fully coupled
quartic sector the one--loop flow admits a fixed line, which we showed to be of saddle type.  Cubic
interactions remain relevant near four dimensions and do not lead to interacting fixed points at
one loop, destabilizing the quartic critical manifolds unless tuned away by symmetry.  When
cross--gradients are present, the renormalization structure is governed by six independent angular
weights, which we interpreted as phase--space--weighted populations and coherences of the
direction--dependent ultraviolet eigenmodes.

There are several natural directions for future work.  First, extending the analysis beyond one loop
is essential to address the running of anisotropy matrices themselves and to assess the emergence
(or lack thereof) of effective Lorentz symmetry in the infrared.  Second, it would be interesting to
generalize the present spectral framework to theories with more fields, internal symmetries, or to
fermionic and gauge sectors with multiple effective metrics.  Third, the phase--space interpretation
of the anisotropy weights suggests a close connection to Wilsonian momentum--shell RG, which could be
made explicit in non--minimal schemes or in real--space RG formulations.  Finally, the framework
developed here provides a controlled setting for studying anisotropic critical phenomena and
multicritical points in coupled systems, both in high--energy and condensed--matter contexts, where
direction--dependent propagation and kinetic mixing are generic rather than exceptional.

The authors are grateful to Irina Aref'eva and Andrei Kataev for useful comments and fruitful discussions.
This work has been supported by Russian Science Foundation Grant No. 24-72-00121, \href{https://rscf.ru/project/24-72-00121/}{https://rscf.ru/project/24-72-00121/}.
\appendix

\section{Explicit evaluations of master integrals}
\label{app:tad}

This appendix collects the explicit and step-by-step regularization of two master-integrals from section \ref{sec:master}. We begin with the  evaluation of tadpole master integral \eqref{eq:T begin}, when turn to bubble one.
In dimensional regularization, massless vacuum integrals at zero external momentum are scaleless.
To isolate the UV $1/\epsilon$ poles in the MS scheme we evaluate the master integrals with an auxiliary
IR regulator mass $\delta_{\rm IR}>0$, implemented by the shift
\be
\omega^2\mathbf 1+\mathbf K(\mathbf p)\;\longrightarrow\;
\omega^2\mathbf 1+\mathbf K(\mathbf p)+\delta_{\rm IR}^2\mathbf 1.
\ee
Equivalently, each propagator denominator becomes $\omega^2+\kappa_a(\mathbf p)+\delta_{\rm IR}^2$ in the
spectral representation. This makes all Schwinger/proper-time integrals convergent at large proper time
and removes the $|\mathbf p|\to0$ IR region from the radial integrals. The MS pole parts are UV-local and
therefore independent of $\delta_{\rm IR}$; all $\delta_{\rm IR}$-dependence resides in finite (IR) terms
that we drop. At the end we may safely send $\delta_{\rm IR}\to0$.



\subsection{Tadpole integral}
To be consistent and make all evaluations clear, we repeat, that first we have to work out the integration over energy $\omega$ in \eqref{eq:T begin}:
\begin{align}
\int_{-\infty}^{\infty} \frac{d \omega}{2 \pi}\left(\omega^2 \mathbf{1}+\mathbf{K}\right)^{-1}=\frac{1}{2} \mathbf{K}^{-1 / 2} ,
\end{align}
and let us introduces intermediate $\delta_{IR}$ regulator to have explicit control on UV/IR convergence
\begin{align}
\int_{-\infty}^{\infty} \frac{d \omega}{2 \pi}\left(\omega^2 \mathbf{1}+\mathbf{K}+\delta_{\rm IR}^2\mathbf 1\right)^{-1}
=\frac{1}{2} \left(\mathbf{K}+\delta_{\rm IR}^2\mathbf 1\right)^{-1 / 2}.
\end{align}
where $\mathbf{K}$ is given by \eqref{K}.
After introduction of such regulator into \eqref{eq:T begin} the original integral 
$$
\mathbf{T}=\frac{1}{2} \int \frac{d^d \mathbf{p}}{(2 \pi)^d} \mathbf{Z}_t^{-1 / 2} \mathbf{K}(\mathbf{p})^{-1 / 2} \mathbf{Z}_t^{-1 / 2}.
$$
turns  to the regulated one
\begin{align}
\mathbf{T}(\delta_{\rm IR})
&\equiv \int \frac{d \omega}{2 \pi} \frac{d^d \mathbf{p}}{(2 \pi)^d}\,
\mathbf{Z}_t^{-1 / 2}\left(\omega^2 \mathbf{1}+\mathbf{K}(\mathbf{p})+\delta_{\rm IR}^2\mathbf 1\right)^{-1} \mathbf{Z}_t^{-1 / 2}.
\end{align}
It is more convenient to rewrite the integral over momentum in spherical coordinates as
\begin{align}
\int \frac{d^d \mathbf{p}}{(2 \pi)^d}=\frac{1}{(2 \pi)^d} \int_0^{\infty} d \rho \rho^{d-1} \int_{S^{d-1}} d \Omega_{d-1}=\frac{\Omega_{d-1}}{2(2 \pi)^d} \int_0^{\infty} d x x^{\frac{d}{2}-1}\langle\cdots\rangle_{\hat{\mathbf{n}}},
\end{align}
where $\mathbf{p}\equiv \rho \hat{\mathbf{n}}$ with $\rho \equiv |\mathbf{p}| \geq 0$, and $x \equiv \rho^2; \Omega_{d-1}$ is the solid angle on the unit sphere and $\Omega_{d-1}=2 \pi^{d / 2} / \Gamma\left(\frac{d}{2}\right)$. We also remind, that $\langle \ldots\rangle_{\hat{\mathbf{n}}}$ is given by \eqref{eq:n average}.
Hence, we arrive at the regulated representation
\begin{align}
\mathbf{T}(\delta_{\rm IR})
=\frac{1}{4} \frac{\Omega_{d-1}}{(2 \pi)^d}
\int_0^{\infty} d x \;x^{\frac{d}{2}-1}
\left\langle\mathbf{Z}_t^{-1 / 2} \big(\mathbf{K}(x, \hat{\mathbf{n}})+\delta_{\rm IR}^2\mathbf 1\big)^{-1 / 2} \mathbf{Z}_t^{-1 / 2}\right\rangle_{\hat{\mathbf{n}}}.
\end{align}
Our aim is to regularize the divergences  in the latter expression. To this end let us use Schwinger parametrization 
\begin{align}
\label{gamma int}
    \frac{1}{A^n} = \frac{1}{\Gamma(n)} \int_0^{\infty}  u^{n-1} e^{-u A} d u,
\end{align}
so
\be 
\left(\mathbf{K}(x,\hat{\mathbf{n}})+\delta_{\rm IR}^2\mathbf 1\right)^{-1 / 2}
=\frac{1}{\sqrt{\pi}} \int_0^{\infty}  t^{-1 / 2} e^{-t\delta_{\rm IR}^2}\,e^{-t \mathbf{K}(x,\hat{\mathbf{n}})}\,d t.
\ee
Recalling that due to \eqref{K in S and L}, we have $\mathbf{K}(\mathbf{p})=|\mathbf{p}|^2 \mathbf{S}(\hat{\mathbf{n}})+\mathbf{L}$ we  obtain
\begin{align}
&\mathbf{T}(\delta_{\rm IR})
=\frac{1}{4} \frac{\Omega_{d-1}}{(2 \pi)^d\sqrt{\pi}}
\int_0^{\infty}  t^{-1 / 2} d t
\int_0^{\infty}  x^{\frac{d}{2}-1}d x\,
\left\langle\mathbf{Z}_t^{-1 / 2}\,
e^{-t\delta_{\rm IR}^2}\,e^{-t(x\mathbf S+\mathbf L)}\,
\mathbf{Z}_t^{-1 / 2}\right\rangle_{\hat{\mathbf{n}}}.
\end{align}
The factor $e^{-t\delta_{\rm IR}^2}$ makes the $t$-integral absolutely convergent at $t\to\infty$.
Therefore we may use the Dyson expansion in $\mathbf L$ under the integral sign to extract the UV pole,
which comes entirely from the short-proper-time region $t\to0$.
As a next step we perform the Dyson expansion for the exponent $e^{-t(x \mathbf{S}+\mathbf{L})}$ with small mass matrix $\mathbf{L}$; this is legitimate, since we extract the UV divergency at large momenta, i.e. large eigenvalues of $\mathbf{S}$ and large $x$. Hence,
\begin{align}
    e^{-t(x \mathbf{S}+\mathbf{L})}=e^{-t x \mathbf{S}}-\int_0^t d \tau e^{-(t-\tau) x \mathbf{S}} \mathbf{L} e^{-\tau x \mathbf{S}} + \mathcal{O}(\mathbf{L}^2),
\end{align}
thus, up to $\mathcal{O}(\mathbf{L}^2)$ (we omit $\mathcal{O}(\mathbf{L}^2)$ in what follows)
\begin{gather}
\label{T two terms}
\mathbf{T}(\delta_{\rm IR})\equiv \mathbf{T}_0(\delta_{\rm IR})+\mathbf{T}_1(\delta_{\rm IR})=\\
=\frac{1}{4} \frac{\Omega_{d-1}}{(2 \pi)^d\sqrt{\pi}}\Bigg[
\int_0^{\infty}  t^{-1 / 2} e^{-t\delta_{\rm IR}^2} d t
\int_0^{\infty}  x^{\frac{d}{2}-1}d x\,
\left\langle\mathbf{Z}_t^{-1 / 2} e^{-tx \mathbf{S}} \mathbf{Z}_t^{-1 / 2}\right\rangle_{\hat{\mathbf{n}}}-
\nonumber\\
-\int_0^{\infty}  t^{-1 / 2} e^{-t\delta_{\rm IR}^2} d t
\int_0^{\infty}  x^{\frac{d}{2}-1}d x\int_0^t d \tau \,
\left\langle\mathbf{Z}_t^{-1 / 2} e^{-(t-\tau) x \mathbf{S}} \mathbf{L} e^{-\tau x \mathbf{S}} \mathbf{Z}_t^{-1 / 2}\right\rangle_{\hat{\mathbf{n}}}
\Bigg].
\end{gather}
Beginning with $\mathbf{T}_0$, we use the decomposition  $\mathbf{S}(\hat{\mathbf{n}})=\sum_{a=1}^2 s_a(\hat{\mathbf{n}}) \boldsymbol{\Pi}_a(\hat{\mathbf{n}})$ given by \eqref{S} and take $x$ integral, then obtain
\begin{align}
\mathbf{T}_0(\delta_{\rm IR})
&=\frac{1}{4} \frac{\Omega_{d-1}}{(2 \pi)^d\sqrt{\pi}}\,
\Gamma\!\left(\frac d2\right)\nonumber\\
&\times\int_0^{\infty}  t^{-1 / 2} e^{-t\delta_{\rm IR}^2} d t\;
t^{-d/2}\,
\sum_a\left\langle
s_a(\hat{\mathbf n})^{-d/2}\,
\mathbf{Z}_t^{-1 / 2} \mathbf{\Pi}_a(\hat{\mathbf{n}})\mathbf{Z}_t^{-1 / 2}
\right\rangle_{\hat{\mathbf{n}}},
\end{align}
and after some algebra reducing to
\begin{align}
\mathbf{T}_0(\delta_{\rm IR})
&=\frac{1}{2^{d+1}\pi^{d/2}\sqrt{\pi}}
\int_0^{\infty}  t^{-\frac{(d+1)}{2}} e^{-t\delta_{\rm IR}^2} d t \;
\sum_a\left\langle s_a^{-\frac{d}{2}}\,
\mathbf{Z}_t^{-1 / 2} \mathbf{\Pi}_a(\hat{\mathbf{n}}) \mathbf{Z}_t^{-1 / 2}
\right\rangle_{\hat{\mathbf{n}}},
\end{align}
where we use $\Omega_{d-1}=2 \pi^{d / 2} / \Gamma\left(\frac{d}{2}\right)$.
Using $d=3-\epsilon$, the $L=0$ contribution $\mathbf T_0$ is a scaleless
power-divergent integral and vanishes in pure dimensional regularization. With the auxiliary IR mass
$\delta_{\rm IR}>0$ kept finite one finds $\mathbf T_0(\delta_{\rm IR})\propto \delta_{\rm IR}^2$
(and may contain a pole $\propto \delta_{\rm IR}^2/\epsilon$), hence it vanishes as
$\delta_{\rm IR}\to0$ and does not contribute to the MS counterterms of the original theory\footnote{I.e. in $d=3-\epsilon$ we use $\int_0^{\infty} d t t^{-\frac{d+1}{2}} e^{-t \delta_{\mathrm{IR}}^2}=\left(\delta_{\mathrm{IR}}^2\right)^{\frac{d-1}{2}} \Gamma\left(\frac{1-d}{2}\right)=\left(\delta_{\mathrm{IR}}^2\right)^{1-\epsilon / 2} \Gamma\left(-1+\frac{ \epsilon}{2}\right) \sim-\frac{2}{\epsilon} \delta_{\mathrm{IR}}^2+\mathcal{O}\left(\delta_{\mathrm{IR}}^2 \ln \delta_{\mathrm{IR}}^2\right)$.}.
Therefore the MS pole relevant for renormalization starts at $\mathcal O(\mathbf L)$, i.e.\ comes from
$\mathbf T_1$.

Turning to the second term $\mathbf{T}_1$ from \eqref{T two terms}, we use \eqref{S} and rewrite exponent as
\begin{align}
    e^{-(t-\tau) x \mathbf{S}} \mathbf{L} e^{-\tau x \mathbf{S}}=\sum_{a, b} e^{-x\left[(t-\tau) s_a+\tau s_b\right]} \mathbf{\Pi}_a \mathbf{L} \mathbf{\Pi}_b,
\end{align}
thus
\begin{align}
\label{T_1 inter 1}
\mathbf{T}_1(\delta_{\rm IR})
&=-\frac{1}{4} \frac{\Omega_{d-1}}{(2 \pi)^d\sqrt{\pi}}
\int_0^{\infty}  t^{-1 / 2} e^{-t\delta_{\rm IR}^2} d t
\int_0^{\infty}  x^{\frac{d}{2}-1}d x\int_0^t d \tau \\
&\qquad\times
\left\langle \sum_{a, b} e^{-x\left[(t-\tau) s_a+\tau s_b\right]}
\mathbf{Z}_t^{-1 / 2}  \mathbf{\Pi}_a \mathbf{L} \mathbf{\Pi}_b \mathbf{Z}_t^{-1 / 2}
\right\rangle_{\hat{\mathbf{n}}} . \nonumber
\end{align}
Take an integral over $x$, using \eqref{gamma int}, we obtain
\begin{gather}
\mathbf{T}_1(\delta_{\rm IR})
=-\frac{1}{4} \frac{\Omega_{d-1}}{(2 \pi)^d\sqrt{\pi}}\,
\Gamma\!\left(\frac d2\right)
\int_0^{\infty}  t^{-1 / 2} e^{-t\delta_{\rm IR}^2} d t
\int_0^t d \tau \times\\\times\,
\left\langle \sum_{a, b} \left[(t-\tau) s_a+\tau s_b\right]^{-d / 2}
\mathbf{Z}_t^{-1 / 2}  \mathbf{\Pi}_a \mathbf{L} \mathbf{\Pi}_b \mathbf{Z}_t^{-1 / 2}
\right\rangle_{\hat{\mathbf{n}}}.
\end{gather}

Introducing new variable $\tilde{\tau}\equiv \frac{\tau}{t}$, so  that 
\begin{align}
    (t-\tau) s_a+\tau s_b=t\left[(1-\tilde{\tau}) s_a+\tilde{\tau} s_b\right],
\end{align}
so that integrals over $t$ and $\tilde{\tau}$ are not entangled. The integral over $\tilde{\tau}$ for $d=3$ is
\begin{align}
\label{int over tilda}
    \int_0^1     \frac{d\tilde{\tau}}{\left[(1-\tilde{\tau}) s_a+\tilde{\tau} s_b\right]^{d / 2}} =\frac{2}{s_a \sqrt{s_b} + s_b \sqrt{s_a}},
\end{align}
which is finite. The integral over $t$ reads
\begin{align}
\int_0^{\infty}  t^{(1-d) / 2} e^{-t\delta_{\rm IR}^2} d t
&=\int_0^{\infty}  t^{-1+\frac{\epsilon}{ 2}} e^{-t\delta_{\rm IR}^2} d t
=(\delta_{\rm IR}^2)^{-\epsilon/2}\Gamma\!\left(\frac{\epsilon}{2}\right)
=\frac{2}{\epsilon}-\gamma_E-\ln(\delta_{\rm IR}^2)+\mathcal{O}(\epsilon).
\end{align}

Collecting all together, the divergent part of tadpole type integral comes from $\mathbf{T}_1$ and it is
\begin{align}
&[\mathbf{T}]_{\text{pole}}=-\frac{1}{4\pi^2 \epsilon} \left\langle \sum_{a, b} \frac{\mathbf{Z}_t^{-1 / 2}  \mathbf{\Pi}_a \mathbf{L} \mathbf{\Pi}_b \mathbf{Z}_t^{-1 / 2}}{s_a \sqrt{s_b} + s_b \sqrt{s_a}} \right\rangle_{\hat{\mathbf{n}}}, 
\end{align}
what is \eqref{T answer}.

\subsection{Bubble integrals}
Bubble integral with non-zero momentum is defined as
\begin{align}\label{app:Bk}
    &\mathbf{B}_{i j ; k l}(k) = \int\frac{d\omega}{2\pi}\frac{d^d \mathbf{p}}{(2 \pi)^d} \mathbf{G}_{i k}(p) \mathbf{G}_{j l}(p+k) \nonumber\\
    &=\int\frac{d\omega}{2\pi}\frac{d^d \mathbf{p}}{(2 \pi)^d}  \Big[\mathbf{Z}_t^{-1 / 2}\left(\omega^2 \mathbf{1}+\mathbf{K}(\mathbf{p})\right)^{-1} \mathbf{Z}_t^{-1 / 2}\Big]_{ik}\nonumber\\
    &\times\Big[\mathbf{Z}_t^{-1 / 2}\left((\omega+\Omega)^2 \mathbf{1}+\mathbf{K}(\mathbf{p+k})\right)^{-1} \mathbf{Z}_t^{-1 / 2}\Big]_{jl},
\end{align}
In the one-loop $D=4-\epsilon$ expansion the bubble diagram is logarithmically divergent, hence its UV pole is local. Therefore the MS pole is independent of the external momentum and can be extracted from the $k=0$ integral.

To define the massless vacuum integral at $k=0$ in dimensional regularization
and to unambiguously separate UV from possible IR contributions, we introduce
the same auxiliary IR regulator as in the tadpole calculation
\be
\omega^2\mathbf 1+\mathbf K(\mathbf p)\;\longrightarrow\;
\omega^2\mathbf 1+\mathbf K(\mathbf p)+\delta_{\rm IR}^2\mathbf 1,
\qquad \delta_{\rm IR}>0.
\ee
which produces shift in  the eigenvalues
$\kappa_a(\mathbf p)\to \kappa_a(\mathbf p)+\delta_{\rm IR}^2$ but does not change the eigenvectors
or the projectors.

We consider the expression corresponding bubble integral at zero momentum with IR regulator introduced
\begin{gather}
\mathbf B_{ij;kl}(0;\delta_{\rm IR})
\equiv \int\frac{d\omega}{2\pi}\frac{d^d\mathbf p}{(2\pi)^d}\,
\mathbf G_{\delta,ik}(p)\,\mathbf G_{\delta,jl}(p),
\\
\mathbf G_\delta(p)=\mathbf Z_t^{-1/2}\big(\omega^2\mathbf 1+\mathbf K(\mathbf p)+\delta_{\rm IR}^2\mathbf 1\big)^{-1}\mathbf Z_t^{-1/2}.
\end{gather}
Using the spectral representation,
\be
\mathbf G_\delta(p)=\sum_a\frac{\tilde{\mathbf P}_a(\mathbf p)}{\omega^2+\kappa_a(\mathbf p)+\delta_{\rm IR}^2},
\ee
and the  integral (for $A,B>0$)
\be
\int_{-\infty}^{\infty}\frac{d\omega}{2\pi}\,
\frac{1}{(\omega^2+A)(\omega^2+B)}
=\frac{1}{2\sqrt{A}\sqrt{B}\,(\sqrt{A}+\sqrt{B})},
\qquad A,B>0,
\ee
hence
\be
\mathbf B_{ij;kl}(0;\delta_{\rm IR})
=\frac{1}{2}\int\frac{d^d\mathbf p}{(2\pi)^d}\sum_{a,b}
\tilde{\mathbf P}_a(\mathbf p)_{ik}\,\tilde{\mathbf P}_b(\mathbf p)_{jl}\,
\mathcal K\!\big(\kappa_a(\mathbf p)+\delta_{\rm IR}^2,\;\kappa_b(\mathbf p)+\delta_{\rm IR}^2\big),
\ee
where we denote
\be
\mathcal K(A,B)\equiv \frac{1}{\sqrt{A}\sqrt{B}\,(\sqrt{A}+\sqrt{B})}.
\ee

For the MS pole it is sufficient to keep the UV asymptotics
$\kappa_a(\mathbf p)=|\mathbf p|^2 s_a(\hat{\mathbf n})+\mathcal O(1)$ and
$\tilde{\mathbf P}_a(\mathbf p)=\mathbf P_a(\hat{\mathbf n})+\mathcal O(1/|\mathbf p|^2)$.
All $\mathcal O(1)$ and $\mathcal O(1/|\mathbf p|^2)$ corrections produce UV-finite contributions.
Thus the pole part equals the pole part of the UV-approximated integral

Switching to $x\equiv |\mathbf p|^2$ and use $d^d\mathbf p=\frac{\Omega_{d-1}}{2}\,dx\,x^{\frac d2-1}$:
\be
\mathbf B^{\rm UV}_{ij;kl}(0;\delta_{\rm IR})
=\frac{\Omega_{d-1}}{4(2\pi)^d}\int_0^\infty dx\,x^{\frac d2-1}\,
\Big\langle\sum_{a,b}\mathbf P_a(\hat{\mathbf n})_{ik}\,\mathbf P_b(\hat{\mathbf n})_{jl}\,
\mathcal I_{ab}(x;\delta_{\rm IR})\Big\rangle_{\hat{\mathbf n}},
\ee
where
\be
\mathcal I_{ab}(x;\delta_{\rm IR})
\equiv \mathcal{K}(x s_a+\delta_{\rm IR}^2,x s_b+\delta_{\rm IR}^2).
\ee
Next, we use the identity (for $A,B>0$)
\be
\int_0^1 d\tau\,\big[(1-\tau)A+\tau B\big]^{-3/2}
=\frac{2}{\sqrt A\sqrt B(\sqrt A+\sqrt B)},
\ee
which implies
\be
\mathcal I_{ab}(x;\delta_{\rm IR})
=\frac12\int_0^1 d\tau\,
\big(x\big[(1-\tau)s_a+\tau s_b\big]+\delta_{\rm IR}^2\big)^{-3/2}.
\ee
The $x$-integral is then just
\begin{align}
&\int_0^\infty dx\,x^{\frac d2-1}\big(xc+\delta_{\rm IR}^2\big)^{-3/2}
=c^{-d/2}\,\delta_{\rm IR}^{\,d-3}\,
\frac{\Gamma(d/2)\Gamma\big(\frac{3-d}{2}\big)}{\Gamma(3/2)}\nonumber\\
&=c^{-d/2}\,\delta_{\rm IR}^{-\epsilon}\,
\frac{\Gamma(d/2)\Gamma(\epsilon/2)}{\Gamma(3/2)},
\end{align}
where $c\equiv (1-\tau)s_a+\tau s_b$.
Taking the pole part and setting $d\to 3$ inside
the finite angular/$\tau$ integrals we obtain
\be
[\mathbf B_{ij;kl}(0)]_{\rm pole}
=\frac{1}{4\pi^2\epsilon}
\Big\langle\sum_{a,b}\mathcal K(s_a,s_b)\,\mathbf P_a(\hat{\mathbf n})_{ik}\,\mathbf P_b(\hat{\mathbf n})_{jl}\Big\rangle_{\hat{\mathbf n}},
\ee
where
\be
\mathcal K(s_a,s_b)=\frac{1}{\sqrt{s_a}\sqrt{s_b}(\sqrt{s_a}+\sqrt{s_b})},
\ee
which coincides with eq.~\eqref{bubble answer} in the main text.
Now let us comment about bubble integral with non-zero external momentum \eqref{app:Bk}.

For the purposes of MS renormalization in $D=4-\epsilon$, it is sufficient to know the UV pole part of the
bubble diagram. The one--loop bubble is superficially logarithmically divergent, hence any UV divergence is
local. Expanding the integrand of \eqref{app:Bk} in the external momentum $k$ produces additional inverse
powers of the loop momentum, so all terms proportional to $\Omega$ and/or $\mathbf k$ are UV finite.
Therefore the $1/\epsilon$ pole is independent of $k$ and is completely captured by the $k=0$ master
integral computed above. Since this is a standard one--loop locality argument, we do not reproduce the
full finite-$k$ analysis here since it is quite awkward and is not illuminating, being not producing divergent terms.

\section{Angular weights $J_{11}$, $J_{22}$ and $J_{12}$ in the limit $\mathbf Z_t=\mathbf 1$ and $\mathbf Y=0$}
\label{app:Jweights}
In this appendix we highlight calculations concerning $J_{11}$, $J_{22}$ and $J_{12}$ used in section~\ref{sec:partic}
in the limit $\mathbf{Z}_t=1$ and $\mathbf{Y}=0$, and  prove several identities and inequalities
employed there. 
All angular averages are taken over the unit sphere $S^2$ with the normalization
\be
\label{eq:S2avg_app}
\langle f(\hat{\bm n})\rangle_{S^2}\equiv \frac{1}{4\pi}\int_{S^2} d\Omega\, f(\hat{\bm n}),
\qquad \hat{\bm n}\in S^2.
\ee
We also specialize to the limit
\be
\label{eq:limit_app}
\mathbf Z_t=\mathbf 1,\qquad \mathbf Y=0,
\ee
so that the two scalar species propagate independently in the UV, while the spatial anisotropy
is entirely encoded in the two positive--definite $3\times3$ matrices $\mathbf C_1\equiv \mathbf C_\phi$
and $\mathbf C_2\equiv \mathbf C_\chi$.

\subsection{Weights $J_{11}$, $J_{22}$ and $J_{12}$ in the limit $\mathbf Z_t=\mathbf 1,\mathbf Y=0$}
\label{sec:threeweights}
In the limit \eqref{eq:limit_app} the matrix $\mathbf C(\hat{\bm n})$ introduced in the main text becomes
block--diagonal in the field space,
\be
\mathbf C(\hat{\bm n})=
\begin{pmatrix}
\hat{\bm n}^{\top}\mathbf C_1 \hat{\bm n} & 0\\
0 & \hat{\bm n}^{\top}\mathbf C_2 \hat{\bm n}
\end{pmatrix}
\equiv
\begin{pmatrix}
s_1(\hat{\bm n}) & 0\\
0 & s_2(\hat{\bm n})
\end{pmatrix},
\qquad
s_a(\hat{\bm n})\equiv \hat{\bm n}^{\top}\mathbf C_a \hat{\bm n}>0.
\ee
Consequently, the spectral projectors in the field space are constant:
\be
\label{eq:Pi_const_app}
\Pi_1=
\begin{pmatrix}
1&0\\0&0
\end{pmatrix},
\qquad
\Pi_2=
\begin{pmatrix}
0&0\\0&1
\end{pmatrix},
\qquad
\Pi_a\Pi_b=\delta_{ab}\Pi_a,\qquad \Pi_1+\Pi_2=\mathbf 1.
\ee

The general pole part of the bubble master integral at zero external momentum, eq.~\eqref{bubble answer},
reads
\be
[\mathbb B_{ij;kl}(0)]_{\rm pole}
=\frac{1}{4\pi^2\epsilon}
\Big\langle\sum_{a,b}\mathcal K(s_a,s_b)\,\mathbf P_a(\hat{\mathbf n})_{ik}\,\mathbf P_b(\hat{\mathbf n})_{jl}\Big\rangle_{\hat{\mathbf n}},
\ee
where
\be
\mathcal K(s_a,s_b)=\frac{1}{\sqrt{s_a}\sqrt{s_b}(\sqrt{s_a}+\sqrt{s_b})}.
\ee
In the limit \eqref{eq:limit_app} one has $\mathbf P_a=\Pi_a$, hence we have explicitly  
\begin{subequations}
\label{eq:K11K22K12_app}
\begin{align}
&\mathcal K(s_1,s_1)=\frac{1}{2\,s_1^{3/2}},
\qquad
\mathcal K(s_2,s_2)=\frac{1}{2\,s_2^{3/2}},
\\
&\mathcal K(s_1,s_2)=\frac{1}{\sqrt{s_1}\sqrt{s_2}\big(\sqrt{s_1}+\sqrt{s_2}\big)}.
\end{align}
\end{subequations}
It is therefore natural to define
\begin{subequations}
\label{eq:Jdefs_app}
\begin{align}
&J_{11}\equiv \Big\langle s_1(\hat{\bm n})^{-3/2}\Big\rangle_{S^2},
\qquad
J_{22}\equiv \Big\langle s_2(\hat{\bm n})^{-3/2}\Big\rangle_{S^2},
\\
&J_{12}\equiv
2\Big\langle\mathcal K\big(s_1(\hat{\bm n}),s_2(\hat{\bm n})\big)\Big\rangle_{S^2}.
\end{align}
\end{subequations}
With these definitions one immediately obtains, for instance,
\bea
[\mathbb B_{11;11}]_{\rm pole}
&=\frac{1}{4\pi^2\eps}\Big\langle \mathcal K(s_1,s_1)\Big\rangle_{S^2}
=\frac{1}{4\pi^2\eps}\Big\langle \frac{1}{2s_1^{3/2}}\Big\rangle_{S^2}
=\frac{J_{11}}{8\pi^2\eps},\\
[\mathbb B_{22;22}]_{\rm pole}
&=\frac{J_{22}}{8\pi^2\eps},\\
[\mathbb B_{12;12}]_{\rm pole}
&=\frac{1}{4\pi^2\eps}\Big\langle \mathcal K(s_1,s_2)\Big\rangle_{S^2}
=\frac{1}{4\pi^2\eps}\cdot \frac{J_{12}}{2}
=\frac{J_{12}}{8\pi^2\eps},
\eea
and similarly for $[\mathbb B_{21;21}]_{\rm pole}$.
All other components vanish in this limit because the projectors \eqref{eq:Pi_const_app} kill
cross--species structures. This explains why all one--loop RG coefficients in this limit are
weighted only by $J_{11}$, $J_{22}$ and $J_{12}$.

Completely analogously, the general pole part of the tadpole master integral, eq.~\eqref{T answer},
reduces in the limit \eqref{eq:limit_app} to the same three weights \eqref{eq:Jdefs_app}, because
the same kernel $\mathcal K(s_a,s_b)$ appears there. In particular, the diagonal entries are
proportional to $J_{11}$ and $J_{22}$, while the off--diagonal entry is proportional to $J_{12}$.

$\,$

Now turn to the next simplification concerning   
$J_{11}$ and $J_{22}$
\be
\label{eq:key_identity_app}
\Big\langle\big(\hat{\bm n}^{\top}\mathbf C\,\hat{\bm n}\big)^{-3/2}\Big\rangle_{S^2}
=\big(\det\mathbf C\big)^{-1/2},
\qquad \mathbf C=\mathbf C^{\top}>0,\qquad \mathbf C\in\mathbb R^{3\times 3}.
\ee
The proof is elementary and follows from evaluating the same Gaussian integral in two different ways.

Consider
\be
I(\mathbf C)\equiv \int_{\mathbb R^3} d^3p\, e^{-\,p^{\top}\mathbf C\,p}.
\ee
Since $\mathbf C$ is symmetric positive definite, one may write $\mathbf C=\mathbf S^{\top}\mathbf S$
with $\mathbf S=\mathbf C^{1/2}$. Introduce $q=\mathbf S\,p$, so that
\be
p^{\top}\mathbf C\,p = p^{\top}\mathbf S^{\top}\mathbf S\,p = q^{\top}q = |q|^2,
\qquad
d^3p = (\det\mathbf S)^{-1} d^3q = (\det\mathbf C)^{-1/2} d^3q,
\ee
therefore
\be
\label{eq:I_det_app}
I(\mathbf C)=(\det\mathbf C)^{-1/2}\int_{\mathbb R^3} d^3q\,e^{-|q|^2}
=(\det\mathbf C)^{-1/2}\,\pi^{3/2}.
\ee
Now  parametrize $p=r\hat{\bm n}$ with $r\ge0$ and $\hat{\bm n}\in S^2$. Then
\be
p^{\top}\mathbf C\,p = r^2\big(\hat{\bm n}^{\top}\mathbf C\,\hat{\bm n}\big),
\qquad
d^3p = r^2\,dr\,d\Omega.
\ee
Hence
\bea
I(\mathbf C)
&=\int_{S^2} d\Omega\int_0^\infty r^2\,dr\,
\exp\!\Big[-r^2\big(\hat{\bm n}^{\top}\mathbf C\,\hat{\bm n}\big)\Big].
\eea
Using the standard integral (valid for $a>0$)
\be
\int_0^\infty r^2 e^{-a r^2}dr=\frac{\sqrt{\pi}}{4}\,a^{-3/2},
\ee
we obtain
\be
\label{eq:I_ang_app}
I(\mathbf C)=\frac{\sqrt{\pi}}{4}\int_{S^2} d\Omega\,
\big(\hat{\bm n}^{\top}\mathbf C\,\hat{\bm n}\big)^{-3/2}.
\ee

Equating \eqref{eq:I_det_app} and \eqref{eq:I_ang_app} and dividing by $4\pi$, we find
\begin{align}
&\Big\langle\big(\hat{\bm n}^{\top}\mathbf C\,\hat{\bm n}\big)^{-3/2}\Big\rangle_{S^2}
=\frac{1}{4\pi}\int_{S^2} d\Omega\,
\big(\hat{\bm n}^{\top}\mathbf C\,\hat{\bm n}\big)^{-3/2}
=\frac{\pi^{3/2}}{(\sqrt{\pi}/4)\,4\pi}\,(\det\mathbf C)^{-1/2}\nonumber\\
&=(\det\mathbf C)^{-1/2},
\end{align}
which proves \eqref{eq:key_identity_app}.

As an immediate corollary, for the weights \eqref{eq:Jdefs_app} one obtains
\be
\label{eq:Jdiag_det_app}
J_{11}=\big(\det\mathbf C_1\big)^{-1/2},
\qquad
J_{22}=\big(\det\mathbf C_2\big)^{-1/2}.
\ee

Now turn to explicit form of $J_{12}$. 
In this subsection we derive the representation of $J_{12}$ used in the main text:
\be
\label{eq:J12_tau_app}
J_{12}=\int_0^1 d\tau\,
\big(\det[(1-\tau)\mathbf C_1+\tau\mathbf C_2]\big)^{-1/2}.
\ee
We proceed in two steps: first we rewrite the mixed kernel $\mathcal K(s_1,s_2)$ as a one--dimensional
integral over $\tau$, and then we use the identity \eqref{eq:key_identity_app}.

For any $u>0$ and $v>0$ one has the elementary identity
\be
\label{eq:kernel_tau_id_app}
\int_0^1 d\tau\,\big[(1-\tau)u+\tau v\big]^{-3/2}
=
\frac{2}{\sqrt u\,\sqrt v\,(\sqrt u+\sqrt v)}.
\ee
Now apply \eqref{eq:kernel_tau_id_app} pointwise to
\be
u=s_1(\hat{\bm n})=\hat{\bm n}^{\top}\mathbf C_1\hat{\bm n},\qquad
v=s_2(\hat{\bm n})=\hat{\bm n}^{\top}\mathbf C_2\hat{\bm n}.
\ee
Using the definition \eqref{eq:Jdefs_app} of $J_{12}$ we obtain
\begin{align}
&J_{12}
=2\Big\langle \frac{1}{\sqrt{s_1}\sqrt{s_2}(\sqrt{s_1}+\sqrt{s_2})}\Big\rangle_{S^2}\nonumber\\
&=\Big\langle \int_0^1 d\tau\,\big[(1-\tau)s_1(\hat{\bm n})+\tau s_2(\hat{\bm n})\big]^{-3/2}\Big\rangle_{S^2}.
\end{align}
By linearity of the angular average, this becomes
\be
\label{eq:J12_avg_tau_app}
J_{12}
=\int_0^1 d\tau\,\Big\langle
\big[(1-\tau)\hat{\bm n}^{\top}\mathbf C_1\hat{\bm n}+\tau\hat{\bm n}^{\top}\mathbf C_2\hat{\bm n}\big]^{-3/2}
\Big\rangle_{S^2}.
\ee
Defining the interpolating matrix
\be
\label{eq:Ctau_app}
\mathbf C(\tau)\equiv (1-\tau)\mathbf C_1+\tau\mathbf C_2.
\ee
we proceed then with  the expression in brackets in \eqref{eq:J12_avg_tau_app}, which is simply $\hat{\bm n}^{\top}\mathbf C(\tau)\hat{\bm n}$.
Since $\mathbf C(\tau)$ is positive definite for all $\tau\in[0,1]$ (convexity of the cone of positive
definite matrices), we may apply \eqref{eq:key_identity_app} to obtain
\be
\Big\langle \big(\hat{\bm n}^{\top}\mathbf C(\tau)\hat{\bm n}\big)^{-3/2}\Big\rangle_{S^2}
=\big(\det\mathbf C(\tau)\big)^{-1/2}.
\ee
Substituting this into \eqref{eq:J12_avg_tau_app} we arrive at \eqref{eq:J12_tau_app}:
\be
J_{12}=\int_0^1 d\tau\,\big(\det[(1-\tau)\mathbf C_1+\tau\mathbf C_2]\big)^{-1/2}.
\ee
This completes the derivation.

It is sometimes convenient to rewrite \eqref{eq:J12_tau_app} in terms of the generalized eigenvalues of
the pair $(\mathbf C_2,\mathbf C_1)$. Introducing
\be
\mathbf A\equiv \mathbf C_1^{-1/2}\mathbf C_2\mathbf C_1^{-1/2}
=U\,\mathrm{diag}(\gamma_1,\gamma_2,\gamma_3)\,U^{\top},
\qquad \gamma_i>0,
\ee
where $U$ is orthogonal, one finds
\be
J_{12}=\big(\det\mathbf C_1\big)^{-1/2}\int_0^1 d\tau\,
\prod_{i=1}^3\big((1-\tau)+\tau\gamma_i\big)^{-1/2},
\ee
and this could be expressed through Lauricella hypergeometric functions $F_D^{(3)}$ as quoted in the main text.

\subsection{Proof of the inequality $J_{12}^2\le J_{11}J_{22}$}
\label{app:J_ineq}
Finally, we prove an inequality which is useful for the existence analysis of fixed points in the
parity--symmetric sector:
\be
\label{eq:Jineq_app}
J_{12}^2\le J_{11}J_{22}.
\ee
The proof uses only elementary inequalities and holds for any positive functions
$s_1(\hat{\bm n}),s_2(\hat{\bm n})$ (in particular, for $s_a(\hat{\bm n})=\hat{\bm n}^\top \mathbf C_a \hat{\bm n}$).

For any $a>0$ and $b>0$ the arithmetic--geometric mean inequality gives
\be
\sqrt a+\sqrt b \ge 2\,(ab)^{1/4}.
\ee
Therefore, pointwise in $\hat{\bm n}$,
\bea
\frac{1}{\sqrt{s_1}\sqrt{s_2}\big(\sqrt{s_1}+\sqrt{s_2}\big)}
&\le \frac{1}{\sqrt{s_1}\sqrt{s_2}\cdot 2(s_1s_2)^{1/4}}
=\frac{1}{2\,s_1^{3/4}\,s_2^{3/4}}.
\eea

Average both sides over $S^2$ and use Cauchy--Bunyakovsky--Schwarz:
\bea
\Big\langle s_1^{-3/4}s_2^{-3/4}\Big\rangle_{S^2}^2
&\le
\Big\langle s_1^{-3/2}\Big\rangle_{S^2}\;
\Big\langle s_2^{-3/2}\Big\rangle_{S^2}
=J_{11}J_{22}.
\eea
Combining with the definition $J_{12}=2\langle 1/(\sqrt{s_1}\sqrt{s_2}(\sqrt{s_1}+\sqrt{s_2}))\rangle_{S^2}$,
we obtain
\bea
J_{12}
&=2\Big\langle \frac{1}{\sqrt{s_1}\sqrt{s_2}\big(\sqrt{s_1}+\sqrt{s_2}\big)}\Big\rangle_{S^2}
\le 2\Big\langle \frac{1}{2s_1^{3/4}s_2^{3/4}}\Big\rangle_{S^2}
=\Big\langle s_1^{-3/4}s_2^{-3/4}\Big\rangle_{S^2},
\eea
hence
\be
J_{12}^2\le \Big\langle s_1^{-3/4}s_2^{-3/4}\Big\rangle_{S^2}^2\le J_{11}J_{22},
\ee
which proves \eqref{eq:Jineq_app}.

Equality in \eqref{eq:Jineq_app} requires equality both in arithmetic--geometric mean and in Cauchy--Bunyakovsky--Schwarz.
The arithmetic--geometric mean equality forces $s_1(\hat{\bm n})=s_2(\hat{\bm n})$ pointwise on $S^2$, while equality in
Cauchy--Bunyakovsky--Schwarz forces the functions $s_1^{-3/4}$ and $s_2^{-3/4}$ to be proportional.
For $s_a(\hat{\bm n})=\hat{\bm n}^{\top}\mathbf C_a\hat{\bm n}$ this implies
$\hat{\bm n}^{\top}\mathbf C_1\hat{\bm n}=\hat{\bm n}^{\top}\mathbf C_2\hat{\bm n}$ for all $\hat{\bm n}$,
i.e. $\mathbf C_1=\mathbf C_2$.

\section{Linear stability of the generalized Wilson--Fisher fixed point}
\label{app:WF_stability}

In this appendix we analyze the one--loop linear stability of the coupled fixed point $S_-$
(the ``generalized Wilson--Fisher'' point) found in Sec.~\ref{sec:partic} in the
limit $\mathbf Z_t=\mathbf 1$, $\mathbf Y=0$ and in the parity--symmetric quartic sector
$\lambda_{4,5}=0$ (we also set $h_i=0$ throughout this appendix).

\subsection*{Beta functions and a convenient rescaling}

In the sector $\lambda_{4,5}=0$ the one--loop beta functions from Sec.~\ref{sec:partic} read
\be
\label{eq:beta_l123_app}
\begin{aligned}
\beta_{\lambda_1}&=-\eps\,\lambda_1+\frac{3}{16\pi^2}\Big(J_{11}\lambda_1^2+J_{22}\lambda_3^2\Big),\\
\beta_{\lambda_2}&=-\eps\,\lambda_2+\frac{3}{16\pi^2}\Big(J_{22}\lambda_2^2+J_{11}\lambda_3^2\Big),\\
\beta_{\lambda_3}&=-\eps\,\lambda_3+\frac{1}{16\pi^2}\Big(J_{11}\lambda_1\lambda_3+4J_{12}\lambda_3^2+J_{22}\lambda_2\lambda_3\Big).
\end{aligned}
\ee
It is useful to absorb the ``weights'' $J_{11},J_{22}$ by introducing rescaled couplings
\be
\label{eq:rescaled_g_app}
g_1\equiv J_{11}\lambda_1,\qquad
g_2\equiv J_{22}\lambda_2,\qquad
g_3\equiv \sqrt{J_{11}J_{22}}\,\lambda_3,
\qquad
\rho\equiv \frac{J_{12}}{\sqrt{J_{11}J_{22}}}\in(0,1],
\ee
and also the standard one--loop coefficient
\be
a\equiv \frac{1}{16\pi^2},\qquad c\equiv \frac{16\pi^2}{3}\,\eps=\frac{\eps}{3a}.
\ee
In terms of $(g_1,g_2,g_3)$ the system \eqref{eq:beta_l123_app} becomes
\be
\label{eq:beta_g123_app}
\begin{aligned}
\beta_{g_1}&\equiv J_{11}\beta_{\lambda_1}=-\eps\,g_1+3a\Big(g_1^2+g_3^2\Big),\\
\beta_{g_2}&\equiv J_{22}\beta_{\lambda_2}=-\eps\,g_2+3a\Big(g_2^2+g_3^2\Big),\\
\beta_{g_3}&\equiv \sqrt{J_{11}J_{22}}\beta_{\lambda_3}=-\eps\,g_3+a(g_1+g_2)g_3+4a\rho\,g_3^2.
\end{aligned}
\ee
Solving \eqref{eq:beta_g123_app} gives two branches $S_\pm$.
It is convenient to introduce
\be
\label{eq:Delta_def_app}
\Delta \equiv \sqrt{4\rho^2-3},
\qquad
4\rho^2\ge 3\;\;\Longleftrightarrow\;\;\rho\ge \frac{\sqrt 3}{2}.
\ee
The coupled fixed points exist if and only if $\Delta\in\mathbb{R}$, i.e. iff $\rho\ge \sqrt3/2$
(which is the same condition as $4J_{12}^2\ge 3J_{11}J_{22}$ in the original variables).
For the IR--stable branch $S_-$ one finds that
\be
g_1^*=g_2^*=g^*,\qquad
\lambda_1^*=\frac{g^*}{J_{11}},\quad \lambda_2^*=\frac{g^*}{J_{22}},\quad
\lambda_3^*=\frac{g_3^*}{\sqrt{J_{11}J_{22}}}.
\ee
where
\be
\label{eq:fp_values_app}
g_3^*=\frac{c}{2(4\rho^2+1)}\Big(4\rho-\Delta\Big),
\qquad
g^*=\frac{c}{2(4\rho^2+1)}\Big(4\rho^2+3+2\rho\,\Delta\Big).
\ee
At the isotropic point $\rho=1$ this reduces to
\be
(\lambda_1^*,\lambda_2^*,\lambda_3^*)=\left(\frac{24\pi^2}{5}\eps,\frac{24\pi^2}{5}\eps,\frac{8\pi^2}{5}\eps\right),
\ee
i.e. the familiar two--scalar Wilson--Fisher fixed point written in our conventions.

\subsection*{Stability matrix and eigenvalues}

Let $t\equiv \ln\mu$. Linearizing the RG flow around a fixed point $g_a^*$ gives
\be
\frac{d}{dt}\,\delta g_a = \sum_b \mathbb{M}_{ab}\,\delta g_b,
\qquad
\mathbb{M}_{ab}\equiv \left. \frac{\partial \beta_{g_a}}{\partial g_b}\right|_{g=g^*}.
\ee
With our convention $\beta_g=\mu\frac{dg}{d\mu}$, the IR limit corresponds to $t\to -\infty$.
Thus the fixed point is IR--attractive iff all eigenvalues $\omega$ of $\mathbb M$ are positive, because
$\delta g\sim e^{\omega t}$ then decays as $t\to-\infty$.

From \eqref{eq:beta_g123_app} we obtain the Jacobian
\be
\label{eq:Jacobian_general_app}
\mathbb{M}=
\begin{pmatrix}
-\eps+6ag_1 & 0 & 6ag_3\\
0 & -\eps+6ag_2 & 6ag_3\\
ag_3 & ag_3 & -\eps+a(g_1+g_2)+8a\rho g_3
\end{pmatrix}_{g=g^*}.
\ee
At the coupled fixed points with $g_1^*=g_2^*=g^*$ and $g_3^*\neq 0$ one may simplify
the diagonal entries to obtain
\be
\label{eq:Jacobian_simpl_app}
\mathbb{M}=
\begin{pmatrix}
4a(g^*-\rho g_3^*) & 0 & 6ag_3^*\\
0 & 4a(g^*-\rho g_3^*) & 6ag_3^*\\
ag_3^* & ag_3^* & 4a\rho g_3^*
\end{pmatrix}.
\ee
Introduce the combinations
\be
\delta g_\pm \equiv \frac12(\delta g_1\pm \delta g_2),
\ee
so that $\delta g_-$ is the ``antisymmetric'' (short notation``(as)'') deformation (splitting $g_1$ and $g_2$),
while $\delta g_+$ is the ``symmetric'' (short notation ``(sym)'') deformation (changing them equally).
From \eqref{eq:Jacobian_simpl_app} one immediately sees that $\delta g_-$ decouples because the third row
contains $\delta g_1+\delta g_2$ only. Explicitly,
\be
\frac{d}{dt}\delta g_- = \omega_-^{(\text{as})}\,\delta g_-,
\qquad
\omega_-^{(\text{as})}=4a\big(g^*-\rho g_3^*\big).
\ee
Using \eqref{eq:fp_values_app} and $ac=\eps/3$, this eigenvalue becomes
\be
\label{eq:omega_as_app}
\omega_-^{(\text{as})}
=\frac{2\eps}{4\rho^2+1}\Big(1+\rho\,\Delta\Big)
\;>\;0
\qquad(\rho>\sqrt3/2).
\ee
In the $(\delta g_+,\delta g_3)$ sector one gets a $2\times2$ matrix,
\be
\label{eq:Jacobian_sym_app}
\frac{d}{dt}
\begin{pmatrix}
\delta g_+\\ \delta g_3
\end{pmatrix}
=
\begin{pmatrix}
4a(g^*-\rho g_3^*) & 6ag_3^*\\
2ag_3^* & 4a\rho g_3^*
\end{pmatrix}
\begin{pmatrix}
\delta g_+\\ \delta g_3
\end{pmatrix}.
\ee
The eigenvalues are obtained by solving the characteristic polynomial. A convenient shortcut is to note
that the trace equals
\be
\mathrm{tr}\,\mathbb{M}_{\rm sym}=4ag^*,
\ee
and after some algebra one finds that one eigenvalue is \emph{exactly}
\be
\label{eq:omega_eps_app}
\omega_+^{(\text{sym})}=\eps,
\ee
independently of $\rho$. The second symmetric eigenvalue is
\be
\label{eq:omega_sym2_app}
\omega_-^{(\text{sym})}
=\frac{\eps}{3(4\rho^2+1)}\Big(3-4\rho^2+4\rho\,\Delta\Big).
\ee
In the allowed range $\rho\in(\sqrt3/2,1]$ one has $\omega_-^{(\text{sym})}>0$, and moreover
$\omega_-^{(\text{sym})}\to 0^+$ as $\rho\downarrow \sqrt3/2$, which reflects the merger of the two
branches at the boundary of existence \eqref{eq:Delta_def_app}.
Collecting \eqref{eq:omega_as_app}, \eqref{eq:omega_eps_app} and \eqref{eq:omega_sym2_app}, the three
(one--loop) stability eigenvalues of the coupled fixed point $S_-$ in the subspace
$(\lambda_1,\lambda_2,\lambda_3)$ are
\begin{subequations}
\label{eq:omegas_final_app}
\begin{align}
&\omega_1=\eps,\\
&\omega_2=\frac{\eps}{3(4\rho^2+1)}\Big(3-4\rho^2+4\rho\sqrt{4\rho^2-3}\Big),\\
&\omega_3=\frac{2\eps}{4\rho^2+1}\Big(1+\rho\sqrt{4\rho^2-3}\Big). 
\end{align}
\end{subequations}

Therefore, for $\eps>0$ and $\rho>\sqrt3/2$ the fixed point $S_-$ is IR--attractive within the quartic
parity--symmetric sector, in agreement with the statement in Sec.~\ref{sec:partic}.
For $\rho=1$ one recovers the familiar values $\{\omega_1,\omega_2,\omega_3\}=\{\eps,\eps/5,4\eps/5\}$.

\subsection*{Stability with respect to parity--breaking quartics $\lambda_{4,5}$}

Still at $h_i=0$, consider infinitesimal parity--breaking couplings $\lambda_{4,5}$ around the
parity--symmetric fixed point $(\lambda_4^*,\lambda_5^*)=(0,0)$.
Linearizing the beta functions from Sec.~\ref{sec:partic} gives
\be
\frac{d}{dt}
\begin{pmatrix}\lambda_4\\ \lambda_5\end{pmatrix}
=
\mathbb{M}_{45}
\begin{pmatrix}\lambda_4\\ \lambda_5\end{pmatrix},
\qquad
\mathbb{M}_{45}=
-\eps\,\mathbf 1
+\frac{3}{16\pi^2}
\begin{pmatrix}
J_{11}\lambda_1^*+2J_{12}\lambda_3^* & J_{22}\lambda_3^*\\
J_{11}\lambda_3^* & J_{22}\lambda_2^*+2J_{12}\lambda_3^*
\end{pmatrix}.
\ee
Using $g_1^*=g_2^*=g^*$ and $g_3^*\neq 0$ one finds that the diagonal entries coincide and
simplify to $\eps/2$, while the off--diagonal product yields $\sqrt{(\mathbb M_{45})_{12}(\mathbb M_{45})_{21}}
=3a g_3^*$. Therefore the two eigenvalues in the $(\lambda_4,\lambda_5)$ sector are
\be
\label{eq:omega_45_app}
\omega_{4,5}^{(\pm)}=\frac{\eps}{2}\pm 3a\,g_3^*
=\frac{\eps}{2}\left(1\pm \frac{4\rho-\sqrt{4\rho^2-3}}{4\rho^2+1}\right).
\ee
For $\rho\ge\sqrt3/2$ both $\omega_{4,5}^{(\pm)}$ are non--negative, and for $\rho>\sqrt3/2$ they are
strictly positive. Hence the generalized Wilson--Fisher fixed point $S_-$ is also IR--stable with respect
to turning on small $\lambda_{4,5}$ (at one loop).

\subsection*{Stability of the fully coupled fixed line ($\lambda_i\neq 0$)}\label{app:stability_full}

In this appendix we analyze the linear stability of the fully interacting sector
in the $\mathbf{Z}_t=1,\;\mathbf{Y}=0$ limit, where all quartic couplings $\lambda_{1,\dots,5}$ are
allowed to be nonzero. The one--loop beta functions read
\begin{align}
\beta_{\lambda_1}&=-\epsilon\lambda_1+\frac{3}{16\pi^2}\Big(J_{11}\lambda_1^2+2J_{12}\lambda_4^2+J_{22}\lambda_3^2\Big),\nonumber\\
\beta_{\lambda_2}&=-\epsilon\lambda_2+\frac{3}{16\pi^2}\Big(J_{22}\lambda_2^2+2J_{12}\lambda_5^2+J_{11}\lambda_3^2\Big),\nonumber\\
\beta_{\lambda_3}&=-\epsilon\lambda_3+\frac{1}{16\pi^2}\Big(J_{11}(\lambda_1\lambda_3+2\lambda_4^2)+4J_{12}\lambda_3^2+J_{22}(\lambda_2\lambda_3+2\lambda_5^2)+2J_{12}\lambda_4\lambda_5\Big),\nonumber\\
\beta_{\lambda_4}&=-\epsilon\lambda_4+\frac{3}{16\pi^2}\Big(J_{11}\lambda_1\lambda_4+2J_{12}\lambda_3\lambda_4+J_{22}\lambda_3\lambda_5\Big),\nonumber\\
\beta_{\lambda_5}&=-\epsilon\lambda_5+\frac{3}{16\pi^2}\Big(J_{22}\lambda_2\lambda_5+2J_{12}\lambda_3\lambda_5+J_{11}\lambda_3\lambda_4\Big).
\label{eq:beta_lambda_full}
\end{align}
It is convenient to absorb the anisotropy weights by defining
\begin{align*}
&g_1\equiv J_{11}\lambda_1,\qquad 
g_2\equiv J_{22}\lambda_2,\qquad 
g_3\equiv \sqrt{J_{11}J_{22}}\,\lambda_3,\\
&g_4\equiv \sqrt{J_{11}J_{12}}\,\lambda_4,\qquad
g_5\equiv \sqrt{J_{22}J_{12}}\,\lambda_5,
\end{align*}
together with the single ratio
\begin{equation}
\rho\equiv \frac{J_{12}}{\sqrt{J_{11}J_{22}}}.
\end{equation}
In terms of $g_i$ the beta functions depend on the anisotropy only through $\rho$:
\begin{align}
\beta_{g_1}&=-\epsilon g_1+\frac{3}{16\pi^2}\Big(g_1^2+g_3^2+2g_4^2\Big),\nonumber\\
\beta_{g_2}&=-\epsilon g_2+\frac{3}{16\pi^2}\Big(g_2^2+g_3^2+2g_5^2\Big),\nonumber\\
\beta_{g_3}&=-\epsilon g_3+\frac{1}{16\pi^2}\Big((g_1+g_2)g_3+4\rho\,g_3^2+\frac{2}{\rho}(g_4^2+g_5^2)+2g_4g_5\Big),\nonumber\\
\beta_{g_4}&=-\epsilon g_4+\frac{3}{16\pi^2}\Big(g_1g_4+2\rho\,g_3g_4+g_3g_5\Big),\nonumber\\
\beta_{g_5}&=-\epsilon g_5+\frac{3}{16\pi^2}\Big(g_2g_5+2\rho\,g_3g_5+g_3g_4\Big).
\label{eq:beta_g_full}
\end{align}
Introducing $c\equiv \frac{16\pi^2}{3}\epsilon$ and $u_i\equiv g_i/c$, the flow becomes
$\beta_{u_i}=\epsilon\,\widehat\beta_i(u)$ with
\begin{align}
\widehat\beta_1&=-u_1+u_1^2+u_3^2+2u_4^2,\nonumber\\
\widehat\beta_2&=-u_2+u_2^2+u_3^2+2u_5^2,\nonumber\\
\widehat\beta_3&=-u_3+\frac13\Big((u_1+u_2)u_3+4\rho u_3^2+\frac{2}{\rho}(u_4^2+u_5^2)+2u_4u_5\Big),\nonumber\\
\widehat\beta_4&=-u_4+u_1u_4+2\rho u_3u_4+u_3u_5,\nonumber\\
\widehat\beta_5&=-u_5+u_2u_5+2\rho u_3u_5+u_3u_4.
\label{eq:beta_u_full}
\end{align}

Assuming $u_4,u_5\neq 0$, define $r\equiv u_5/u_4$. The conditions
$\widehat\beta_4=\widehat\beta_5=0$ give
\begin{equation}
u_1=1-u_3(2\rho+r),\qquad
u_2=1-u_3\Big(2\rho+\frac1r\Big).
\label{eq:u1u2_in_terms}
\end{equation}
Using $\widehat\beta_1=\widehat\beta_2=0$ together with $u_5^2=r^2u_4^2$ yields
\begin{equation}
u_3=\frac{1}{2\rho+r+\frac1r}.
\label{eq:u3_fixedline}
\end{equation}
Writing $D\equiv r^2+2\rho r+1$, the fixed line can be parameterized as
\begin{align}
\label{eq:fixedline_u}
&u_1^*=\frac{1}{D},\qquad
u_2^*=\frac{r^2}{D},\qquad
u_3^*=\frac{r}{D},\qquad
u_4^*=\sigma\,\frac{\sqrt{\rho r}}{D},\qquad
u_5^*=\sigma\,\frac{\sqrt{\rho r^3}}{D},
\end{align}
where
\begin{align}
    \sigma=\pm1,
\end{align}
and where $\widehat\beta_3=0$ enforces that $u_4^*$ and $u_5^*$ have the same sign
(i.e.\ the common $\sigma$ above).

Linearizing $u_i=u_i^*+\delta u_i$ gives
\begin{equation}
\frac{d}{d\ln\mu}\,\delta u_i=\epsilon\sum_{j}\Omega_{ij}\,\delta u_j,
\qquad
\Omega_{ij}\equiv \left.\frac{\partial \widehat\beta_i}{\partial u_j}\right|_{u=u^*}.
\end{equation}
A direct differentiation of \eqref{eq:beta_u_full} yields
\begin{align}
\label{eq:Omega_full}
&\Omega=\\
&=\begin{pmatrix}
-1+2u_1 & 0 & 2u_3 & 4u_4 & 0\\
0 & -1+2u_2 & 2u_3 & 0 & 4u_5\\
\frac{u_3}{3} & \frac{u_3}{3} & -1+\frac{u_1+u_2}{3}+\frac{8\rho u_3}{3} &
\frac{1}{3}\Big(\frac{4u_4}{\rho}+2u_5\Big) &
\frac{1}{3}\Big(\frac{4u_5}{\rho}+2u_4\Big)\\
u_4 & 0 & 2\rho u_4+u_5 & -1+u_1+2\rho u_3 & u_3\\
0 & u_5 & 2\rho u_5+u_4 & u_3 & -1+u_2+2\rho u_3
\end{pmatrix}_{u=u^*}\nonumber.
\end{align}
 Since $\widehat\beta(u)=-u+Q(u)$ with $Q$ homogeneous
quadratic, Euler's theorem implies $\Omega u^*=u^*$, giving an eigenvalue
$\widehat\omega=1$ (hence $\omega=\epsilon$ in the original variables).
Moreover, because the fixed points form a line $u^*(r)$, differentiating
$\widehat\beta(u^*(r))=0$ gives $\Omega\,\partial_r u^*(r)=0$, i.e.\ a marginal
eigenvalue $\widehat\omega=0$.

The remaining three eigenvalues can be extracted from trace invariants.
Substituting \eqref{eq:fixedline_u} into \eqref{eq:Omega_full} and evaluating
$\mathrm{tr}\,\Omega^n$ gives the universal results
\begin{equation}
\mathrm{tr}\,\Omega=-\frac{5}{3},\qquad
\mathrm{tr}\,\Omega^2=\frac{31}{9},\qquad
\mathrm{tr}\,\Omega^3=-\frac{35}{27},
\end{equation}
independent of $r$, $\rho$, and $\sigma$. Writing the spectrum as
$\{1,0,\alpha,\beta,\gamma\}$, Newton identities yield
\begin{equation}
\alpha+\beta+\gamma=-\frac{8}{3},\qquad
\alpha\beta+\alpha\gamma+\beta\gamma=\frac{7}{3},\qquad
\alpha\beta\gamma=-\frac{2}{3},
\end{equation}
so that
\begin{equation}
\widehat\omega^3+\frac{8}{3}\widehat\omega^2+\frac{7}{3}\widehat\omega+\frac{2}{3}
=0
\quad\Longleftrightarrow\quad
(\widehat\omega+1)^2(3\widehat\omega+2)=0.
\end{equation}
Therefore the full stability spectrum along the fixed line is
\begin{equation}
\widehat\omega\in\left\{1,\;0,\;-1,\;-1,\;-\frac{2}{3}\right\}
\qquad\Longrightarrow\qquad
\omega\in\left\{\epsilon,\;0,\;-\epsilon,\;-\epsilon,\;-\frac{2}{3}\epsilon\right\}.
\label{eq:stability_spectrum_full}
\end{equation}
Hence the fully coupled fixed line is a saddle line: one irrelevant direction
($+\epsilon$), one marginal direction ($0$), and three relevant directions
($-\epsilon,-\epsilon,-2\epsilon/3$). In particular, generic trajectories are
repelled from the line, leading to runaway flow in the fully coupled sector.

\subsection*{Cubic and mass perturbations around the fully coupled fixed line}
\label{app:cubic-mass-fixedline}

In this appendix we analyze the linearized RG flow of the cubic couplings \(h_i\) and
the quadratic sector around the fully coupled fixed line found in the quartic sector
(defined by \(\mathbf{Z}_t=\mathbf 1\) and \(\mathbf{Y}=0\)).
Throughout we work in \(D=4-\epsilon\).

In the decoupled-gradient limit the one-loop quartic flow depends on the three
positive angular averages \(J_{11},J_{22},J_{12}\) only through
\be
\rho \;\equiv\; \frac{J_{12}}{\sqrt{J_{11}J_{22}}}\,.
\ee
We also use the rescaled quartic couplings
\begin{align*}
&g_1=J_{11}\lambda_1,\qquad g_2=J_{22}\lambda_2,\qquad
g_3=\sqrt{J_{11}J_{22}}\,\lambda_3,\\
&g_4=\sqrt{J_{11}J_{12}}\,\lambda_4,\qquad
g_5=\sqrt{J_{22}J_{12}}\,\lambda_5,
\end{align*}
and the dimensionless variables
\be
c \;\equiv\; \frac{16\pi^2}{3}\,\epsilon,\qquad u_i\;\equiv\;\frac{g_i}{c}\,.
\ee
The fully coupled fixed line may be parameterized by \(r>0\) as
\be
u_1^\ast=\frac{1}{D},\qquad
u_2^\ast=\frac{r^2}{D},\qquad
u_3^\ast=\frac{r}{D},\qquad
u_4^\ast=\sigma\,\frac{\sqrt{\rho r}}{D},\qquad
u_5^\ast=\sigma\,\frac{\sqrt{\rho r^3}}{D},
\label{app:ufixedline}
\ee
where
\begin{align}
    D\equiv r^2+2\rho r+1,
\end{align}
with \(\sigma=\pm 1\) the sign choice \((u_4,u_5)\to-(u_4,u_5)\).  In particular, the
fixed-line variables obey the identities
\begin{subequations}
\label{app:ufixedline-identities}
\begin{align}
&u_2=r^2 u_1,\qquad u_3=r u_1,\qquad u_5=r u_4,\\
&u_1+u_2+2\rho u_3=1,\qquad u_4^2=\rho\,u_1u_3.
\end{align}
\end{subequations}

\label{app:cubic-sector}

\paragraph{One-loop flow and matrix form.}
In the \(\mathbf{Z}_t=\mathbf 1\), \(\mathbf{Y}=0\) limit, the one-loop beta functions for the cubic
couplings \(h_1,h_2,h_3,h_4\) read
\be
\beta_{h_1}= -\Bigl(1+\frac{\epsilon}{2}\Bigr)h_1
+\frac{3}{16\pi^2}\Bigl(J_{11}\lambda_1 h_1+2J_{12}\lambda_4 h_3+J_{22}\lambda_3 h_4\Bigr),
\ee
\be
\beta_{h_2}= -\Bigl(1+\frac{\epsilon}{2}\Bigr)h_2
+\frac{3}{16\pi^2}\Bigl(J_{22}\lambda_2 h_2+2J_{12}\lambda_5 h_4+J_{11}\lambda_3 h_3\Bigr),
\ee
\begin{align}
&\beta_{h_3}= -\Bigl(1+\frac{\epsilon}{2}\Bigr)h_3\nonumber\\
&+\frac{1}{16\pi^2}\Bigl(J_{11}(\lambda_1 h_3+2\lambda_4 h_1)
+4J_{12}\lambda_3 h_3
+2J_{12}\lambda_4 h_4
+J_{22}(\lambda_3 h_2+2\lambda_5 h_4)\Bigr),
\end{align}
\begin{align}
&\beta_{h_4}= -\Bigl(1+\frac{\epsilon}{2}\Bigr)h_4\nonumber\\
&+\frac{1}{16\pi^2}\Bigl(J_{22}(\lambda_2 h_4+2\lambda_5 h_2)
+4J_{12}\lambda_3 h_4
+2J_{12}\lambda_5 h_3
+J_{11}(\lambda_3 h_1+2\lambda_4 h_3)\Bigr).
\end{align}
These equations are linear in the \(h_i\) and can be written in matrix form
\be
\beta_{\bm h} \;=\; -\Bigl(1+\frac{\epsilon}{2}\Bigr)\bm h
+\frac{1}{16\pi^2}\,\mathbb{A}_h(\lambda,J)\,\bm h,
\qquad
\bm h\equiv (h_1,h_2,h_3,h_4)^{\sf T}.
\label{app:betah-matrixform}
\ee

It is convenient to remove explicit factors of \(J_{11}\) and \(J_{22}\) that appear as
square-root ratios in \(\mathbb{A}_h\).  Define the rescaled cubic variables
\be
H_1 \equiv h_1,\qquad
H_2 \equiv \sqrt{\frac{J_{12}J_{22}}{J_{11}^2}}\,h_2,\qquad
H_3 \equiv \sqrt{\frac{J_{12}}{J_{11}}}\,h_3,\qquad
H_4 \equiv \sqrt{\frac{J_{22}}{J_{11}}}\,h_4,
\label{app:H-rescaling}
\ee
so all together it reads
\begin{align}
    \bm H\equiv(H_1,H_2,H_3,H_4)^{\sf T}.
\end{align}
Since this is a \(\mu\)-independent linear transformation, it only conjugates the
linear flow matrix and does not affect its eigenvalues.
In terms of the quartic variables \(u_i=g_i/c\), the cubic flow becomes
\be
\beta_{\bm H}
\;=\;
-\Bigl(1+\frac{\epsilon}{2}\Bigr)\bm H
+\frac{\epsilon}{3}\,\mathbb{M}_h(u,\rho)\,\bm H,
\label{app:betaH}
\ee
with
\be
\mathbb{M}_h(u,\rho)
=
\begin{pmatrix}
3u_1 & 0 & 6u_4 & 3u_3\\[2pt]
0 & 3u_2 & 3u_3 & 6\rho\,u_5\\[2pt]
2u_4 & u_3 & u_1+4\rho u_3 & 2\rho u_4+2u_5\\[2pt]
u_3 & 2u_5/\rho & 2u_5+2u_4/\rho & u_2+4\rho u_3
\end{pmatrix}.
\label{app:Mh-matrix}
\ee

\paragraph{Evaluation on the fixed line and eigenvalues of \(\mathbb{M}_h\).}
We now set \(u=u^\ast(r)\) from \eqref{app:ufixedline}.
The fixed-line identities \eqref{app:ufixedline-identities} imply that
\(\mathbb{M}_h(u^\ast,\rho)\) has two independent null vectors, hence two zero
eigenvalues.  Concretely, using \eqref{app:ufixedline-identities} one checks
\be
\mathbb{M}_h(u^\ast,\rho)\,
\begin{pmatrix}
-2\sqrt{\rho r}\\[2pt]
-1/r\\[2pt]
1\\[2pt]
0
\end{pmatrix}
=0,
\qquad
\mathbb{M}_h(u^\ast,\rho)\,
\begin{pmatrix}
-r\\[2pt]
-2\rho\,u_4^\ast/u_3^\ast\\[2pt]
0\\[2pt]
1
\end{pmatrix}
=0,
\label{app:Mh-nullvectors}
\ee
so \(\mathrm{spec}(\mathbb{M}_h)\) contains \(\{0,0,\cdots\}\).

To determine the remaining two eigenvalues it is enough to compute two invariants.
First, the trace of \eqref{app:Mh-matrix} is
\be
\mathrm{tr}\,\mathbb{M}_h
=
3u_1+3u_2+(u_1+4\rho u_3)+(u_2+4\rho u_3)
=
4(u_1+u_2+2\rho u_3)
=4,
\label{app:Mh-trace}
\ee
where we used \(u_1+u_2+2\rho u_3=1\) on the fixed line.
Second, we compute \(\mathrm{tr}\,\mathbb{M}_h^2=\sum_{i,j}(\mathbb{M}_h)_{ij}(\mathbb{M}_h)_{ji}\).
From \eqref{app:Mh-matrix} one finds
\bea
\mathrm{tr}\,\mathbb{M}_h^2
&=&
(3u_1)^2+(3u_2)^2+(u_1+4\rho u_3)^2+(u_2+4\rho u_3)^2
\nonumber\\
&&
+ (6u_4)(2u_4) + (3u_3)(u_3) + (3u_3)(u_3) + (6\rho u_5)(2u_5/\rho)
\nonumber\\
&&
+ (2\rho u_4+2u_5)(2u_5+2u_4/\rho)\,.
\label{app:Mh-trace2-expanded}
\eea
Now imposing the fixed-line relations \eqref{app:ufixedline-identities} (in particular
\(u_2=r^2u_1\), \(u_3=ru_1\), \(u_5=ru_4\), and \(u_4^2=\rho u_1u_3\)) reduces
\eqref{app:Mh-trace2-expanded} to the constant
\be
\mathrm{tr}\,\mathbb{M}_h^2 \;=\; 10.
\label{app:Mh-trace2}
\ee
Let the two nonzero eigenvalues be \(\alpha,\beta\).  From \eqref{app:Mh-trace} and
\eqref{app:Mh-trace2} we obtain
\be
\alpha+\beta=4,\qquad \alpha^2+\beta^2=10
\quad\Longrightarrow\quad
\alpha\beta=\frac{(\alpha+\beta)^2-(\alpha^2+\beta^2)}{2}=\frac{16-10}{2}=3,
\ee
so \(\alpha,\beta\) are the roots of \(x^2-4x+3=0\), i.e.
\be
\mathrm{spec}\bigl(\mathbb{M}_h(u^\ast,\rho)\bigr)=\{0,0,1,3\}.
\label{app:Mh-spectrum}
\ee
Equivalently, the characteristic polynomial factorizes as
\be
\det\!\bigl(\mathbb{M}_h-x\mathbf 1\bigr)=x^2(x-1)(x-3).
\ee
Importantly, the spectrum \eqref{app:Mh-spectrum} is independent of the line parameter \(r\)
and of \(\rho\).

Linearizing \eqref{app:betaH} around \(\bm H=0\) on the fixed line, an eigenmode of
\(\mathbb{M}_h\) with eigenvalue \(x\in\{0,0,1,3\}\) scales with RG exponent
\be
\omega_h(x)
=
-\Bigl(1+\frac{\epsilon}{2}\Bigr) + \frac{\epsilon}{3}\,x,
\qquad (x\in\{0,0,1,3\}).
\label{app:omega-h}
\ee
Thus the four cubic eigenvalues are
\be
\omega_h=\Bigl\{-1-\frac{\epsilon}{2},\; -1-\frac{\epsilon}{2},\; -1-\frac{\epsilon}{6},\; -1+\frac{\epsilon}{2}\Bigr\}.
\label{app:omega-h-list}
\ee
All are negative for small \(\epsilon>0\), so any nonzero cubic couplings are relevant
perturbations of the quartic fixed line in \(4-\epsilon\) dimensions.  In particular, there
is no perturbative fixed point with \(h_i\neq 0\) controlled by the \(4-\epsilon\) expansion:
the equation \(\beta_{\bm H}=0\) admits only \(\bm H=0\) at this order.

\label{app:mass-sector}

To discuss stability it is natural to use dimensionless mass parameters
\be
m_{ij}^2 \;\equiv\; \frac{M_{ij}^2}{\mu^2}.
\ee
The one-loop RG equations in the \(\mathbf{Z}_t=\mathbf 1\), \(\mathbf{Y}=0\) limit can be written as
\begin{align}
\label{app:beta-m2-general}
&\beta_{m_{ij}^2}
=
-2\,m_{ij}^2
-\frac{1}{16\pi^2}\Bigl[\text{(linear terms in \(m^2\) proportional to \(\lambda\))}\Bigr]\nonumber\\
&+\frac{1}{16\pi^2}\Bigl[\text{(terms quadratic in \(h\))}\Bigr].
\end{align}
For the stability matrix around the critical manifold \(m^2=0\), \(h=0\) we may drop
the inhomogeneous \(h^2\) contribution and keep only the linear terms in \(m^2\).

As in the cubic sector, we perform a constant rescaling that removes square-root ratios of the
\(J\)'s.  Define the vector
\be
\bm{\mathcal{M}}
\;\equiv\;
\begin{pmatrix}
\mathcal{M}_1\\ \mathcal{M}_2\\ \mathcal{M}_3
\end{pmatrix}
\;\equiv\;
\begin{pmatrix}
m_{11}^2\\[2pt]
\sqrt{\frac{J_{22}}{J_{11}}}\,m_{22}^2\\[6pt]
\sqrt{2}\sqrt{\frac{J_{12}}{J_{11}}}\,m_{12}^2
\end{pmatrix}.
\label{app:m-rescaling}
\ee
Then the linearized mass flow takes the form
\be
\beta_{\bm{\mathcal{m}}}
=
-2\,\bm{\mathcal{m}}
-\frac{\epsilon}{3}\,\mathbb{M}_m(u,\rho)\,\bm{\mathcal{m}}
\qquad\text{(at linear order in \(m^2\) and with \(h=0\))},
\label{app:beta-matrix-m}
\ee
where the \(3\times 3\) matrix \(\mathbb{M}_m\) is
\be
\mathbb{M}_m(u,\rho)
=
\begin{pmatrix}
u_1 & u_3 & \sqrt{2}\,u_4\\[2pt]
u_3 & u_2 & \sqrt{2}\,u_5\\[2pt]
\sqrt{2}\,u_4 & \sqrt{2}\,u_5 & 2\rho\,u_3
\end{pmatrix}.
\label{app:Mm-matrix}
\ee

Setting \(u=u^\ast(r)\) and using \eqref{app:ufixedline-identities}, we first exhibit two null
vectors:
\be
\mathbb{M}_m(u^\ast,\rho)\,
\begin{pmatrix}
-r\\[2pt]
1\\[2pt]
0
\end{pmatrix}
=0,
\qquad
\mathbb{M}_m(u^\ast,\rho)\,
\begin{pmatrix}
-\sqrt{2}\sqrt{\rho r}\\[2pt]
0\\[2pt]
1
\end{pmatrix}
=0.
\label{app:Mm-nullvectors}
\ee
Hence \(\mathbb{M}_m\) has two zero eigenvalues.  The remaining eigenvalue is fixed by the trace:
\be
\mathrm{tr}\,\mathbb{M}_m
=
u_1+u_2+2\rho u_3
=1,
\label{app:Mm-trace}
\ee
so
\be
\mathrm{spec}\bigl(\mathbb{M}_m(u^\ast,\rho)\bigr)=\{0,0,1\},
\qquad
\det(\mathbb{M}_m-x\mathbf 1)=x^2(x-1).
\label{app:Mm-spectrum}
\ee
Again, the spectrum is independent of \(r\) and \(\rho\).

Combining \eqref{app:beta-matrix-m} and \eqref{app:Mm-spectrum}, the three mass eigenvalues are
\be
\omega_m(x)=-2-\frac{\epsilon}{3}x,
\qquad x\in\{0,0,1\},
\ee
i.e.
\be
\omega_m=\Bigl\{-2,\;-2,\;-2-\frac{\epsilon}{3}\Bigr\}.
\label{app:omega-m-list}
\ee
Therefore all quadratic deformations are relevant near the quartic fixed line, as expected for a
critical manifold in \(4-\epsilon\) dimensions.

For completeness we note that the full one-loop mass beta functions contain inhomogeneous terms
quadratic in the cubic couplings (schematically \(\beta_{m^2}\supset +h^2\)), so the subspace
\(m^2=0\) is not invariant if \(h\neq 0\).  Consequently, remaining on the critical surface of the
quartic fixed line requires tuning both \(m^2\) and \(h\) to zero.



\bibliographystyle{JHEP}

\end{document}